\documentclass[reprint,amsmath,amssymb,aps,prx,superscriptaddress, floatfix]{revtex4-2}
\usepackage{graphicx}
 \usepackage{enumerate}
\usepackage{gensymb}
\usepackage{mathtools}
\usepackage{float}
\usepackage{placeins}
\usepackage{wrapfig}

\newcommand{\ph}{{\hat{\rho}}}
\newcommand{\qzi}{{q_{z}^i}}
\newcommand{\qzij}{{q_{z,j}^i}}
\newcommand{\qz}{{q_{z}}}
\newcommand{\qzk}{{q_{z,k}}}

\newcommand{\fho}{\hat{f_1}}
\newcommand{\fht}{\hat{f_2}}

\newcommand{\eb}[1]{e^{2\pi i#1}}
\newcommand{\qp}{q_{||}}
\newcommand{\qpk}{q_{||,k}}
\newcommand{\qzit}{\tfrac{\qzi}{2}}
\newcommand{\mpp}{m_{||}}
\newcommand{\mz}{m_z}
\newcommand{\npp}{n_{||}}
\newcommand{\nz}{n_z}

\newcommand{\rp}{r_{||}}
\newcommand{\rz}{{r_{z}}}
\newcommand{\kp}{k_{||}}
\newcommand{\kz}{{k_{z}}}
\DeclareMathOperator*{\argmin}{arg\,min}
\usepackage{algpseudocode}
\usepackage{algorithm}
\usepackage{mathrsfs}
\DeclareMathSymbol{\shortminus}{\mathbin}{AMSa}{"39}
\usepackage{multirow}
\usepackage{pgfplots}
\usepgfplotslibrary{groupplots}
\pgfplotsset{compat=1.15}

\usepackage{mathrsfs}
\usepackage{siunitx}
\usepackage{xr}
\usepackage{microtype}
\allowdisplaybreaks
\renewcommand{\thetable}{\Roman{table}}
\usepackage{array}
\usepackage{booktabs}
\usepackage{url}

\usepackage{graphicx}

\begin{document}

\title{Nonuniform Iterative Phasing Framework and Sampling Requirements for 3D Dynamical Inversion from Coherent Surface Scattering Imaging}

\author{Jeffrey J. Donatelli}
\email{jjdonatelli@lbl.gov}
\affiliation{Mathematics Department, Lawrence Berkeley National Laboratory, Berkeley, CA USA 94720}
\affiliation{Center for Advanced Mathematics for Energy Research Applications, Lawrence Berkeley National Laboratory, Berkeley, CA USA 94720}
\author{Miaoqi Chu}
\affiliation{X-ray Science Division, Advanced Photon Source, Argonne National Laboratory, Lemont, IL USA 60439}
\author{Zixi Hu}
\affiliation{Mathematics Department, Lawrence Berkeley National Laboratory, Berkeley, CA USA 94720}
\affiliation{Center for Advanced Mathematics for Energy Research Applications, Lawrence Berkeley National Laboratory, Berkeley, CA USA 94720}
\author{Zhang Jiang}
\affiliation{X-ray Science Division, Advanced Photon Source, Argonne National Laboratory, Lemont, IL USA 60439}
\author{Nicholas Schwarz}
\affiliation{X-ray Science Division, Advanced Photon Source, Argonne National Laboratory, Lemont, IL USA 60439}
\author{Jin Wang}
\affiliation{X-ray Science Division, Advanced Photon Source, Argonne National Laboratory, Lemont, IL USA 60439}
\author{James A. Sethian}
\affiliation{Mathematics Department, Lawrence Berkeley National Laboratory, Berkeley, CA USA 94720}
\affiliation{Center for Advanced Mathematics for Energy Research Applications, Lawrence Berkeley National Laboratory, Berkeley, CA USA 94720}
\affiliation{Department of Mathematics, University of California, Berkeley, CA USA 94720}

\begin{abstract}
Coherent surface scattering imaging (CSSI) is an emerging experimental technique uniquely suited to probing the structure of thin nanostructures. In these experiments, a specimen is placed on a substrate, and a series of X-ray diffraction patterns is collected at grazing incidence angles as the specimen is rotated. However, reconstructing the specimen’s 3D structure from the data is challenging due to dynamical scattering effects induced by the experimental geometry and the lack of direct phase measurements. Specifically, the data involves nonuniformly sampled Fourier-transform values of the specimen density, and failure to effectively address this nonuniformity can lead to errors or degraded performance. Here we introduce a mathematical inversion framework that combines iterative-projection-based phasing techniques with new fast nonuniform Fourier inversion methods to efficiently reconstruct isolated 3D structures from their CSSI rotation-series data. We also analyze the theoretical properties of CSSI reconstruction to derive requirements on experimental parameters and characterize solution uniqueness. We validate our approach using CSSI data simulated from a conical Siemens star and a porous medium, demonstrating that high-resolution 3D structures can be reconstructed even in the presence of significant dynamical scattering, from data collected at as few as one or two incident angles. More broadly, the presented nonuniform reconstruction framework provides a foundation for solving challenging generalizations of the phase problem in which measurements involve nonlinear combinations of nonuniformly sampled Fourier values. 
\end{abstract}

\maketitle

\section{Introduction}
 
Coherent surface scattering imaging (CSSI), sometimes referred to as coherent grazing-incidence small-angle X-ray scattering, is an emerging experimental technique for probing nanoscale surface patterns. In this method, a highly coherent and monochromatic X-ray beam is directed onto a specimen in a reflection geometry at incident angles close to the substrate’s total external reflection angle. The resulting diffraction patterns are collected and subsequently processed using computational phase-retrieval algorithms to reconstruct the real-space structure of the specimen \cite{sun2012three}. 
 
Due to the unique geometry and short wavelength of X-rays, CSSI measurements are very sensitive to small surface variations in the specimen, making CSSI an ideal technique to probe the structure of thin material and biological nanostructures, such as planar nanoelectronics \cite{akinwande2014two}, polymer films \cite{tan2022x}, thin-film quantum dots \cite{brown2014energy}, biological membranes \cite{paracini2023structural}, and biofilms \cite{azulay2022multiscale}. Furthermore, CSSI data can be collected in a series of rotations at multiple incident angles to achieve isotropic in-plane resolution, capture 3D information, and improve depth sensitivity to resolve buried structures \cite{jiang2025design}. Recent advances in brightness and coherence at synchrotron light-source facilities are making high-resolution CSSI experiments feasible \cite{kerby2023advanced, shi2025measurements, chapman2023fourth, hellert2024status}.
 
However, reconstructing 3D specimen structure from CSSI is challenging due to missing phase information in the data and the presence of strong dynamical scattering effects induced by the grazing-incidence reflection geometry. For a sufficiently thin specimen, the dynamical scattering effects in CSSI can be described through the distorted-wave Born approximation (DWBA) \cite{sinha1988x}, which models the measured intensity as the squared magnitude of the sum of four combinations of reflection/refraction events from the substrate and specimen. When these dynamical scattering effects are sufficiently strong, the Born approximation is no longer valid, preventing the use of the iterative phasing reconstruction algorithms used in conventional coherent diffractive imaging \cite{gerchberg1972practical, fienup1978reconstruction, millane1990phase, millane2006recent, luke2004relaxed, elser2003phase, elser2007searching}. 
 
Additional practical complications arise from the inherent nonuniform sampling of the specimen's Fourier transform in the DWBA and of the data, due to the curvature of the Ewald sphere and rotation-series sampling. Treating this nonuniformity via conventional interpolation approaches can lead to errors that may limit the resolution of the reconstruction, while directly inverting nonuniform Fourier transforms through conventional iterative approaches can be computationally prohibitive.

In this paper, we focus on CSSI reconstruction of isolated 3D specimens that are sufficiently thin (less than a few hundred nanometers) and weakly scattering for the DWBA to be valid. This thickness range covers the principal science cases targeted by CSSI. For thicker or more strongly scattering specimens, it may be necessary to incorporate additional dynamical-scattering effects to accurately model the data \cite{venkatakrishnan2016multi, myint2023multislice, chu2023probing}. We also assume that the specimen is much smaller than the X-ray beam width. When a specimen exceeds the beam footprint, ptychographic CSSI \cite{myint2024three, sung2025automatic} can be deployed to collect diffraction patterns as the X-ray probe scans the specimen. Extensions of the framework presented in this paper to these regimes will be studied in a future manuscript.

Several approaches have been proposed for reconstructing isolated objects using grazing-incidence reflection geometries \cite{marathe2010coherent, roy2011lensless}. For reflection geometries where the diffraction signal is dominated by single scattering, conventional iterative phasing schemes were shown to be effective, including reconstruction of 3D nanoislands from simulated rotation-series data \cite{yefanov2009three} and reconstruction of an elongated 3D test object from experimental data acquired at a single specimen orientation and multiple incident angles \cite{sun2012three}. In \cite{kim2021inversion}, a convolutional neural network was used to directly predict the structure of 2D pentagonal geometric shapes from simulated CSSI datasets. Alternatively, \cite{liu2018} proposed estimating the Born approximation from DWBA intensities using a simplification that neglects cross-terms, which is valid in regimes where the dynamical contributions are sufficiently limited.

More recently, \cite{yang2023three} developed a new 3D iterative phasing algorithm that incorporates the full DWBA to invert CSSI data collected in a rotation series from several different incident angles. Their approach maps the rotation/incident-angle series data onto a uniform grid via standard interpolation and decouples the DWBA terms through a direct matrix inversion within the iterative procedure, resulting in a computational complexity of $O(N_xN_yN_z^2)$ for reconstruction on an $N_x \times N_y \times N_z$ grid. Consequently, this method requires sufficiently dense rotational sampling to control interpolation error and a relatively large number of incident angles to ensure the matrix inversion is well-conditioned (e.g., 32 angles in their simulations).
 
 Here, we develop a new nonuniform iterative phasing framework that is capable of inverting the full DWBA formulation to recover the electron density of an isolated specimen from its CSSI data in $O(N_xN_yN_z\log(N_xN_yN_z))$ time from a limited number of incident angles (typically requiring just one or two) and without introducing interpolation errors due to the use of nonuniform Fourier transforms. This approach utilizes nonuniform fast Fourier transforms (NUFFTs) \cite{dutt1993fast,greengard2004accelerating} coupled with advanced projection operators to decouple the multiple scattering events from the data in a way that directly generalizes established iterative phasing methods. We also develop a new linear data reduction algorithm, based on a generalization of the Wiener-Khinchin theorem \cite{wiener1930generalized}, that can efficiently denoise CSSI data and reduce it to a compact representation that can further accelerate the reconstruction. Furthermore, we study the theoretical properties of CSSI reconstruction to derive experimental parameter requirements and establish uniqueness properties of the solution. 

The rest of the paper is organized as follows. In Section~\ref{sec:background}, we describe the basic geometry of CSSI experiments, define our notation, and provide background on nonuniform Fourier transforms and iterative phasing techniques. In Section~\ref{sec:overview}, we provide a high-level overview of our proposed CSSI reconstruction framework. In Section~\ref{sec:reduction}, we introduce a linear data reduction approach that can be used to filter noise and accurately resample nonuniformly sampled CSSI rotation-series data onto computationally efficient 3D staggered-grid representations that improves reconstruction stability. In Section~\ref{sec:acceleration}, we develop approaches to efficiently directly invert nonuniform Fourier transforms, stabilize inversion from incomplete Fourier information, and accelerate both the forward and inverse NUFFT on staggered grids. In Section~\ref{sec:NIP}, we present a new iterative phasing algorithm that couples direct NUFFT inversion with advanced projection operations that utilize the full DWBA formulation to accurately and efficiently reconstruct 3D structure from CSSI rotation-series data collected at one or more incident angles. In Section~\ref{sec:uniqueness}, we derive experimental parameter requirements and uniqueness properties for CSSI reconstruction. In Section~\ref{sec:results}, we demonstrate the performance of these new techniques on CSSI data simulated using several different sets of parameters. 
 
 \section{Background}\label{sec:background}
 Here we describe the basic geometry and notation used throughout the paper, formulate the relation between the imaged specimen and the observed CSSI intensity data, and review background on nonuniform Fourier transforms and iterative phasing algorithms, which we build upon in this paper.
 
 \subsection{Basic Geometry and Notation}\label{sec:basic}
 
 \begin{figure}[h!] 
 \centering
\includegraphics[width = \columnwidth]{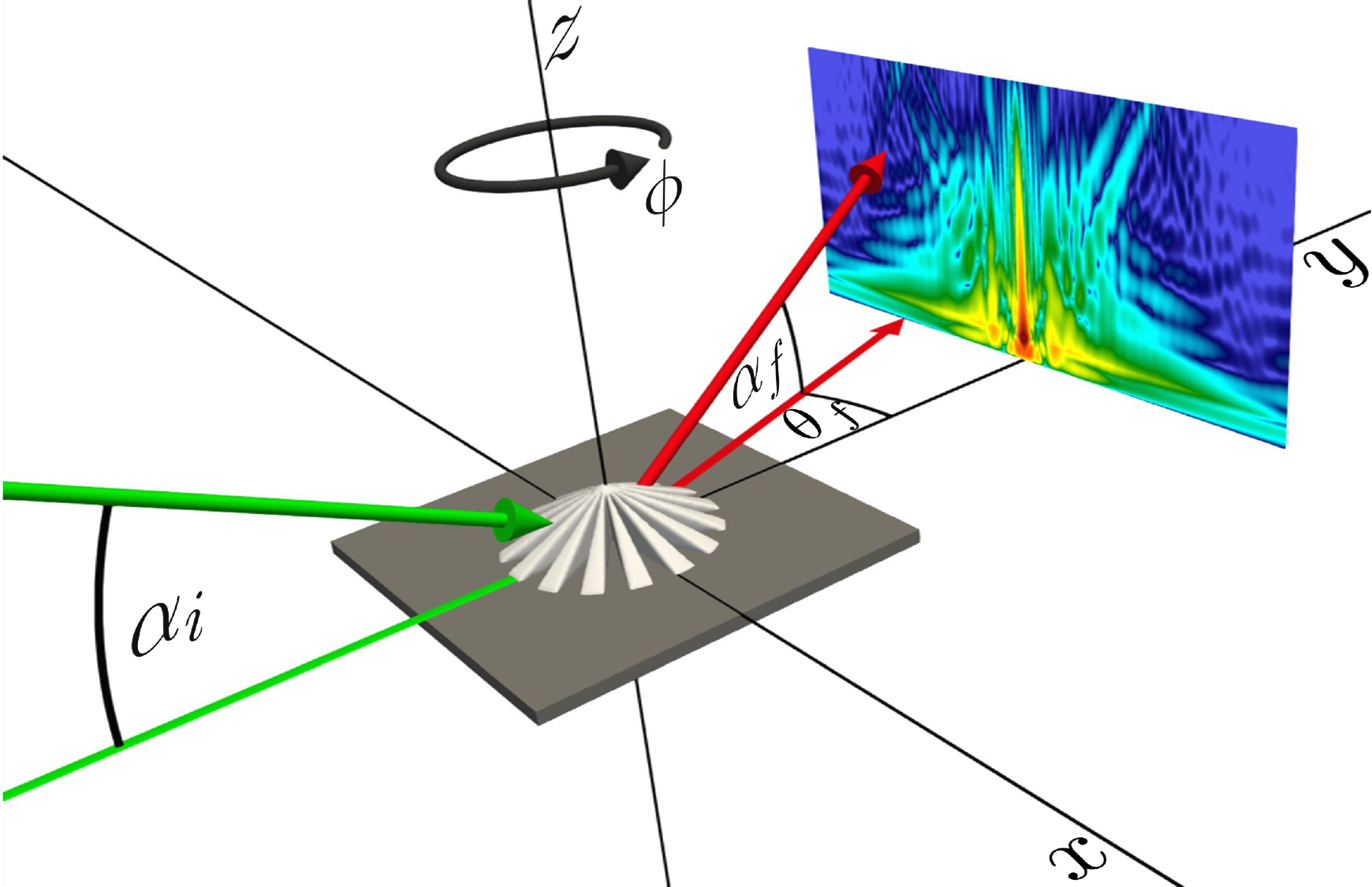}
\caption{Experimental CSSI geometry. $\alpha_i$ is the angle between the substrate surface and the incident beam. Each pixel in a detector is described by two angles: $\alpha_f$, which is the exit angle between the reflected beam and the substrate, and $\theta_f$, which is the scattering angle between the reflected beam and the $yz$ plane. Multiple diffraction patterns are collected in a rotation series by applying an in-plane rotation of angle $\phi$ to the specimen. }\label{fig:geometry}
\end{figure}
 
 Here we work in a coordinate system where the $x$ and $y$ axes are parallel to the substrate surface, with $y$ being parallel to the incident beam projected along the surface, and the $z$ axis is normal to the surface. The scattering geometry is described in Fig.~\ref{fig:geometry} and is defined in terms of four angles: 1) $\alpha_i \geq 0$ is the angle between the incident beam and the $xy$ plane; 2) $\alpha_f \geq 0$ is the angle between the reflected beam (i.e.\@ the vector pointing from the beam-specimen interaction point to a pixel in the detector) and the $xy$ plane; 3) $\theta_f$ is the signed angle between the reflected beam and the $yz$ plane; and 4) $\phi$ is the angle of the specimen in-plane rotation. The collection of pixels in a single CSSI diffraction pattern samples a range of different $\alpha_f$ and $\theta_f$, $\phi$ is varied via in-plane rotation of the specimen, and $\alpha_i$ is varied by changing the direction of the incident beam relative to the specimen.
 
To simplify the forthcoming notation and derivations, we use the crystallography convention to define the incident wavevector for a beam with wavelength $\lambda$ as
 \begin{equation}\label{eq:ki}
 \boldsymbol\kappa^i = (\kappa^i_x,\, \kappa^i_y,\, \kappa^i_z) = \frac{1}{\lambda}\big(0,\, \cos(\alpha_i),\, -\sin(\alpha_i)\big),
 \end{equation}
 and we define the reflected wavevector for a pixel with exit angle $\alpha_f$ and azimuthal scattering angle $\theta_f$ as
 \begin{equation}\label{eq:kf}
 \begin{aligned}
\boldsymbol \kappa^f &= (\kappa^f_x,\, \kappa^f_y,\, \kappa^f_z)\\
 &= \frac{1}{\lambda}\big(\sin(\theta_f)\cos(\alpha_f), \cos(\theta_f)\cos(\alpha_f), \sin(\alpha_f)\big).
 \end{aligned}
 \end{equation}
 
We denote 3D coordinates in real space as $\mathbf r$ and in Fourier space (also known as reciprocal space) as $\mathbf q$. To simplify some of the forthcoming expressions, we separate out the $z$-component by expressing $\mathbf q = (\qp,\qz)$, where $\qp = (q_x,q_y)$. Each pixel of a diffraction pattern has an associated 3D coordinate in Fourier space, which, for an in-plane specimen rotation of $\phi$, is defined via
 \begin{equation}\label{eq:q}
\qp = \begin{bmatrix} \cos(\phi) & -\sin(\phi) \\ \sin(\phi) & \cos(\phi)\end{bmatrix}\left(\kappa_{||}^f - \kappa_{||}^i\right),\quad \qz = \kappa_z^f - \kappa_z^i.
\end{equation}
For a fixed $\phi$ and $\alpha_i$, Eq.~\ref{eq:q} traces out a spherical surface in Fourier space, referred to as the Ewald sphere, as $\alpha_f$ and $\theta_f$ are varied.

Given a function $f(\mathbf r)$, we define the set-theoretic support ${\rm supp}(f)$ as the set of inputs where $f$ is nonzero:
\begin{equation}
    {\rm supp}(f) = \{\mathbf r: f(\mathbf r) \neq 0\}.
\end{equation}

We will often use notation of the form $f[k]$ and $f[\mathbf m]$ to emphasize that the function $f$ has integer scalar $k$ and integer vector $\mathbf m$ inputs, respectively. We typically reserve $\mathbf m$ and $\mathbf n$ for real-space inputs and $k$ for Fourier-space inputs.

We use $D_d$ to denote an upper bound on the specimen extent in the $d$-th dimension and $D_{||}$ to denote an upper bound on twice the horizontal radius of the specimen, i.e.,
\begin{equation}
    D_{||} \geq 2 \min_{c\in \mathbb R^2}\max_{\mathbf r\in {\rm supp}(\rho)}|r_{||}-c|.
\end{equation}

To fix notation, we take all grid sizes to be odd. Given an upper bound $q_d^{\max}$ on the Fourier-coordinates sampled in the $d$-th dimension, we choose the specimen grid size $N_d$ for that dimension to be the smallest odd integer greater than the corresponding number of Nyquist elements within the bound (i.e., the minimum number of grid points needed to resolve the Fourier transform up to $q_d^{\max}$ without aliasing):
\begin{equation}
    N_d = 2\lceil{q}^{\max}_d{D}_d\rceil +1.
\end{equation}

To avoid introducing additional notation, we use the same symbols to denote coordinates in both physical and nondimensionalized units. In particular, we assume all computational coordinates have been nondimensionalized so that $-\tfrac{1}{2} \leq q_d \leq \tfrac{1}{2}$ for $d=x,y,z$. Given dimensional quantities $q_d^{\rm phys}$, $r_d^{\rm phys}$, and $D_d^{\rm phys}$ expressed in consistent physical units, the corresponding nondimensionalized quantities are given by 
\begin{equation}
\begin{aligned}
q_d &= \frac{q_d^{\rm phys} D_d^{\rm phys}}{N_d-1},\\
D_d &= N_d-1,\\
r_d &= \frac{r_d^{\rm phys} (N_d-1)}{D_d^{\rm phys}}.
\end{aligned}
\end{equation}
$\lambda$ can also be nondimensionalized similarly to $r_d$, but this has the inconvenience of requiring a different value for each dimension, unless the grid resolutions $D^{\rm phys}_d/(N_d-1)$ are the same for each dimension. However, we do not require a nondimensionalized $\lambda$ for any of our computational coordinates.

We denote the real-space electron density of a given specimen as $\rho(\mathbf r)$. This paper focuses mainly on the case where $\rho$ is real-valued, allowing us to exploit additional symmetries and simplifications, but the core theory and algorithms can also be generalized to a complex-valued density.

We define the Fourier transform $\hat{\rho}(\mathbf q)$ using the convention
  \begin{equation}\label{eq:Fourier_continuous}
 \hat{\rho}(\mathbf q) = \int_{\mathbb R^3}\rho(\mathbf r)e^{-2\pi i \mathbf q \cdot \mathbf r}d\mathbf r.
 \end{equation}
When $\rho$ is defined on an $N_x \times N_y \times N_z$ uniform grid, this notation instead denotes the discrete Fourier transform, i.e.,
\begin{equation}\label{eq:Fourier_discrete}
\hat{\rho}(\mathbf q) = \sum_{\mathbf m} \rho[\mathbf m]e^{-2\pi i \mathbf q \cdot \mathbf m},
\end{equation}
where $\mathbf m = (m_x, m_y, m_z)$ is an integer vector whose components take values in the ranges $0 \leq m_d \leq N_d-1$, where $d \in \{x,y,z\}$. As an illustration of the nondimensionalization discussed above, note that Eq.~\ref{eq:Fourier_continuous} is independent of the coordinate units, whereas Eq.~\ref{eq:Fourier_discrete} requires $\mathbf q$ to be nondimensionalized since $\mathbf m$ is unitless.

The Nyquist length for the $d$-th dimension is $1/D_d$, which is a tight upper bound on the Fourier-space sampling interval required to resolve the Fourier transform of an object with extent $D_d$ in that dimension. We refer to a Nyquist element/pixel/voxel as the 1D interval/2D rectangle/3D box in Fourier space whose side lengths are equal to the Nyquist lengths in each dimension.

Given two functions $f$ and $g$, we denote their convolution and cross-correlation as
\begin{equation}
\begin{aligned}
f \ast g (\mathbf r) &= \int_{\mathbb R^3} f(\mathbf t)g(\mathbf r - \mathbf t)d\mathbf t,\\
f \star g (\mathbf r) &= \int_{\mathbb R^3} \overline{f(\mathbf t)}g(\mathbf r + \mathbf t)d\mathbf t.
\end{aligned}
\end{equation}
For functions defined on a grid, we use discrete analogs of the above convolution and cross-correlation by replacing the integration with a summation.

An important identity that we refer to later is the Wiener-Khinchin theorem \cite{wiener1930generalized}, which states that the Fourier transform of the autocorrelation of a function is equal to the squared magnitude of that function's Fourier transform, i.e.,
\begin{equation}
    \widehat{\rho \star \rho} = |\hat{\rho}|^2.
\end{equation}
The Wiener-Khinchin theorem holds for both continuous and discrete functions.

 \subsection{CSSI Scattering Formulation and Reconstruction Problem}
 
\begin{figure}[h!] 
 \centering
\includegraphics[width = .475\textwidth]{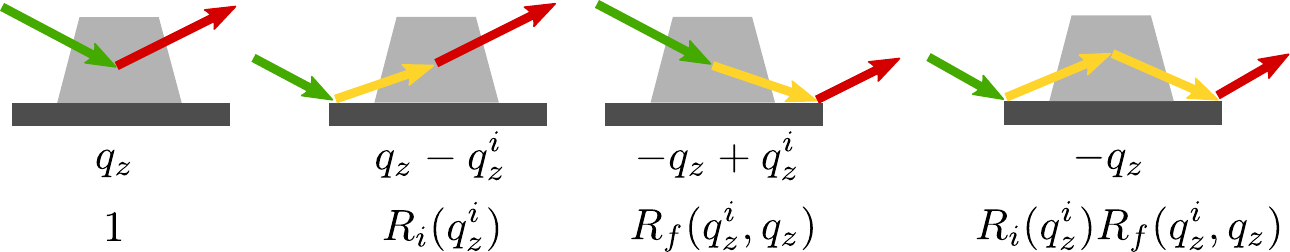}
\caption{Four scattering events (Top) modeled by DWBA, with corresponding $z$ components (Middle) and reflection coefficients (Bottom). The incident beam (green) is scattered off a combination of the specimen (light gray) and substrate (dark gray), resulting in intermediate reflections (yellow) that influence the scattered beam (red). }\label{fig:DWBA}
\end{figure}

 The grazing-incidence reflection geometry of CSSI experiments can induce strong dynamical scattering effects in the data, which can be described via combinations of reflection and refraction events for both the specimen and substrate. For a sufficiently thin specimen, there are four main scattering events that have significant contributions to the CSSI intensity data (see Fig.~\ref{fig:DWBA}). In this case, the data can be accurately modeled by the distorted-wave Born approximation (DWBA) \cite{vineyard1982grazing, sinha1988x}, which we now describe.

 The scattering predicted by DWBA depends on the complex refractive index $n_r$ of the substrate at the given X-ray wavelength. In particular, the refractive index is used to define the vertical components of the incident and reflected transmitted waves through the substrate as
 \begin{equation}\label{eq:k}
 \tilde{\kappa}_z^i = -\frac{1}{\lambda}\sqrt{n_r^2 - \cos^2(\alpha_i)} \text{ and } \tilde{\kappa}_z^f = \frac{1}{\lambda}\sqrt{n_r^2 - \cos^2(\alpha_f)},
 \end{equation}
where the complex square root is taken to be the principal branch. 

\begin{figure}
 \centering
 \includegraphics[width = \columnwidth]{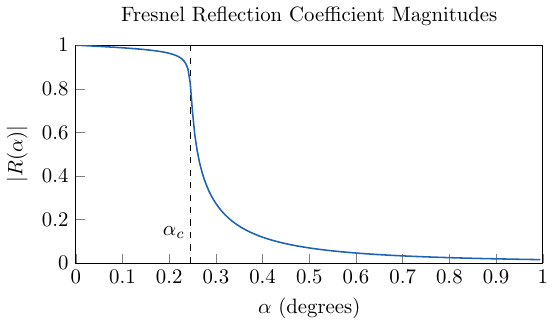}
\caption{Example of the Fresnel reflection coefficient magnitude $|R(\alpha)|$ as a function of the incident/exit angle $\alpha$ for a silicon substrate at $\lambda = 1.7$ \AA. The magnitude of the reflection coefficient decreases rapidly near the critical angle $\alpha_c$ (dashed line).  }\label{fig:Ri}
\end{figure}

The Fresnel reflection coefficients associated with these components are given by 
 \begin{equation}\label{eq:r}
  R_i = \frac{\kappa_z^i - \tilde{\kappa}_z^i}{\kappa_z^i + \tilde{\kappa}_z^i} \text{ and }  R_f = \frac{\kappa_z^f - \tilde{\kappa}_z^f}{\kappa_z^f + \tilde{\kappa}_z^f}.
 \end{equation}
In general, $|R_i|$ and $|R_f|$ are approximately $1$ for very small $\alpha$, decay very slowly as $\alpha$ increases until it approaches the critical angle $\alpha_c=\sqrt{2\text{Re}(1-n_r)}$ radians (where Re refers to the real part), drop rapidly near the critical angle, and then gradually decay towards zero as $\alpha$ is further increased (see Fig.~\ref{fig:Ri} for an example).

In Eq.~\ref{eq:r}, the dependence on $\alpha$ has been suppressed to simplify notation. However, it will be convenient to express these dependencies as functions of the vertical coordinate $q_z$ and incident-angle dependent shift $\qzi = -2\kappa_z^i$. Specifically, we use $R_i(q_z^i)$ and $R_f(q_z^i, q_z)$ to denote the reflection coefficients calculated using  $\alpha_i = \sin^{-1}(\lambda \qzi/2)$ and $\alpha_f = \sin^{-1}(\lambda(2q_z - \qzi)/2)$ in Eqs.~\ref{eq:ki}-\ref{eq:kf} and \ref{eq:k}-\ref{eq:r}. When considering a fixed incident angle, we suppress the dependence on $\qzi$, simplifying the above notation to $R_i$ and $R_f(\qz)$.
 
The DWBA describing the CSSI intensity data $I$ for a specimen with electron density $\rho$ can now be formulated as
\begin{equation}\label{eq:CSSI}
\begin{aligned}
 I(\qp,\qz,\qzi) = &\big|\ph(\qp,\qz) + R_i(\qzi)\ph(\qp,\qz - \qzi) \\
 &+ R_f(\qzi,\qz)\ph(\qp,-\qz + \qzi)\\
 &+ R_i(\qzi)R_f(\qzi,\qz)\ph(\qp,-\qz)\big|^2,
\end{aligned}
\end{equation}
where, for each CSSI diffraction pattern, the $\qp$ and $\qz$ are determined by the pixel positions, in-plane orientation, and incident angle via Eqs.~\ref{eq:ki}-\ref{eq:q}, and $\qzi$ depends only on the incident angle via Eq.~\ref{eq:ki}. The first term in the summation is the Born approximation term, describing scattering directly off the specimen, while the remaining terms describe the three other scattering events in Fig.~\ref{fig:DWBA}, consisting of combinations of reflection and refraction for both the specimen and substrate.  

Eq.~\ref{eq:CSSI} describes a series of 3D CSSI diffraction volumes (each indexed by $\qp$ and $\qz$) for different incident beam angles (indexed by $\qzi$). Each of these diffraction volumes can be sampled within a cylindrical region in Fourier space by collecting the CSSI data in a rotation series. However, the set of $\qz$ that can be sampled could potentially be restricted by the incident angle and detector geometry via Eqs.~\ref{eq:ki}-\ref{eq:q}. 
 
To condense our notation, we assume the data is collected at $J$ discrete incident angles $\alpha_{i,j}$, $j=1,\hdots, J$, and index the corresponding incident-angle-dependent shifts as $\qzij = 2\sin(\alpha_{i,j})/\lambda$ and the CSSI diffraction volumes as $I_j(\qp,\qz) = I(\qp,\qz,\qzij)$. It will also be convenient to index the four components in Eq.~\ref{eq:CSSI} by $s=1,\hdots,4$:
\begin{equation}\label{eq:4J}
\hat{\rho}_{j,s}(\qp,\qz) = \begin{cases}\hat{\rho}(\qp,\qz), & \text{ if }s = 1, \\
\hat{\rho}(\qp,\qz-\qzij), &  \text{ if }s = 2, \\
\hat{\rho}(\qp,-\qz+\qzij), &  \text{ if }s = 3, \\
\hat{\rho}(\qp,-\qz), &  \text{ if }s = 4,
\end{cases}
\end{equation}
\begin{equation}\label{eq:4JR}
R_{j,s}(\qz) = \begin{cases}1, & \text{ if }s = 1, \\
R_i(\qzij), &  \text{ if }s = 2, \\
R_f(\qzij,\qz), &  \text{ if }s = 3, \\
R_i(\qzij)R_f(\qzij,\qz),&  \text{ if }s = 4.
\end{cases}
\end{equation}
To simplify notation, the dependencies on $j$ and $s$ are suppressed in cases where we are considering a fixed incident angle or referring to Fourier values directly as functions of $\mathbf q$ by itself.

The CSSI reconstruction problem is to recover a density $\rho$ that is consistent with the measured $I_j(\qp,\qz)$ (i.e., satisfies Eq.~\ref{eq:CSSI}) and satisfies known physical constraints, such as compact support. This results in a challenging problem since 1) phases are not measured; 2) the Fourier values need to be decoupled from the four-term DWBA summations; 3) the problem is high-dimensional and nonconvex with numerous local minima; and 4) the measured data does not lie on any natural structured grid due to the rotation-series sampling, curvature of the Ewald sphere, potential use of multiple incident angles, and shift operations in the DWBA summation. 

Unlike in the classical phase problem \cite{fienup1978reconstruction}, where only the Born approximation term is significant, the CSSI intensity in Eq.~\ref{eq:CSSI} is generally not invariant to translations of the density along the $z$ dimension. Therefore, it is essential to ensure that the grid-centering conventions used for the computational grid in the reconstruction and in any utilized fast Fourier transform (FFT) and NUFFT packages are consistent with the experimental geometry. In particular, the grids should be centered so that their vertical extent ranges from $z=0$ to $z=D_z$.

In Sections~\ref{sec:overview}-\ref{sec:NIP}, we describe a new algorithmic reconstruction framework that can meet these challenges and solve the CSSI reconstruction problem. This new framework is based on efficient forward and inverse calculations of nonuniform Fourier transforms, as well as iterative phasing techniques, which we review in the following subsections.

\subsection{Nonuniform Discrete Fourier Transform and Nonuniform Fast Fourier Transform}

In this paper, it will be important to compute the Fourier transform at a set of nonuniformly sampled Fourier coordinates from a function defined on an $N_x\times N_y \times N_z$ uniform grid, a process known as the nonuniform discrete Fourier transform (NDFT). Given a set of Fourier coordinates $\{\mathbf q_k\}_{k=1}^K$, the NDFT $F$ of a function $\rho$ defined on the grid is
\begin{equation}\label{eq:NUDFT}
(F\rho)[k] = \sum_{\mathbf m} \rho[\mathbf m]e^{-2\pi i \mathbf q_k\cdot\mathbf m }.
\end{equation}
Throughout this paper, we use the term ``set'' loosely to allow for repeated elements. In particular, the Fourier coordinates $\{\mathbf q_k\}_{k=1}^K$ in Eq.~\ref{eq:NUDFT} need not be distinct.

The corresponding adjoint NDFT operator $F^*$ of a function $b[k]$ is given by
\begin{equation}\label{eq:NUDFT*}
    (F^*b)[\mathbf m] = \sum_k b[k]e^{2\pi i  \mathbf q_k\cdot\mathbf m}.
\end{equation}
Note that $F^*$ is not proportional to the inverse of $F$ when the set of Fourier coordinates $\{\mathbf q_k\}_{k=1}^K$ does not form a grid uniformly spaced in each dimension.

Although directly computing Eqs.~\ref{eq:NUDFT}-\ref{eq:NUDFT*} has complexity $\mathcal O(KN_xN_yN_z)$, they can be efficiently approximated to within a relative error of $\epsilon$, by using a nonuniform fast Fourier transform (NUFFT) \cite{dutt1993fast,fessler2003nonuniform,greengard2004accelerating}, which has complexity $\mathcal O(N_xN_yN_z\log(N_xN_yN_z) + K\log^3(1/\epsilon))$ in three dimensions.

There are two main types of NUFFT: 1) type-1, which maps a nonuniform point set to a uniform grid, and 2) type-2, which maps a uniform grid to a nonuniform point set. Both transform types can represent either a transformation from real to Fourier space or vice versa by changing the sign of the exponential in Eq.~\ref{eq:NUDFT}. For example, Eq.~\ref{eq:NUDFT} can be approximated using a type-2 NUFFT with a negative sign in the exponential, and Eq.~\ref{eq:NUDFT*} can be approximated using a type-1 NUFFT with a positive sign in the exponential.

The type-2 NUFFT is computed by dividing the real-space grid values by the Fourier transform of a spreading kernel (most commonly Gaussian, Kaiser-Bessel, or exponential of a semicircle), computing the fast Fourier transform (FFT) of the result on the uniform grid, and then evaluating a truncated convolution between the FFT output and the spreading kernel on the nonuniform grid. The type-1 transform is computed by reversing the order of these operations. All NUFFT calculations in this paper are performed using the Flatiron Institute NUFFT (FINUFFT) library \cite{barnett2019parallel,barnett2021aliasing}.

\subsection{Nonuniform Fourier Inversion and Multilevel Toeplitz Acceleration}\label{sec:INUDFT}

In addition to efficiently computing the NDFT, we will need fast methods to invert it. However, note that, unlike Fourier data defined on uniform grids, a general function defined on a nonuniform set of Fourier points is not necessarily the NDFT of a real-space function defined on a finite uniform grid, i.e., the NDFT may not be an invertible linear transformation. Therefore, in this context, inversion is instead defined as finding a least-squares solution, assuming that sufficient $\mathbf q_k$ are sampled for the NDFT to have full rank. More specifically, given a set of Fourier values $b$ and weights $w$ defined on a nonuniform set of coordinates in Fourier space, the inverse NDFT is defined as the solution to 
\begin{equation}\label{eq:Fp-b}
    \argmin_\rho ||F\rho - b||_w,
\end{equation}
where $||f||_w = \sqrt{\sum_k |f[k]|^2 w[k]}$ is the weighted $\ell^2$ norm. 

The weights can, for example, be chosen to approximate the inverse sampling density to improve convergence of linear solvers applied to Eq.~\ref{eq:Fp-b}, set to the inverse noise variance to downweight noisy data, or simply set to unity. Also, as shown in \cite{kircheis2023fast} and Section~\ref{sec:DirectInversion}, the weights can be optimized to simplify and accelerate the inversion.

Eq.~\ref{eq:Fp-b} is equivalent to solving the normal equations
\begin{equation}\label{eq:normal}
    F^*WF\rho = F^*Wb,
\end{equation}
where $W={\rm diag}(w[1],w[2],\hdots,w[K])$ is the diagonal matrix of weights.

Eq.~\ref{eq:normal} can be efficiently solved by applying iterative linear solvers based on Krylov subspace methods, such as conjugate gradient. These methods require repeated multiplication by $F^*WF$. Although this multiplication can be computed with two NUFFTs, it can be significantly accelerated by using multilevel Toeplitz acceleration \cite{dutt1993fast,Wajer2001,ou2017gnufftw, wang2013fourier}. In particular, $F^*WF$ has multilevel Toeplitz symmetry, which means that the corresponding matrix-vector multiplication can be represented as a discrete convolution that can be efficiently calculated using the convolution theorem coupled with FFTs. In particular, given $\rho$ defined on an $N_x \times N_y \times N_z$ grid, the normal equations in Eq.~\ref{eq:normal} can be efficiently represented as follows. 

First, define the weighted point-spread function $Q$ for the set of nonuniformly sampled Fourier coordinates as
\begin{equation}\label{eq:Q}
    Q[\mathbf n] = \sum_k w[k]e^{2\pi i \mathbf q_k\cdot\mathbf n  }, \quad n_d = -N_d+1,\hdots, N_d-1.
\end{equation}
Note that $Q$ is defined on a larger grid with dimensions $(2N_x-1) \times (2N_y-1) \times (2N_z-1)$. The left-hand side of Eq.~\ref{eq:normal} can then be represented as a convolution:
\begin{equation}\label{eq:F^*F}
\begin{aligned}
    F^*WF\rho[\mathbf n] &= \sum_{\mathbf m}\rho[\mathbf m]Q[\mathbf n - \mathbf m]\\
    &= (\rho \ast Q)[\mathbf n],
    \end{aligned}
\end{equation}
where $n_d = 0,\hdots, N_d-1$. 

Eq.~\ref{eq:F^*F} can be computed in $\mathcal O(N'\log N')$ time via the convolution theorem, where $N'=(2N_x-1)(2N_y-1)(2N_z-1)$. Specifically, this is accomplished by zero-padding $\rho$ from an $N_x \times N_y \times N_z$ grid to a grid of size at least $(2N_x-1)\times(2N_y-1)\times(2N_z-1)$, computing the FFT of the zero-padded $\rho$, multiplying the resulting Fourier values pointwise by the FFT of $Q$, computing the inverse FFT (IFFT) of the product, and then restricting the result back to the original grid, i.e.,
\begin{equation}\label{eq:FCT}
\rho \ast Q = \operatorname{Res}\Big[\operatorname{IFFT}\Big(\operatorname{FFT}\big[\operatorname{Pad}(\rho)\big] \odot \operatorname{FFT}\big[Q\big]\Big)\Big],
\end{equation}
where $\odot$ denotes element-wise multiplication, and Pad and Res are the corresponding zero-padding and restriction operations described above.

Since $Q$ depends only on the Fourier coordinates, its FFT can be precomputed and reused for different Fourier data $b$ as long as the sample coordinates remain fixed. When solving Eq.~\ref{eq:normal}, $F^*Wb$ only needs to be computed via an NUFFT once at the start of the iterative solver. In three dimensions, computing $F^*WF$ via Eqs.~\ref{eq:F^*F}-\ref{eq:FCT} is typically a few orders of magnitude faster than computing it using two NUFFT operations, for example, when $K \gtrsim N_xN_yN_z$ and $\epsilon$ is on the order of machine precision.

Although the above theory holds for a complex-valued $\rho$, since we assume $\rho$ to be real-valued in this paper, it is useful to have a complex-to-real nonuniform Fourier inversion to help constrain the problem and speed up the computation. Note that $\rho$ being real valued is equivalent to its Fourier transform satisfying Friedel symmetry, i.e., $\hat{\rho}(-\mathbf q) = \overline{\hat{\rho}(\mathbf q)}$. This symmetry implies that for each sample point $\mathbf q_k$, we also have Fourier data at $-\mathbf q_k$. Therefore, we can take our sampling set to be the union $\{\mathbf q_k\}_k \bigcup \{-\mathbf q_k\}_k$ and set the weights for Friedel pairs to be equal, each being half of the original weight values. This is equivalent to setting $Q$ and $F^*Wb$ to be the real parts
\begin{equation}
\begin{aligned}
    Q[\mathbf n] &= {\rm Re}\left(\sum_k w[k]e^{2\pi i \mathbf q_k\cdot\mathbf n  }\right), \\
    (F^*Wb)[\mathbf m] &= {\rm Re}\left(\sum_k b[k]w[k]e^{2\pi i \mathbf q_k\cdot\mathbf m }\right),
    \end{aligned}
\end{equation}
where ${\rm Re}(z)$ is the real part of the complex number $z$. In this case, real-to-complex FFTs and complex-to-real IFFTs can be used in Eq.~\ref{eq:FCT} to speed up the convolution calculation by a factor of two.

In addition to iterative methods, several direct methods have been proposed for nonuniform Fourier inversion. Direct approximation methods \cite{choi1998analysis, pipe1999sampling, rasche2002resampling, greengard2007fast} are among the fastest approaches, but typically lack sufficient accuracy when embedded within a larger iterative reconstruction algorithm, where accumulated errors can cause instabilities. Several exact direct methods \cite{dutt1995fast, gelb2014frame, selva2018efficient, kircheis2019direct, wilber2025superfast} are highly accurate but are either limited to 1D or have unfavorable computational cost (e.g., greater than log-linear complexity) in higher dimensions. More recently, \cite{kircheis2023fast} proposed a direct inversion approach for sufficiently oversampled Fourier data (e.g., by a factor of two) based on computing the weights $w$ such that the left-hand side of Eq.~\ref{eq:normal} simplifies to the identity matrix, allowing Eq.~\ref{eq:Fp-b} to be solved with a single weighted adjoint NUFFT. 

In Section~\ref{sec:acceleration}, we extend this weight-optimization approach to direct NDFT inversion to incorporate i) additional pre-weighting factors that improve convergence, ii) multilevel Toeplitz acceleration of the weight computation, iii) frequency-filling stabilization to handle undersampled Fourier data, and iv) further acceleration strategies tailored to CSSI reconstruction.

\subsection{The Classical Phase Problem and Iterative Phasing Algorithms on Uniform Grids}\label{sec:IPA}

Our approach to CSSI reconstruction is based on generalizing a class of techniques known as iterative phasing algorithms, which are popular for solving the classical phase problem where the Born approximation is valid, meaning that $I(\mathbf q) = |\hat{\rho}(\mathbf q)|^2$. In this case, the solution is typically constrained to be zero outside a specified support region to make the problem well-posed. More specifically, given an intensity $I$ and support region $S$, the classical phase problem can be formulated as recovering a density $\rho$ that satisfies two constraints: 1) the support constraint $\rho(\mathbf r) = 0 \text{ for all } \mathbf r \not\in S$, and 2) the magnitude constraint $|\hat{\rho}(\mathbf q)|^2 = I(\mathbf q)$ for all sampled $\mathbf q$. Other constraints on the solution, such as nonnegativity, symmetry, and density histograms, can also be enforced. 

Here it is typically assumed that the intensity is sampled on a uniform grid, which allows the use of FFTs to efficiently compute and invert the Fourier transform. Later, we show how these techniques can be extended to nonuniform data.

Apart from the one-dimensional case, the phase problem has a unique solution up to translation, multiplication by an arbitrary unitary scalar, and reflection through the origin as long as the intensity function is oversampled by a factor of at least two and the solution is not a homometric structure, i.e., it is not the convolution between two noncentrosymmetric functions \cite{hayes1982reconstruction}. Note that computing the oversampled intensity using an FFT requires the real-space density grid to be zero-padded by a factor of at least two in each dimension.

One widely used approach to solving the phase problem is based on iterative phasing algorithms. These algorithms are based on formulating Bregman projection operators \cite{bregman1967relaxation} for each constraint to be enforced and then applying these projections in an iteration to seek a solution consistent with all the constraints. Given a constraint $C$, the corresponding Bregman projection operator $P_C$ maps a vector/function to the closest object satisfying $C$, i.e.,
\begin{equation}\label{eq:PC}
    P_Cf = \argmin_{g \text{ satisfies } C} ||f-g||,
\end{equation}
where the norm $||\cdot||$ is typically chosen as the $\ell^2$ norm, but it can be chosen differently to simplify the computation or improve its robustness.

In particular, the projection operators $P_S$ and $P_M$ for the support and magnitude constraint, respectively, of the phase problem have closed forms given by
\begin{equation}\label{eq:PS_PM}
\begin{aligned}
    &P_S \rho(\mathbf r) = \begin{cases}\rho(\mathbf r), & \text{ if } \mathbf r \in S, \\ 0, & \text{ if } \mathbf r \not\in S, \end{cases} \\
    &P_M\rho = F_{\rm uni}^*\widehat{P_M}F_{\rm uni}\rho, \text{ where } \widehat{P_M}\hat{\rho}(\mathbf q) = \frac{\hat{\rho}(\mathbf q)}{|\hat{\rho}(\mathbf q)|}\sqrt{I(\mathbf q)},
\end{aligned}
\end{equation}
where $F_{\rm uni}$ denotes the uniform discrete Fourier transform (DFT), which can be efficiently computed with an FFT. If $\hat{\rho}(\mathbf q) =0$, then Eq.~\ref{eq:PC} has multiple solutions, in which case $\widehat{P_M}\hat{\rho}(\mathbf q)$ is typically taken as $\sqrt{I(\mathbf q)}$.

To solve the phase problem, these projections are applied in an iterative scheme, which is typically initialized with a constant or random density $\rho^{(0)}$, and produce the density estimate $\rho^{(n)}$ after $n$ iterations. One popular locally optimizing scheme is the error reducing (ER) method \cite{gerchberg1972practical}, which consists of alternating between the two projections, i.e.,
\begin{equation}\label{eq:ER}
    \rho^{(n+1)} = P_SP_M\rho^{(n)}.
\end{equation}
ER is an effective local optimizer as it reduces error monotonically, but it tends to get trapped in local minima. Therefore, ER is often supplemented with a globally optimizing scheme that can escape these local minima. 

An example of a global optimization iterative scheme is the hybrid input-output (HIO) method \cite{fienup1978reconstruction}, which is typically expressed as
\begin{equation}\label{eq:HIO}
    \rho^{(n+1)}(\mathbf r) = \begin{cases}(P_M\rho^{(n)})(\mathbf r),& \text{ if } \mathbf r \in S, \\ 
                                          \rho^{(n)}(\mathbf r) - \beta (P_M\rho^{(n)})(\mathbf r),& \text{ if } \mathbf r \not\in S, \end{cases}
\end{equation}
where $\beta \in (0,1]$ is a constant. We refer to the update in the second row of Eq.~\ref{eq:HIO} as the negative feedback update. This negative feedback prevents HIO from becoming trapped in fixed points (where $\rho^{(n+1)} = \rho^{(n)}$) that do not satisfy the constraints, since $|\rho^{(n+1)}(\mathbf r)-\rho^{(n)}(\mathbf r)| = \beta|P_M\rho^{(n)}(\mathbf r)| > 0$ for $\mathbf r \not\in S$ if $\rho^{(n)}$ is not a solution to the phase problem. 

Later in the manuscript, it will be convenient to express HIO in an equivalent coordinate-free form as
\begin{equation} \label{eq:hio_coord_free}
\begin{aligned}
    \rho^{(n+1)} &= P_SP_M\rho^{(n)} + \eta^{(n+1)},\\
    \eta^{(n+1)} &= \eta^{(n)}-\beta(P_M\rho^{(n)} - P_SP_M\rho^{(n)}),
\end{aligned}
\end{equation}
where $\eta^{(n)}$ denotes the negative feedback update, which is zero on $S$ and equals the second row of Eq.~\ref{eq:HIO} outside of $S$, and is initialized as $\eta^{(0)} = \rho^{(0)}-P_S\rho^{(0)}$. In practice, one typically alternates between applying several iterations of ER with several iterations of HIO until convergence to a density consistent with all enforced constraints. 

Other popular global optimization phasing schemes include the difference-map (DM)  \cite{elser2003phase,elser2007searching}, hybrid-projection-reflection (HPR) \cite{bauschke2003hybrid}, relaxed averaged-alternating-reflections (RAAR) \cite{luke2004relaxed}, and relaxed-reflect-reflect (RRR) \cite{elser2018benchmark} algorithms. 

Typically, the support $S$ is not known precisely a priori, but it can be estimated during reconstruction, for example, using the Shrinkwrap algorithm \cite{marchesini2003x}. More specifically, after several iterations of the above schemes, one creates a new support estimate $S$ from the density after the $n$-th iteration $\rho^{(n)}$ by computing a Gaussian blur of the density and then applying a threshold, i.e.,
\begin{equation}
    S = \left\{\mathbf r: (\rho^{(n)}\ast G_\sigma)(\mathbf r) \geq \tau \right\},
\end{equation}
where $G_\sigma(\mathbf r)$ is a diagonal multivariate normal distribution with mean 0 and standard deviations $\sigma_d$, and $\tau$ is a chosen threshold. Good heuristics for the parameter selection include choosing each $\sigma_d$ to be the pixel length of the real-space grid in the $d$-th dimension (i.e.\@ $\sigma_d = 1/N_d$) and $\tau$ to be a few percent (e.g.\@ 1-5\%) of the maximum value of $\rho^{(n)}$ \cite{donatelli2015iterative}. This new support estimate is then used in the support projection for the next phasing cycle, and the process is repeated until convergence is reached.

In this paper, we develop a new approach that generalizes the above iterative phasing framework to handle nonuniformly sampled Fourier values and intensity data described by the DWBA, which is needed to solve the CSSI reconstruction problem.

\section{Overview of the Proposed CSSI Reconstruction Framework}\label{sec:overview}

\begin{figure*}[t]
    \centering
    \includegraphics[width=\linewidth]{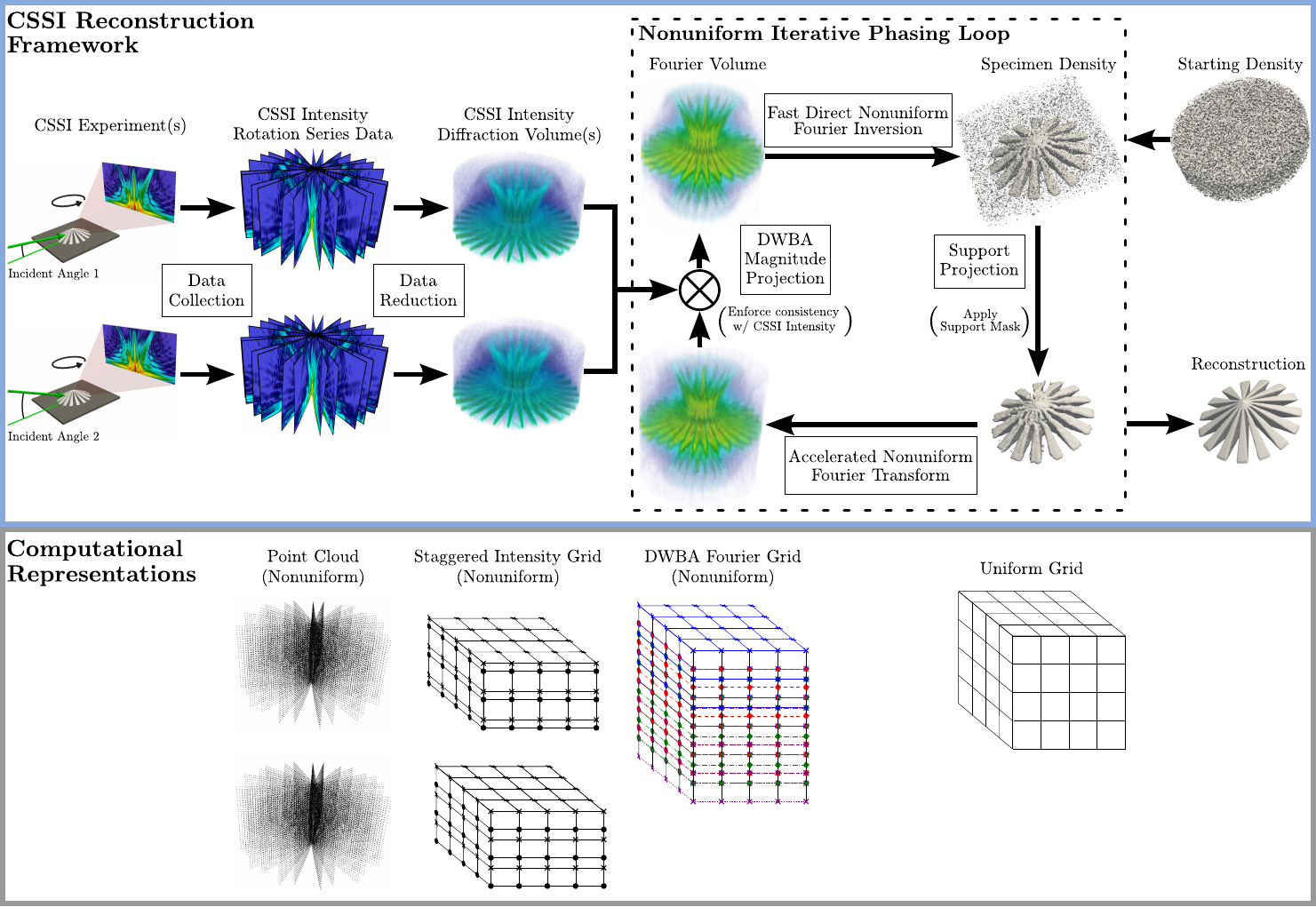}
    \caption{High-level illustration of the proposed CSSI reconstruction framework, with computational representations shown beneath the corresponding components. CSSI intensities from a rotation series (represented as a point cloud) at one or more incident angles are reduced to compressed diffraction volumes (defined on a nonuniform staggered grid). Starting with a random density, iterative projections alternately enforce real-space constraints (e.g., finite support) and Fourier-space constraints (consistency with reduced intensities) on the density (defined on a uniform grid) and its Fourier transform (defined on a nonuniform DWBA grid) until convergence to a reconstruction of the specimen. The nonuniform grids introduce additional oversampling and redundancies that align with symmetries in the DWBA and Fourier transforms to improve reconstruction stability and greatly simplify and accelerate the projection operations and the forward and inverse nonuniform Fourier transforms.
 }\label{fig:High_Level}
\end{figure*}

Here we outline the CSSI reconstruction framework developed in Sections~\ref{sec:reduction}-\ref{sec:NIP}. Fig.~\ref{fig:High_Level} provides a high-level overview of the reconstruction workflow and the computational representations associated with each step.

The input to our reconstruction framework consists of one or more rotation series of CSSI intensity measurements $ I_j(\mathbf q) $, which we treat as being sampled on a point cloud of 3D Fourier coordinates $\mathbf q$. The sampling locations are determined by the experimental geometry via Eqs.~\ref{eq:ki}-\ref{eq:q} for incident angles $\alpha_{i,j}$.

Section~\ref{sec:reduction} introduces an efficient, noise-robust linear reduction procedure that reduces oversampled intensity measurements defined on a point cloud to greatly compressed representations. From these compressed representations, a noise-filtered intensity can then be sampled at any prescribed set of coordinates for subsequent analysis. In particular, Section~\ref{sec:Cartesian} shows how the data can be resampled onto a specially designed structured, yet nonuniform, staggered grid that improves both reconstruction speed and stability when solving for the specimen density. Alternatively, the intensity can be re-evaluated at the original measurement coordinates to produce filtered CSSI diffraction images.

Note that, even if the CSSI intensities are resampled onto a structured grid, evaluating them via the DWBA (Eq.~\ref{eq:CSSI}) generally involves a linear combination of nonuniformly sampled Fourier values, unless the coordinate shifts $\pm \qzij$ are multiples of the grid spacing. Although it is possible to choose a single uniform grid that aligns with these shifts when using a single incident angle, this is not feasible for reconstruction using multiple incident angles without imposing unnecessary constraints on the angles. 
Moreover, even when using a single incident angle, enforcing such alignment may require a substantial increase in grid points, e.g., when the shifts are either much smaller or slightly larger than the Nyquist length. Additionally, if the real-space density grid is chosen so that data on the intensity grid can be computed via FFTs, such alignment can complicate comparisons across data from even small variations in X-ray energies and experimental geometries, since each configuration may then require a different grid to represent the specimen reconstruction.

Although NUFFTs can efficiently compute the nonuniformly sampled Fourier values in the DWBA calculation, inverting the associated NDFT using the conventional iterative schemes reviewed in Section~\ref{sec:INUDFT} is prohibitively computationally expensive to apply repeatedly within a larger reconstruction loop. Section~\ref{sec:acceleration} shows how this limitation can be overcome via a fast direct NDFT inversion approach in which weights are precomputed so that the inverse NDFT can be computed using a single weighted adjoint NUFFT. Additionally, a frequency-filling stabilization method is introduced to enable robust direct NDFT inversion when the Fourier data is incomplete or undersampled, as is common for CSSI reconstruction. 

Section~\ref{sec:CartAccel} introduces the DWBA Fourier grid, which includes a grid point for each DWBA term for each intensity sample on the staggered intensity grid. In particular, it is shown how symmetries in the DWBA and Fourier transform can be exploited on the DWBA Fourier grid to reduce the NUFFT operations on this grid to faster complex-weighted combinations of FFTs when solving the CSSI reconstruction problem from staggered-grid CSSI data.

Section~\ref{sec:NIP} presents a new nonuniform iterative phasing algorithm for \textit{ab initio} 3D reconstruction from CSSI intensity data collected at one or more incident angles. This approach establishes a new framework that generalizes the classical iterative phasing algorithms in Section~\ref{sec:IPA} to handle nonuniformly sampled Fourier values. This is accomplished by reformulating the reconstruction updates in Fourier space and leveraging the fast NDFT forward and inverse operations described in Section~\ref{sec:INUDFT}. In addition, Section~\ref{sec:PD} formulates a new Bregman projection operator that extends the magnitude projection in Eq.~\ref{eq:PS_PM} to enforce consistency with intensities described by the DWBA, and shows that it can be computed efficiently via a simple closed-form expression by exploiting redundancies in the DWBA grid.

Finally, Section~\ref{sec:uniqueness} describes the experimental-parameter requirements and uniqueness properties of CSSI reconstruction, and Section~\ref{sec:results} demonstrates the effectiveness of the algorithmic components and the overall reconstruction pipeline described above.

\section{CSSI Data Reduction and Resampling onto a Staggered Grid}\label{sec:reduction}
When collected from a rotation series, the CSSI diffraction volume measurements in Eq.\@ \ref{eq:CSSI} are often highly oversampled by the pixels along the detector dimensions and, due to the polar sampling, this sampling density further increases at smaller $\mathbf q$ values. Here we develop a new linear reduction technique that transforms the oversampled raw data into a compressed representation in a way that avoids physical information loss and removes noise. This compressed representation can then be used to compute a denoised resampling of the data onto a specialized grid, significantly improving the convergence, speed, and memory efficiency of the 3D specimen reconstruction step. 

Our data reduction approach is based on a new generalization of the Wiener-Khinchin identity that allows us to represent the oversampled raw data as a linear transformation of two compact real-space functions. To compute this compact real-space representation, we introduce a generalization of the multilevel-Toeplitz acceleration, discussed in Section\@ \ref{sec:INUDFT}, to efficiently solve the corresponding linear system. Once found, this real-space representation can then later be transformed back to Fourier space to obtain either a filtered version of the CSSI diffraction images or a convenient discretization of the diffraction volume on a structured grid. Specifically, we resample the data on a special nonuniform grid that stabilizes the CSSI reconstruction process and accelerates the nonuniform Fourier calculations required by the reconstruction algorithm, as discussed in Section \ref{sec:CartAccel}.

Compared to conventional techniques for resampling data, our approach circumvents common interpolation errors by accurately computing the CSSI intensity from the reduced real-space representations using NUFFTs, and it can leverage additional real-space constraints (e.g., the compact size of the specimen) to robustly suppress noise. Furthermore, our reduction scheme facilitates direct on-the-fly compression of CSSI diffraction data into an intermediate compressed state, which can be processed later by iterative methods to accurately interpret the compressed data.

Since combining data from multiple incident angles is an inherently nonlinear process, here we formulate the reduction for a CSSI rotation series collected from a single fixed incident angle $\alpha_i$, and accordingly suppress the $\alpha_i$ dependence of $R_i$ and $R_f$. To handle multiple incident angles, this data-reduction scheme can be applied independently to the data for each incident angle, and the resulting reduced data representations can be subsequently combined to determine the 3D structure via the nonlinear iterative reconstruction algorithm presented in Section \ref{sec:NIP}.

\subsection{A Generalized Wiener-Khinchin Identity}\label{sec:WK}

As we now show, the CSSI data from each incident angle can be represented as a linear transformation of compact real-space functions with various symmetries. By solving the associated linear system, we can recover these real-space functions, which provide a reduced representation of the data that can be used to compress and filter the raw measurements. 

To formulate the linear reduction, we seek to express the CSSI intensity in Eq.~\ref{eq:CSSI} as a linear transformation of real-space functions $f_1$ and $f_2$ with as few linear degrees of freedom as possible.

To this end, we represent the sum of the first two DWBA terms as
\begin{equation}\label{eq:g}
\hat{g}(\qp,\qz) = \ph(\qp,\qz) + R_i\ph(\qp,\qz - \qzi),
\end{equation}
which has the real-space representation
\begin{equation}
  g(\rp,\rz) = \left(1 + R_i e^{2\pi i \qzi \rz}\right)\rho(\rp,\rz).
\end{equation}
Rewriting Eq.~\ref{eq:CSSI} in terms of $\hat{g}$ and expanding the squared modulus yields
\begin{equation}\label{eq:gintoI}
\begin{aligned}
I(\qp,\qz) &= |\hat{g}(\qp,\qz) + R_f(\qz)\hat{g}(\qp,-\qz+\qzi)|^2\\[.5em]
&= |\hat{g}(\qp,\qz)|^2 + |R_f(\qz)|^2|\hat{g}(\qp,-\qz+\qzi)|^2\\[.5em]
&+ \overline{R_f(\qz)}\hat{g}(\qp,\qz)\overline{\hat{g}(\qp,-\qz+\qzi)}\\
&+ R_f(\qz)\overline{\hat{g}(\qp,\qz)}\hat{g}(\qp,-\qz+\qzi).
\end{aligned}
\end{equation}
Although Eq.~\ref{eq:gintoI} depends nonlinearly on $\hat{g}$, we can express it as a linear transformation of the functions
\begin{equation}\label{eq:fs}
\begin{aligned}
\fho(\qp,\qz) &= |\hat{g}(\qp,\qz)|^2,\\
\fht(\qp,\qz) &= \hat{g}(\qp,\qz+\qzit)\overline{\hat{g}(\qp,-\qz+\qzit)}e^{2\pi i \qz D_z},
\end{aligned}
\end{equation}
where the purpose of the coordinate shift by $\qzi/2$ is to induce the symmetry $\fht(\qp,\qz) = \overline{\fht(\qp,-\qz)}$, and the phase shift by $D_z$ centers the real-space representation $f_2$ about the origin, allowing us to numerically represent it on the same grid as $f_1$. 

The real-space representations of these functions are given by
\begin{equation}\label{eq:fsreal}
\begin{aligned}
f_1(\rp, \rz) &= (g\star g)(\rp,\rz),\\
f_2(\rp,\rz) &= \big((\mathcal R_z g_e)\star g_e\big)(\rp,\rz+D_z),
\end{aligned}
\end{equation}
where $g_e(\rp,\rz) = g(\rp,\rz)e^{-\pi i \qzi\rz}$ and $\mathcal R_z$ is reflection through the $xy$ plane, e.g., $\mathcal R_zg_e(\mathbf r)=g_e(\rp, -\rz)$. The supports satisfy ${\rm supp}(f_1),\, {\rm supp}(f_2) \subseteq ([-D_x,D_x]\times[-D_y,D_y]\times[-D_z,D_z]) \bigcap \{\mathbf r: |\rp| \leq D_{||}\}$, and we have the symmetries $f_1(\rp,\rz) = \overline{f_1(-\rp,-\rz)}$ and $f_2(\rp, \rz) = \overline{f_2(-\rp, \rz)}$.

We can now express the CSSI intensity as a linear transformation of $f_1$ and $f_2$: 
\begin{equation}\label{eq:linred}
\begin{aligned}
I(\qp,\qz) &= \fho(\qp,\qz) + |\tilde{R}_f(\qz)|^2\fho(\qp,-\qz+\qzi) \\
&+ \overline{\tilde{R}_f(\qz)}\fht(\qp,\qz-\qzit)\\
&+ \tilde{R}_f(\qz)\fht(\qp,-\qz+\qzit).
\end{aligned}
\end{equation}
where we define $\tilde{R}_f(\qz) = R_f(\qz)e^{2\pi i (\qz - \qzi/2) D_z}$ to simplify notation. Eq.~\ref{eq:linred} shows that $f_1$ and $f_2$ form a reduced representation of the CSSI intensity, in the sense that the intensity can be reconstructed from $f_1$ and $f_2$, which together generally require far fewer voxels than the number of pixels measured in the full rotation-series dataset.

Eqs.~\ref{eq:fsreal}-\ref{eq:linred} form a generalization of the Wiener-Khinchin theorem for DWBA intensity data. In fact, when the Born approximation is valid, we have $R_i = R_f = 0$, which causes Eqs.~\ref{eq:fsreal}-\ref{eq:linred} to reduce to the original Wiener-Khinchin identity described in Section \ref{sec:basic}.

\subsection{CSSI Linear Data Reduction}\label{sec:linear_reduction}

A reduced representation of the CSSI intensity data can be obtained by discretizing the linear system in Eq.\@ \ref{eq:linred} and solving for the functions $f_1$ and $f_2$. These functions are discretized on a Cartesian grid with dimensions $(2N_x-1)\times (2N_y-1) \times (2N_z-1)$ and indexed by $\mathbf m$ = $(\mpp, \mz)$ = $(m_x, m_y, m_z)$ with $-N_d+1 \leq m_d \leq N_d-1$. We collect them into the block vector $\mathbf f = \left[\begin{smallmatrix} f_1 \\ f_2 \end{smallmatrix}\right]$. Also, suppose that the supports of the discretized functions are known to be contained within $\mathcal D$ (e.g., via the support bounds for $f_1$ and $f_2$ described in the previous subsection).

Given CSSI intensity values $I(\mathbf q_k)$ sampled at $K$ Fourier coordinates $\mathbf q_k = (q_{||,k}, q_{z,k})$, define $\mathbf q^1_k = (q_{||,k}, q_{z,k})$, $\mathbf q^2_k = (q_{||,k}, -q_{z,k}+\qzi)$, $\mathbf q^3_k = (q_{||,k}, q_{z,k}-\qzit)$, and $\mathbf q^4_k = (q_{||,k}, -q_{z,k}+\qzit)$. Let $F_s$ be the NDFT in Eq.\@ \ref{eq:NUDFT} for the coordinates $\{\mathbf q^s_k\}_k$. The discretization of the linear operator $A$ in Eq.~\ref{eq:linred} can be expressed as
\begin{equation}\label{eq:A}
A\mathbf f = A\begin{bmatrix} f_1 \\ f_2 \end{bmatrix}= F_1 f_1 + |\tilde{R}_f|^2F_2 f_1 + \overline{\tilde{R}_f}F_3 f_2 + \tilde{R}_fF_4 f_2,
\end{equation}
where here $\tilde{R}_f$, $\overline{\tilde{R}_f}$, $|\tilde{R}_f|^2$ denote the diagonal matrices with $k$-th diagonal entries $\tilde{R}_f(q_{z,k})$, $\overline{\tilde{R}_f(q_{z,k})}$, and $|\tilde{R}_f(q_{z,k})|^2$, respectively.

The corresponding adjoint operator is given by
\begin{equation}\label{eq:A*}
A^*b = \begin{bmatrix} F_1^*b &+&\hspace*{-.5em} F_2^*|\tilde{R}_f|^2b \\ F_3^* \tilde{R}_f b\hspace*{-.5em} &+& F_4^*\overline{\tilde{R}_f}b \end{bmatrix}.
\end{equation}

To enforce the symmetries and supports of $f_1$ and $f_2$, define the operator $\mathcal S \mathbf f = \left[\begin{smallmatrix} \mathcal S_1 f_1 \\ \mathcal S_2 f_2 \end{smallmatrix}\right]$ with components
\begin{equation}
\begin{aligned}
(\mathcal S_1f_1)[\mathbf m] &= \frac{\chi_{\mathcal D}[\mathbf m]}{2}\left(f_1[\mpp, \mz] + \overline{f_1[-\mpp, -\mz]}\right),\\
(\mathcal S_2f_2)[\mathbf m] &= \frac{\chi_{\mathcal D}[\mathbf m]}{2}\left(f_2[\mpp, \mz] + \overline{f_2[-\mpp, \mz]}\right),\\
\end{aligned}
\end{equation}
where $\chi_{\mathcal D}[\mathbf m] = 1$ if $\mathbf m \in \mathcal D$ and $\chi_{\mathcal D}[\mathbf m] = 0$ otherwise. Enforcing the symmetries and supports of $f_1$ and $f_2$ is equivalent to requiring $\mathcal S \mathbf f = \mathbf f$. Note that $\mathcal S$ is not complex-linear, but when $\mathbf f$ is represented as a real vector of its real and imaginary components, $\mathcal S$ becomes real-linear and symmetric.

We can now formulate the linear reduction as finding a weighted least-squares solution to $A \mathbf f = I$ satisfying $\mathcal S \mathbf f = \mathbf f$, i.e.,
\begin{equation} \label{eq:constrained}
    \min_{\mathbf f = \mathcal S \mathbf f} ||A \mathbf f - I||^2_w, 
\end{equation}
where the weight $w[k] = |\qpk|(1+|R_f(\qzk)|^2)^{-1}$ is used to compensate for the polar-sampling density in the $\qp$ dimensions and the factors of $|R_f(\qzk)|$ in the $\qz$ direction.

Eq.\@ \ref{eq:constrained} is equivalent to solving the symmetrized normal equations
\begin{equation}\label{eq:Anormal}
\mathcal SA^*WA\mathcal S \mathbf f = \mathcal S A^*WI,
\end{equation}
where $W={\rm diag}(w[1],w[2],\hdots, w[K])$, and then applying $\mathcal S$ to the solution. 
 
 Eq.\@ \ref{eq:Anormal} can be solved using iterative linear solvers, such as conjugate gradient. However, since $\mathcal S$ is only real-linear, the system must be treated as a real-linear system when applying iterative linear solvers. Complex operators may still be used to perform the arithmetic by separating the real and imaginary parts of the output into different components. 
 
 Since the data typically cannot fully sample the Fourier transforms of $f_1$ and $f_2$, Eq.\@ \ref{eq:Anormal} will typically be ill-conditioned. In practice, this can cause the iterative solver to eventually stall. However, our numerical tests indicate that the iterations eventually reach a stable region with a relative residual error between approximately $10^{-6}$ and $10^{-8}$. Additional regularization and preconditioning strategies could potentially improve the convergence and residual error.

The solution $\mathbf f$ to Eq.\@ \ref{eq:Anormal} provides a reduced representation of the measured intensity data, which can be used to compute a filtered estimate of the intensity at any desired $\mathbf q$ coordinates via Eq.\@ \ref{eq:linred}. For example, this filtered intensity estimate can be evaluated at the coordinates sampled by the original data to produce a filtered version of the original CSSI images, or it can be evaluated on a convenient computational grid to facilitate further analysis, which we discuss in Section~\ref{sec:Cartesian}. The intensity estimate will be accurate near regions sampled by the original data, but it may deviate from the true intensity in regions far from where the data is sampled. This reduction process allows both the size and noise of the original data to be reduced by a factor approximately equal to the number of intensity measurements divided by the number of voxels in the enforced support for $f_1$ and $f_2$, i.e., the compression ratio scales approximately as $K/((2N_x-1)(2N_y-1)(2N_z-1))$.

Note that iterative linear solvers applied to Eq.~\ref{eq:Anormal} will typically require repeated application of $A^*WA$, which can be very computationally demanding even if the matrix-vector multiplications for $A$ and $A^*$ are performed using NUFFTs. In the following subsection, we show that $A^*WA$ fortunately has a special structure that can be leveraged to massively speed up its corresponding matrix-vector multiplication. 

\subsection{Generalized Multilevel Toeplitz Acceleration}\label{sec:GMTA}
Here we show that the normal equations in Eq.~\ref{eq:Anormal} have rich algebraic structure that can be leveraged to significantly accelerate the corresponding repeated matrix-vector multiplication needed in iterative linear solvers, such as conjugate gradient. Our approach is based on a generalization of the Toeplitz acceleration method described in Section~\ref{sec:INUDFT}. In particular, the $A^*WA$ operator in Eq.~\ref{eq:Anormal} can be represented as a combination of $8$ different convolutions, which we now describe (see Section S1 of the Supplemental Material for a derivation). 

In this case, a total of $16$ point-spread functions are needed (one for each pair of terms in Eq.~\ref{eq:linred}), but they can be combined into the following 6 functions for $l = 1,\hdots,6$:
\begin{equation}\label{eq:Qij}
\begin{aligned}
    Q_{l}[\mathbf n] = \sum_k\big(&C_{l,1}[k]\eb{(\qpk,\qzk-\qzi/2)\cdot(\npp,\nz)}\\
    + &C_{l,2}[k]\eb{(\qpk,-\qzk+\qzi/2)\cdot(\npp,\nz)}\big)w[k],
\end{aligned}
\end{equation}
where the $C_{l,t}$ for $t=1,2$ are given by powers of $\tilde{R}_f$ and its conjugate via
\begin{equation}\label{eq:Rjl}
C_{l,t}[k] = \tilde{R}_f^{a_{l,t}}(\qzk)\overline{\tilde{R}_f^{b_{l,t}}(\qzk)},
\end{equation}
and the exponents $a_{l,t}$ and $b_{l,t}$ are given in Table~\ref{table:exponents}.

\begin{table}[h]
\begin{center}
\begin{tabular}{l c c c c c c} 
 \midrule
 &\multicolumn{6}{c}{$l$}\\ 
 Exponents & $1$ & $2$ & $3$ & $4$ & $5$ & $6$ \\ 
 \midrule
$(a_{l,1},b_{l,1})$ & $(0,0)$ & $(1,1)$ & $(0,1)$ & $(1,0)$ & $(2,1)$ & $(2,0)$ \\[.5ex] 
 $(a_{l,2},b_{l,2})$ & $(2,2)$ & $(1,1)$ & $(2,1)$ & $(1,2)$ & $(0,1)$ & $(0,2)$\\
 \midrule
\end{tabular}
\end{center}
\vspace*{-1em}
\caption{Reflection coefficient exponents used in Eq.~\ref{eq:Rjl}}\label{table:exponents}
\end{table}

The $Q_l$ terms depend only on the sampled $\mathbf q$ points and, therefore, can be precomputed via an NUFFT and stored to be reused throughout the linear solver iterations. The matrix $A^*WA$ can then be expressed in terms of eight convolutions, reflection $\mathcal R_z$ through the $xy$-plane, and multiplication by complex exponentials $Ef_t[n_{||},n_z] = f_t[n_{||},n_z]e^{-\pi i \qzi n_z}$ and $\overline{E}f_t[n_{||},n_z] = f_t[n_{||},n_z]e^{\pi i \qzi n_z}$ as
\begin{gather}\label{eq:GMLToeplitz}
A^*WA\begin{bmatrix} f_1 \\ f_2 \end{bmatrix} = \begin{bmatrix} h_1 \\ h_2    \end{bmatrix},\\
\nonumber\begin{aligned}
h_1 &= \overline{E}\big((Ef_1) \ast Q_{1} + (\mathcal R_z E f_1) \ast Q_{2} + f_2 \ast Q_{3}\\
&+ (\mathcal R_z f_2) \ast Q_{4}\big),\\
h_2 &=  (Ef_1) \ast Q_{4} + (\mathcal R_z Ef_1) \ast Q_{5} + f_2 \ast Q_{2} \\
&+  (\mathcal R_z f_2) \ast Q_{6}.
\end{aligned}
\end{gather}

Similarly to the original Toeplitz acceleration approach discussed in Section~\ref{sec:INUDFT}, the $Q_l$ functions and $\mathcal S A^*WI$ can be efficiently precomputed using NUFFTs, the FFTs for $\hat{Q_l}$ can be precomputed, and the only operation that needs to be computed iteratively to solve Eq.~\ref{eq:Anormal} using the conjugate gradient method is $\mathcal SA^*WA\mathcal S$.  Fortunately, the symmetry operator $\mathcal S$ is trivial to compute, and the $A^*WA$ operation can be accelerated by applying the convolution theorem (Eq.~\ref{eq:FCT}) to evaluate Eq.~\ref{eq:GMLToeplitz}. In particular, since the reflections and the four-term summations for $h_1$ and $h_2$ can be performed in Fourier space, computing $A^*WA$ requires only a total of four FFT operations: two for the Fourier transforms of $Ef_1$ and $f_2$, and two for the inverse Fourier transforms of the four-term summations that were computed in Fourier space. An additional twofold speedup could be achieved by leveraging the symmetry of $f_1$ and $f_2$ to reduce the number of operations needed to compute the above Fourier transforms.

The complexity of solving Eq.~\ref{eq:Anormal} using the above generalization of multilevel Toeplitz acceleration is $O(N' \log N' + K\log^3(1/\epsilon))$ for precomputation and $O(N' \log N')$ for each iteration of the linear solver, where $N'=(2N_x-1)(2N_y-1)(2N_z-1)$, $K$ is the total number of pixel measurements in the CSSI data, and $\epsilon$ is the selected error tolerance for the NUFFT. This results in a significant speedup compared to directly computing $A$ and $A^*$ using NUFFTs, which would have complexity $O(N' \log N' + K\log^3(1/\epsilon))$ per iteration of the linear solver.

Furthermore, to reduce memory requirements, the NUFFT operations for $Q_l$ and $A^*WI$ can be streamed by partitioning the data into chunks that fit into memory, computing the corresponding NUFFTs for each chunk, and then averaging the results across all chunks. This streaming of $A^*WI$ could be used to potentially enable on-the-fly compression of the CSSI rotation-series data into a single reduced 3D volume, from which $f_1$ and $f_2$ can later be retrieved by solving Eq.~\ref{eq:Anormal} after all data has been collected.

\subsection{Resampling CSSI Data onto a Staggered Intensity Grid}\label{sec:Cartesian}

Once the reduced real-space representation, consisting of $f_1$ and $f_2$, has been found, Eq.~\ref{eq:linred} can then be used to evaluate the corresponding CSSI diffraction volume $I$ at any desired set of $\mathbf q$ coordinates near those sampled by the original data. This procedure thus enables accurate resampling of the original CSSI rotation-series data onto a more convenient set of coordinates. Here we define and motivate the use of a special nonuniform staggered intensity grid that is able to exploit symmetries inherent in the DWBA to increase oversampling and improve reconstruction stability with almost no additional computational cost.

Although it is possible to simply resample the intensity data onto a standard uniform grid, this approach introduces instabilities in CSSI reconstruction when $\qzi$ is far from an integer multiple of the grid spacing. See Section S5 in the Supplemental Material for a demonstration of these instabilities. As explained in Section~\ref{sec:overview}, simply choosing a grid spacing that makes $\qzi$ an integer multiple of it is not always possible when using multiple incident angles and can introduce other inconveniences when comparing results between different incident angles. 

To understand this instability, note that since the intensity cannot be measured for $\qz < \qzi/2$, the first DWBA term $\hat{\rho}(\qp,\qz)$ is generally unable to sample the Fourier transform of the specimen near $\qz = 0$, which is important for reconstruction stability as these Fourier values determine the bulk structure. Therefore, the second DWBA term $\hat{\rho}(\qp, \qz - \qzi)$ or the third DWBA term $\ph(\qp,-\qz+\qzi)$ is needed to sample this region. However, if $I(\qp,\qz)$ is sampled on a standard uniform grid, then the second or third terms can only sample $\hat{\rho}(\qp, 0)$ if $\qz - \qzi \approx 0$ for some point $(\qp,\qz)$ on the grid, i.e., when $\qzi$ is close to a multiple of the grid spacing. 

To resolve the instabilities described above, we instead resample the intensity onto a nonuniform grid that is the union of a standard uniform grid and an additional uniform grid shifted in the vertical direction by $\qzi$ minus an integer multiple of the grid spacing. This configuration ensures that the Fourier transform of the specimen is sampled at $\qz = 0$ by the second and third DWBA terms, as long as the data is collected with minimum and maximum exit angles satisfying $\alpha_f^{\rm min}\leq \alpha_i \leq \alpha_f^{\rm max}$. We refer to this combined nonuniform grid layout as a ``staggered CSSI intensity grid", which we formulate for a given incident angle as follows. 

\begin{figure}[h] 
 \centering
\includegraphics[width = \columnwidth]{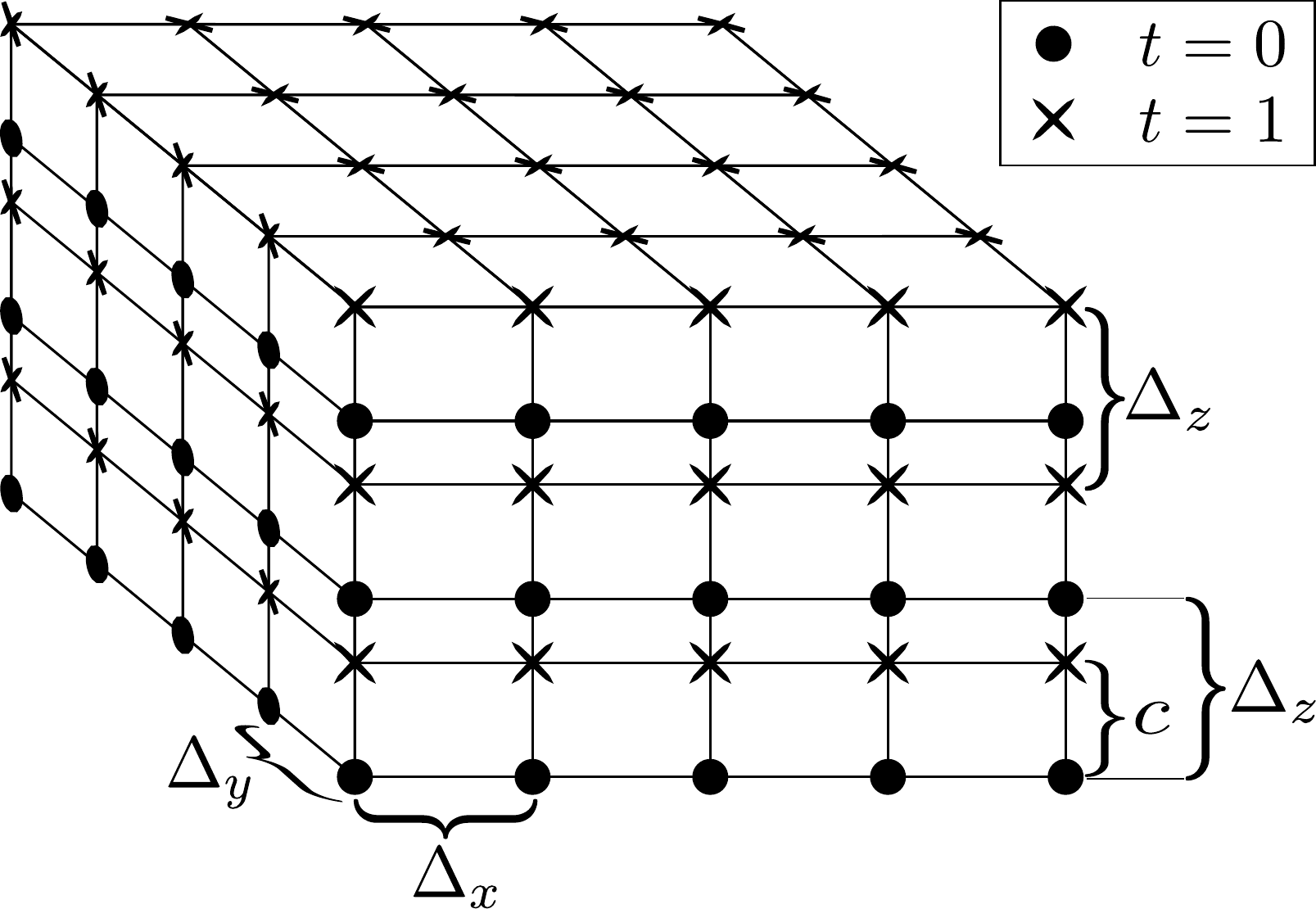}
\caption{Illustration of the staggered CSSI intensity grid. The grid consists of two components ($t=0$ and $t=1$). Within each component, the grid points have a uniform spacing of $\Delta_d = 1/(2N_d-1)$ for $d = x,y,z$. The $t=1$ component is shifted in the $z$ direction by $c = \qzi -\lfloor (2N_z-1)\qzi  \rfloor/(2N_z-1)$. }\label{fig:stag_grid}
\end{figure}

Define the voxel lengths $\Delta_d = 1/(2N_d-1)$ for $d=x,y,z$ and the vertical shift $c = \qzi -\lfloor (2N_z-1)\qzi  \rfloor/(2N_z-1)$. The shift $c$ is the positive offset of $\qzi$ from a uniform grid with $2N_z-1$ points along $\qz$. The staggered CSSI intensity grid is given by the set of $\mathbf q$ coordinates $(k_x \Delta_x, k_y \Delta_y, k_z \Delta_z +tc)$ for $k_d = -N_d+1, -N_d+2,\hdots, N_d-2, N_d-1$ for $d=x,y$; $k_z = 0,1\hdots, N_z-1$; and $t =0,1$. See Fig.~\ref{fig:stag_grid} for an illustration of this grid.

The $t=0$ component corresponds to a standard Fourier grid, and the $t=1$ component corresponds to a vertically-shifted grid. The vertical shift $c$ is the remainder when $\qzi$ is divided by the vertical grid spacing $\Delta_z$. This design ensures that the intensity values at $I(\qp,\qzi)$ are sampled on the shifted grid. Consequently, at these grid points, the second and third DWBA terms sample $\ph(\qp, \qzi -\qzi) = \ph(\qp,0)$ and $\ph(\qp,-\qzi +\qzi) = \ph(\qp,0)$ respectively, thus satisfying the stability requirements mentioned earlier. Furthermore, as we show in Section~\ref{sec:CartAccel}, we can exploit symmetries in the DWBA to compute and invert the DWBA terms on the combined grid with nearly the same computational cost as performing these operations on either grid component individually.

To resample CSSI rotation-series data to a staggered intensity grid, we solve Eq.~\ref{eq:Anormal} and then use the solution to evaluate Eq.~\ref{eq:linred} at the $\mathbf q$ coordinates for the staggered grid. 

Note that there will be regions in the staggered grid where the intensity is not sampled by nearby data or cannot be accurately resolved. The unsampled regions can occur: i) at the corners of each $xy$ slice of the volume (due to the rotational sampling), ii) below/above the minimum/maximum $q_z$ values sampled by the incident and exit angles, iii) along large gaps in the detector, and iv) in various regions undersampled by or inaccessible to the experimental geometry, which are discussed in Section~\ref{sec:uniqueness}. Furthermore, if the $\qz$ coordinate of the topmost plane (where $k_z = N_z-1$ and $t=1$) of the staggered grid exceeds the Nyquist limit (i.e., $(N_z-1)\Delta_z+c \geq 1/2$), then the intensity values computed there will be aliased. While Eq.~\ref{eq:linred} can be evaluated within the regions described above, the resulting values will be unreliable and contaminated with numerical artifacts. Therefore, these unsampled regions should be masked out of any subsequent analysis, which will be explained in more detail in Section~\ref{sec:prelim}.

Note that both the data reduction and resampling procedures described above must be executed independently for each incident angle used in the experiment. Consequently, when data from multiple incident angles are used, this process produces several diffraction volumes, each on a different staggered grid. As described in Sections~\ref{sec:NIP} and \ref{sec:incident}, these multiple staggered intensity grids can be combined within a single reconstruction to improve accuracy and robustness. 

\section{Direct Inversion and Acceleration of the Nonuniform Fourier Transform}\label{sec:acceleration}

Our iterative-phasing approach to solving the CSSI reconstruction problem will require computing an NDFT and its inverse several times within an iterative loop. Note that even if the CSSI data is resampled onto a structured grid, e.g., as described in Section~\ref{sec:Cartesian}, the Fourier coordinates used in the DWBA will typically be nonuniform in the $z$ dimension. Although NUFFTs can be used to efficiently compute the NDFT, the iterative inversion techniques in Section~\ref{sec:INUDFT} are too computationally expensive to be used practically within a larger iterative scheme.

To overcome this limitation, here we describe an algorithmic approach to efficiently and directly invert the NDFT with a single weighted adjoint NUFFT call after precomputing a set of optimized weights, making NDFT inversion feasible when repeatedly applied within an iterative reconstruction method. We also discuss how to stabilize this direct inversion approach when Fourier space is insufficiently sampled. Finally, we show how the computation and inversion of the nonuniform Fourier values used in the DWBA can be further accelerated when the CSSI data is mapped to the staggered intensity grid described in Section~\ref{sec:Cartesian}.

\subsection{Direct Nonuniform Fourier Inversion}\label{sec:DirectInversion}
Here we show how to obtain a fast direct NDFT inversion using a single weighted adjoint NUFFT. The weights depend only on the nonuniform Fourier coordinates and can therefore be precomputed and reused as long as these coordinates remain fixed. Our approach is related to \cite{kircheis2023fast}, but additionally incorporates pre-weighting factors that allow particular Fourier samples to be emphasized or de-emphasized (e.g., to compensate for sampling density, sample importance, or noise levels), and exploits the multilevel Toeplitz structure in Section~\ref{sec:INUDFT} to accelerate the weight computation.

To describe our inversion process, we first express our weight function as $w = uv$, where $v$ is any chosen pre-weight factor (e.g., based on sampling density, noise variance, data masking, reflection coefficient magnitudes, or simply all ones) and $u$ is a to-be-determined bandlimited function of the form $u=\tilde{F}h$, where $h$ is a compact real-space function defined on a $(2N_x-1) \times (2N_y-1) \times (2N_z-1)$ grid, and $\tilde{F}$ refers to the NDFT that maps real-space values on this grid to Fourier values on $\{\mathbf q_k\}_k$. This bandlimited ansatz on $u$ serves to stabilize the resulting nonuniform Fourier inversion operator by preventing large oscillations in the weights and, as we show below, allows us to leverage an additional multilevel Toeplitz structure to accelerate the weight calculation.  

Here we denote the dependency of the point spread function in Eq.~\ref{eq:Q} on $h$ as
\begin{equation}\label{eq:Qg}
Q_h[\mathbf n] = \sum_k (\tilde{F}h)[k]v[k]e^{2\pi i  \mathbf q_k\cdot\mathbf n},
\end{equation}
where $\mathbf n = (n_x, n_y, n_z)$ is defined on a $(2N_x-1) \times (2N_y-1) \times (2N_z-1)$ grid with $-N_d+1 \leq n_d \leq N_d-1$.

We now seek an $h$ such that $Q_h$ becomes a Kronecker delta function, i.e.,
\begin{equation}\label{eq:Q=delta}
Q_h[\mathbf n] = \delta[\mathbf n].
\end{equation}
where $\delta[\mathbf 0] = 1$ and $\delta[\mathbf n] = 0$ for $\mathbf n \neq \mathbf 0$. If Eq.~\ref{eq:Q=delta} can be enforced then, by Eq.~\ref{eq:F^*F}, we have that
\begin{equation}
F^*WF\rho = \rho \ast Q_h = \rho \ast \delta = \rho,
\end{equation}
and hence the normal equations in Eq.~\ref{eq:normal} simplify to
\begin{equation}\label{eq:complex_inverse}
\rho = F^*Wb,
\end{equation}
i.e., $F^*W$ becomes the pseudoinverse for the weighted linear system. Therefore, we can directly invert nonuniform Fourier data with a single application of $F^*W$ instead of using an iterative linear solver as described in Section~\ref{sec:INUDFT}.

To solve Eq.~\ref{eq:Q=delta}, note that Eq.~\ref{eq:Qg} can be expressed in matrix notation as $Q_h = \tilde{F}^*V\tilde{F}h$, where $V={\rm diag}(v[1],v[2],\hdots, v[K])$. Hence $h$ is a solution to the linear system
\begin{equation}\label{eq:delta}
\tilde{F}^*V\tilde{F}h = \delta,
\end{equation}
which can be solved in $\mathcal O(N'\log N')$ time using the Toeplitz-acceleration methods discussed in Section~\ref{sec:INUDFT}. 

Eq.~\ref{eq:delta} admits a unique and stable solution if and only if the subset of $\{\mathbf q_k\}_k$ where $v[k]$ is not vanishingly small adequately oversamples Fourier space. The details of this requirement will be explored in more detail in a later manuscript, but, in general, it suffices for this subset to oversample the Fourier volume by a factor of two, which is also a general requirement for solving the phase problem \cite{hayes1982reconstruction}. When this condition is met, the convergence of any iterative linear solver applied to Eq.~\ref{eq:delta} is best when the $v$-weighted sampling density is approximately uniform and worst when it contains sharp variations, e.g., peaks near the origin and/or poles for cylindrical or spherical-polar grids. A strategy for handling situations where the oversampling condition is not satisfied is discussed in the next subsection.

Once $h$ is obtained, we compute the final weights as $w[k] = (\tilde{F}h)[k]v[k]$, which essentially approximate $v$ multiplied by the inverse of the $v$-weighted sampling density. With this weight choice, by Eq.~\ref{eq:complex_inverse}, the weighted least squares solution to $F\rho=b$ becomes $\rho = F^*Wb$, which can be computed with a single type-1 NUFFT. We note that these weights only need to be computed once and can then be used throughout an iterative reconstruction algorithm as long as the nonuniform Fourier coordinates remain fixed.

A direct complex-to-real inversion is obtained by applying Friedel symmetry to the $\mathbf q_k$ coordinates and pre-weights $v$ when solving Eq.~\ref{eq:delta}, as discussed in Section~\ref{sec:INUDFT}. In this case, $h$ is treated as a real-valued function, and the pseudoinverse is now computed as
\begin{equation} \label{eq:real_inverse}
\rho = {\rm Re}(F^*Wb).
\end{equation}

An important property of this weighting scheme is that the NDFT $F$ forms an isometry between the set of real-space functions defined on the $N_x \times N_y \times N_z$ grid with the $\ell^2$ norm and the set of nonuniform Fourier-space functions defined on $\{\mathbf q_k\}_k$ with the $w$-weighted $\ell^2$ norm since
\begin{equation}\label{eq:isometry}
||Ff||_w^2 = ||\sqrt{W}Ff||_2^2 = f^*F^*WFf = f^*f = ||f||_2^2.
\end{equation}
This isometry implies that applying a Bregman projection to $f$ in real space using the $\ell^2$ norm is equivalent to applying the projection to its NDFT in Fourier space using the $w$-weighted norm. In Section~\ref{sec:NIP}, this property will allow us to formulate projection operators in nonuniform Fourier space, where they can have simpler expressions, with the guarantee that they are mathematically equivalent to their real-space counterparts.

\subsection{Frequency Filling for Stabilizing Inversion from Incomplete Fourier Data}\label{sec:freqfill}

Insufficient sampling of Fourier space can lead to ill-conditioned linear systems for both the nonuniform Fourier inversion (Eq.~\ref{eq:Fp-b}) and the weight calculation (Eq.~\ref{eq:delta}), resulting in convergence issues and numerical instabilities in the inversion process. While traditional regularization techniques \cite{calvetti2018inverse} are commonly used to address such ill-conditioning, they require fine-tuning solution-dependent regularization weights, can bias the solution by prioritizing reducing the regularization penalty term over directly matching known Fourier data, and are incompatible with the direct NDFT inversion methodology described in Section~\ref{sec:DirectInversion}.

To overcome these limitations, here we develop an alternative solution to stabilize the inversion process based on a frequency-filling approach. This approach utilizes model data generated from a previous solution estimate to fill in insufficiently sampled regions of Fourier space, whose coordinates are often referred to as frequencies. The main idea behind this method is to generalize the concept of allowing unsampled data regions to float, as is commonly done in uniform iterative phasing methods \cite{marchesini2003x}.

Suppose we wish to invert Fourier data $b(\mathbf q_k)$ sampled on a set of nonuniform coordinates $\mathcal V_{\rm data} = \{\mathbf q_k\}_k$ with corresponding pre-weight factors $v[k]$, where the coordinate set and pre-weights do not meet the oversampling requirements, described in the previous subsection, for computing the weights $w$. This means that there are regions of non-negligible size (e.g., with volumes larger than one-eighth of a Shannon voxel) where $\mathcal V_{\rm data}$ has no samples or $v$ is very small relative to its average value, e.g., due to data masking. Furthermore, assume that we have access to some solution estimate $\rho^{\rm prev}$ from either a previous iteration of a reconstruction algorithm or some form of initialization. To stably invert the associated NDFT in this case, we perform the following steps.

We begin by determining a set $\mathcal V_{\rm model}$ of Fourier coordinates such that the union $\mathcal V_f = \mathcal V_{\rm data} \bigcup\mathcal V_{\rm model}$ samples Fourier space at a sufficient rate to yield a stable NDFT inversion. Depending on the application, this region may be known a priori (e.g., a set of points outside a certain radius for cylindrical/spherical sampling) or can be inferred by estimating the $v$-weighted density of the coordinates in $\mathcal V_{\rm data}$. In principle, $\mathcal V_{\rm model}$ could be any set of coordinates that oversamples the missing regions by a factor of at least two, but placing these on a uniform grid sampled at twice the Nyquist rate is a convenient option. We will explore the exact conditions needed for a stable inversion in a future manuscript, but, in general, it is sufficient to choose $\mathcal V_{\rm model}$ so that the resulting $\mathcal V_f$ does not contain unsampled voxels with dimensions $\tfrac{1}{2D_x} \times \tfrac{1}{2D_y} \times \tfrac{1}{2D_z}$.

Once an appropriate $\mathcal V_{\rm model}$ is constructed, we can stably compute the optimal weights $w$ needed for direct nonuniform Fourier inversion by applying the steps in Section~\ref{sec:DirectInversion} to the set of $\mathbf q$ in $\mathcal V_{f}$, with $v(\mathbf q)$ set to a constant $v_{\rm model}$ for all $\mathbf q \in V_{\rm model}$. To avoid discontinuities at the boundary between $\mathcal V_{\rm data}$ and $\mathcal V_{\rm model}$, we suggest setting $v_{\rm model}$ to a value on the order of the average $v$ value in $\mathcal V_{\rm data}$. However, the process is not highly sensitive to this choice. As mentioned earlier, these weights depend only on the Fourier coordinates; therefore, they can be precomputed and then used repeatedly for inverting different Fourier data located on the same set of coordinates.

Once the optimal weights for $\mathcal V_f$ are computed, we can invert Fourier data $b$ sampled on $\mathcal V_{\rm data}$ by augmenting the data with information from $\rho^{\rm prev}$. In particular, we set the Fourier values for the $\mathbf q$ coordinates in $\mathcal V_{\rm model}$ to be the discrete Fourier transform values of $\rho^{\rm prev}$, thus forming the ``frequency-filled" Fourier data $b_f$ on $\mathcal V_{f}$ as
\begin{equation}
b_f(\mathbf q) = 
\begin{cases}
b(\mathbf q), &\text{if }\mathbf{q}\in \mathcal V_{\rm data},\\
\hat{\rho}^{\rm prev}(\mathbf q), &\text{if } \mathbf{q}\in \mathcal V_{\rm model}.
\end{cases}
\end{equation}
We can then apply Eq.~\ref{eq:complex_inverse} or \ref{eq:real_inverse} to $b_f$ to directly pseudoinvert the associated NDFT, which yields a solution $\rho$. This inversion process is equivalent to solving
\begin{equation}\label{eq:freqfill}
\min_{\rho}\!\sum_{\mathbf q \in\mathcal  V_{\rm data}}\hspace*{-.9em}|\hat{\rho}(\mathbf q) - b(\mathbf q)|^2w(\mathbf q) +\hspace*{-.5em} \sum_{\mathbf q \in\mathcal V_{\rm model}}\hspace*{-1em} |\hat{\rho}(\mathbf q) - \hat{\rho}^{\rm prev}(\mathbf q)|^2w(\mathbf q),
\end{equation}
i.e., it seeks to find a solution that is consistent with the available data, while minimizing the perturbation from the previous solution estimate for $\mathbf q$ where no nearby data is available. 

In practice, one can avoid computing Fourier values at the frequency-filled points in $\mathcal V_{\rm model}$ by representing the solution to Eq.~\ref{eq:freqfill} as $\rho = \rho^{\rm prev} + \Delta\rho$, where $\Delta\rho = F^*W(b-F\rho^{\rm prev})$ for complex-to-complex inversion and $\Delta\rho = {\rm Re}(F^*W(b-F\rho^{\rm prev}))$ for complex-to-real inversion. Since $b_f(\mathbf q)-F\rho^{\rm prev}(\mathbf q) = 0$ for $\mathbf q \in \mathcal V_{\rm model}$, the $F^*W$ operation in the expression for $\Delta\rho$ only needs to be computed using $\mathbf q \in \mathcal V_{\rm data}$. However, the weights must still be calculated using all points in $\mathcal V_{f}$.

\subsection{Accelerating the Nonuniform Fourier Transform and Its Inversion on DWBA Grids}\label{sec:CartAccel}
To reconstruct a density from CSSI data resampled onto the staggered grid defined in Section~\ref{sec:Cartesian}, we will need to repeatedly compute the NDFT and its pseudoinverse (defined using the $w$-weighted norm) for the set of nonuniform Fourier coordinates involved in computing the DWBA on the staggered CSSI intensity grid. We refer to this set of coordinates as the DWBA grid. Here we develop a strategy to further accelerate these operations on the DWBA grid. 

To define the DWBA grid, we first express the DWBA for each intensity value on the staggered grid as
\begin{equation}\label{eq:I_staggered_grid}
    I_j[\kp,\kz,t] = \Big|\sum_s R_{j,s}[\kz]\hat{\rho}_{j,s}[\kp,\kz, t]\Big|^2,
\end{equation}
where $\hat{\rho}_{j,s}$ and $R_{j,s}$ are defined in Eqs.~\ref{eq:4J}-\ref{eq:4JR} using $\qp = (k_x\Delta_x, k_y\Delta_y)$ and $\qz = k_z\Delta_z+tc_j$, where $\Delta_d = 1/(2N_d-1)$ and $c_j = \qzij -\lfloor (2N_z-1)\qzij \rfloor/(2N_z-1)$.

\begin{figure*}[t]
    \centering
    \includegraphics[width=\linewidth]{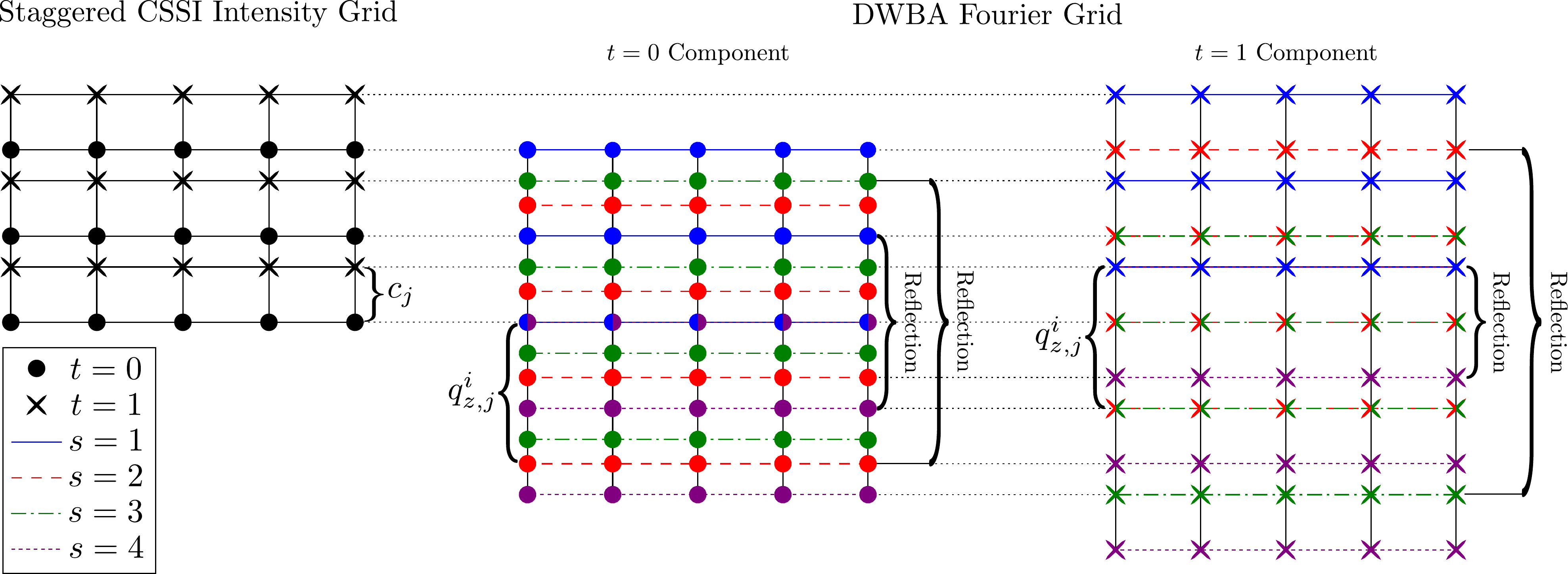}
    \caption{Illustration of the DWBA grid (defined in Eq.~\ref{eq:Fourier_grid}) and comparison to its associated staggered CSSI intensity grid (defined in Section~\ref{sec:Cartesian}) in the $xz$ plane. For each incident angle and each of the $t=0$ and $t=1$ components of the staggered intensity grid, the DWBA grid comprises 4 components ($s=1,2,3,4$) given by the coordinates of the $s$-th term in the DWBA. Each DWBA grid component is either a point on the intensity grid or a shift and/or reflection of one: the $s=1$ component aligns with the intensity grid; the $s=2$ component is shifted by $\qzij$; and the $s=3$ and $s=4$ components are reflections of the $s=2$ and $s=1$ components, respectively, through the $xy$ plane.}
    \label{fig:Fourier_grid}
\end{figure*}

For a density $\rho$ defined on an $N_x \times N_y \times N_z$ grid, we define the corresponding DWBA grid $\{\mathbf q^s_j[\kp,\kz,t]\}_{s,j,\kp,\kz,t}$ as the set of $8J(2N_x-1)(2N_y-1)N_z$ Fourier coordinates needed to evaluate Eq.~\ref{eq:I_staggered_grid}, i.e.,
\begin{equation}\label{eq:Fourier_grid}
\begin{aligned}
    \mathbf q^1_j[\kp,\kz,t] &= (k_x\Delta_x, k_y\Delta_y, k_z\Delta_z+tc_{j})\\
    \mathbf q^2_j[\kp,\kz,t] &= (k_x\Delta_x, k_y\Delta_y, k_z\Delta_z+tc_{j}-\qzij)\\
    \mathbf q^3_j[\kp,\kz,t] &= (k_x\Delta_x, k_y\Delta_y, -k_z\Delta_z-tc_{j}+\qzij)\\
    \mathbf q^4_j[\kp,\kz,t] &= (k_x\Delta_x, k_y\Delta_y, -k_z\Delta_z-tc_{j}),
\end{aligned}
\end{equation}
where $-N_d+1 \leq k_d \leq N_d-1$ for $d=x,y$; $0\leq k_z \leq N_z-1$; $t=0,1$; and $1 \leq j \leq J$. Note that $\mathbf q_j^s[k_{||},k_z,t]$ gives the coordinate of the $s$-th Fourier term in the DWBA for the CSSI intensity at $(k_{||},k_z)$ for the $t$-th component of the staggered intensity grid, i.e., $\hat{\rho}(\mathbf q_j^s[k_{||},k_z,t]) = \hat{\rho}_{j,s}[\kp,\kz, t]$. See Fig.~\ref{fig:Fourier_grid} for an illustration of the DWBA grid.

By design, the DWBA grid contains duplicate points, allowing us to formulate a simple and efficient Bregman projection operator to enforce consistency between Fourier values on this grid and the CSSI intensity data, as shown in Section~\ref{sec:PD}.

Let $F$ be the NDFT for the set of Fourier coordinates on the DWBA grid and denote $(F\rho)_{j,s}[\kp,\kz,t]=\hat{\rho}(\mathbf q_j^s[k_{||},k_z,t]) = \hat{\rho}_{j,s}[\kp,\kz, t]$. Although $F$ can be computed with a NUFFT, we can achieve much better performance by leveraging the various symmetries and redundancies in the DWBA grid to represent the NDFT in terms of $J+1$ FFTs. We derive this acceleration for a real-valued density $\rho$ to achieve greater speedup, though similar expressions can be obtained for complex-valued densities as well.

To obtain this FFT representation, we need to identify each of the $(F\rho)_{j,s}[\kp,\kz,t]$ as an output of an FFT on a uniform grid. Now define $\hat{\rho}_0 = {\rm FFT}({\rm Pad}(\rho))$ and $\hat{\rho}_j = {\rm FFT}({\rm Pad}(\rho_j))$, where $\rho_{j}[m_{||},m_z] = \rho[m_{||},m_z]e^{2\pi i \qzij m_z}$ and ${\rm Pad}(\rho)$ zero-pads the density to a $(2N_x-1) \times (2N_y-1) \times (2N_z-1)$ grid. We can now identify each of the Fourier values on the DWBA grid with values on the uniform grids given by $\hat{\rho}_0$ or one of the $\hat{\rho}_j$ via the following operations/identities:
\begin{enumerate}[1)]
\item If $\mathbf (q_x, q_y,\qz)$ is located at a grid point, then any reflections and translations by an integer multiple of the grid spacing, i.e., $(\pm q_x \pm n_x\Delta_x, \pm q_y \pm n_y\Delta_y, \pm \qz \pm n_z\Delta_z)$, are also on the grid for any integers $n_x,n_y,n_z$, provided the new point remains within the grid boundaries.
\item The Fourier modulation/shift identity gives $\hat{\rho}_{j}(\qp,\qz) = \hat{\rho}(\qp,\qz-\qzij)$. This allows us to evaluate off-grid shifts of $-\qzij$ by evaluating $\hat{\rho}_j$ on a uniform grid.
\item To attain positive shifts by $\qzij$, we can apply the Friedel-symmetry identity and the result in 2) to get $\hat{\rho}(\qp,\qz+\qzij) = \overline{\hat{\rho}(-\qp,-\qz-\qzij)} = \overline{\hat{\rho}_{j}(-\qp,-\qz)}$. 
 \end{enumerate}

By applying the above operations, we can express the NDFT values for the DWBA grid in terms of the previously defined $J+1$ FFTs via
\begin{equation} \label{eq:CartFFT_forward}
(F\rho)_{j,s}[\kp,\kz, t] = 
\begin{cases}\hat{\rho}_0[\kp, \kz], & \text{ if }s = 1, t = 0, \\
\hat{\rho}_j[\kp,\kz], &  \text{ if }s = 2, t = 0, \\
\overline{\hat{\rho}_j}[-\kp,\kz], &  \text{ if }s = 3, t = 0, \\
\hat{\rho}_0[\kp,-\kz], &  \text{ if }s = 4, t = 0,\\
\overline{\hat{\rho}_j}[-\kp,-\kz+k_j],\hspace*{-2em} & \text{ if }s = 1, t = 1, \\
\hat{\rho}_0[\kp,\kz-k_j], &  \text{ if }s = 2, t = 1, \\
\hat{\rho}_0[\kp,-\kz+k_j], &  \text{ if }s = 3, t = 1, \\
\hat{\rho}_j[\kp,-\kz+k_j], &  \text{ if }s = 4, t = 1,
\end{cases}
\end{equation}
where $k_j = (\qzij-c_j)(2N_z-1) = \lfloor (2N_z-1)\qzij \rfloor$ gives the number of voxels between $\qzij$ and $c_j$.

As can be seen from Eq.~\ref{eq:CartFFT_forward}, the $t=1$ terms can be obtained directly from the FFTs needed for the $t=0$ terms. As mentioned in Section~\ref{sec:Cartesian}, this feature was a major motivation behind our design of the nonuniform CSSI intensity grid, i.e., the inclusion of the $t=1$ points can improve reconstruction stability with almost no additional computational cost.

The above acceleration can also be extended to the direct nonuniform Fourier inversion scheme described in Section~\ref{sec:DirectInversion} for Fourier data on a DWBA grid. Given a set of nonuniform Fourier data $b_{j,s}[\kp,\kz,t]$ defined on the DWBA Grid, and the corresponding optimized weights $w_{j,s}[\kp,\kz,t]$ described in Section~\ref{sec:DirectInversion}, denote the weighted Fourier data as $d_{j,s}[\kp,\kz,t] = b_{j,s}[\kp,\kz,t]w_{j,s}[\kp,\kz,t]$. The pseudoinverse in Eq.~\ref{eq:real_inverse} can then be represented as $\rho = \text{Re}(F^*Wb) = \text{Re}(F^*d)$, which can be computed using $J+1$ inverse FFT (IFFT) operations by applying the adjoint of the operations in Eq.~\ref{eq:CartFFT_forward} in reverse order. 

The adjoint operation of Eq.~\ref{eq:CartFFT_forward} maps the $d_{j,s}[\kp,\kz,t]$ values onto $J+1$ different uniform grids: $d_0$ and $d_j$ for $j=1,\hdots, J$, through
\begin{equation}
\begin{aligned}
d_0[\kp,\kz] &= \sum_j\big( d_{j,1}[\kp,\kz,0] + d_{j,2}[\kp,\kz+k_j,1]\\ &+ d_{j,3}[\kp,-\kz+k_j,1] +d_{j,4}[\kp,-\kz,0]\big),\\
d_j[\kp,\kz] &= \overline{d_{j,1}}[-\kp,-\kz+k_j,1] + d_{j,2}[\kp,\kz,0]\\ &+ \overline{d_{j,3}}[-\kp,\kz,0] +d_{j,4}[\kp,-\kz+k_j,1].
\end{aligned}
\end{equation}

The adjoint operation for $\rho\mapsto\hat{\rho_0}$ is the IFFT, followed by a restriction to the $N_x\times N_y\times N_z$ grid. The adjoint for $\rho\mapsto\hat{\rho}_j$ is the IFFT multiplied by $e^{-2\pi i \qzij m_z}$, followed by the above restriction. Applying these adjoint operations to $d_0$ and $d_j$ then allows us to compute the pseudoinverse in Eq.~\ref{eq:real_inverse} on a DWBA grid as
\begin{equation}\label{eq:DINUFTaccel}
\begin{aligned}
\text{Re}(F^*Wb)[\mathbf m] &= \text{Re}(F^*d)[\mathbf m]\\
&= \text{Re}\Big({\rm IFFT}(d_0)[\mathbf m]\!\\
&\hspace{3em} + \sum_je^{-2\pi i \qzij m_z }{\rm IFFT}(d_j)[\mathbf m]\Big)\!,
\end{aligned}
\end{equation}
where $\mathbf m$ is restricted to the $N_x\times N_y\times N_z$ grid.

Additional speedup in computing Eqs.~\ref{eq:CartFFT_forward} and \ref{eq:DINUFTaccel} can be achieved by leveraging the fact that the 2D FFT in the $xy$ dimensions can be shared across the $J+1$ 3D FFTs in each case. In particular, $\hat{\rho}_j$ can be obtained by first taking the 2D FFT of ${\rm Pad}(\rho)$ in the $xy$ dimensions, multiplying the result by $e^{2\pi i \qzij m_z}$, and then taking the 1D FFT of this product in the $z$ dimension. Similarly, Eq.~\ref{eq:DINUFTaccel} can be accelerated by computing the 1D IFFT of each $d_j$ in the $z$ dimension, multiplying the results by $e^{-2\pi i \qzij m_z}$, summing these products with the 1D IFFT of $d_0$, and then computing the 2D IFFT in the $xy$ dimensions. Therefore, the forward and inverse transforms can each be computed using a single 2D FFT and $J+1$ 1D FFTs, reducing the cost from $J+1$ times to $J/3+1$ times that of a single 3D FFT.

As discussed in Section~\ref{sec:Cartesian}, the $t=1$ intensity samples in the topmost plane ($k_z=N_z-1$) of the staggered grid can become aliased if $c_j$ is too large. In this situation, the corresponding Fourier samples ($k_z=N_z-1$, $t=1$, $s=1,2,3, 4$) on the DWBA grid may also become aliased. However, this aliasing does not affect the validity of Eqs.~\ref{eq:CartFFT_forward} and \ref{eq:DINUFTaccel}, and the masking procedure in Section~\ref{sec:prelim} removes any aliasing effects by setting the weights for the aliased grid points to zero.

\section{CSSI Reconstruction via Nonuniform Iterative Phasing}\label{sec:NIP}

Here we develop a new CSSI reconstruction algorithm that can determine the 3D density of a specimen from CSSI rotation-series data collected at one or more incident angles. Our approach recasts conventional iterative phasing techniques (described in Section~\ref{sec:IPA}) in a nonuniform reconstruction framework formulated in Fourier space, allowing Fourier values sampled at arbitrary coordinates to be accurately and efficiently handled. This framework is coupled with a generalization of the magnitude projection for an intensity described by the DWBA to enable reconstruction from CSSI data on either the original measurement images or the resampled staggered grid. 

Although we focus here on results from the staggered grid, the algorithms presented below can also be applied directly to the original unreduced data at the cost of increased compute and memory requirements. Reconstruction directly from the original data could be beneficial if additional physical modeling or data corrections are needed. Additionally, the presented nonuniform iterative phasing framework can be generalized to other forms of the phase problem by substituting the DWBA magnitude projection with the appropriate projection for the data. These extensions will be explored in a future manuscript.

\subsection{Preliminaries}\label{sec:prelim}

Given CSSI intensity data $I_j(\mathbf q)$ for $j=1,\hdots,J$, we formulate our projection operators and iterative phasing schemes in Fourier space on the set of $\hat{\rho}_{j,s}(\mathbf q)$, $j=1,\hdots, J$ and $s=1,\hdots, 4$, defined in Eq.~\ref{eq:4J}, at the same coordinates $\mathbf q$ as the intensity data. Our formulation does not require any assumptions on the set of $\mathbf q$, but the computational and memory requirements are substantially reduced if $I_j$ is resampled on a staggered grid for each $j$, which means that $\hat{\rho}_{j,s}(\mathbf q)$ are on a DWBA Fourier grid. To simplify notation, we refer to the vector of all $\hat{\rho}_{j,s}(\mathbf q)$ as simply $\hat{\rho}$ and suppress $\mathbf q$-dependency for any formulas that hold for all $\mathbf q$.

With the above formulation, we express the CSSI reconstruction problem in Fourier space as follows. Given a set of CSSI intensity values $I_j(\mathbf q)$ and support $S$, we seek a set of Fourier values $\hat{\rho}$ such that
\begin{equation}
\begin{aligned}
    &\left|\sum_{s=1}^4 R_{j,s} \hat{\rho}_{j,s}(\mathbf q) \right| = \sqrt{I_{j}(\mathbf q)}, \text{ for all } \mathbf q \text { and } j, \text{ and }\\
    &\hat{\rho} = F\rho \text{ for some $\rho$ satisfying } \rho(\mathbf r) = 0 \text{ for } \mathbf r \notin S.
\end{aligned}
\end{equation}
Once $\hat{\rho}$ is found, we can recover the real-space density solution $\rho$ by inverting the associated NDFT.

This specific Fourier-space formulation allows us to 1) derive a simple expression for the DWBA magnitude projection in Section~\ref{sec:PD}, 2) formulate the support projection in Fourier space using a projection norm consistent with the DWBA projection, and 3) represent the HIO negative feedback update completely in Fourier space, which enables the use of a smaller unpadded real-space grid to compute the associated nonuniform Fourier transforms.

As mentioned in Section~\ref{sec:Cartesian}, we need to mask out any unsampled or aliased regions from the resampled CSSI diffraction volumes. In particular, define the mask $m_j(\mathbf q) = 1$ if $\mathbf q$ is in a sampled and alias-free region of the $j$-th diffraction volume and $m_j(\mathbf q) = 0$ otherwise. 

We set the pre-weight factors to $v_{j,s}(\mathbf q) = \sqrt{|R_{j,s}(\qz)|}$ when $m_j(\mathbf q) = 1$ and $v_{j,s}(\mathbf q) = 0$ when $m_j(\mathbf q) = 0$. This choice of $v$ ensures that the CSSI reconstruction algorithm reduces to the classical iterative phasing algorithms in Section~\ref{sec:IPA} in the limit as the reflection coefficients tend to zero. Using the square root of the reflection coefficient magnitudes for the pre-weights yields the best empirical performance among similar tested power-law choices, though the improvements are marginal. In general, assigning larger pre-weights to terms that contribute more significantly to the data can improve convergence, but the algorithm is not highly sensitive to the exact weighting.

Here we compute all nonuniform Fourier transforms and their inverses using the acceleration techniques discussed in Section~\ref{sec:acceleration}. To handle any ill-posedness in inversion induced by $\mathbf q$-regions that are masked out or have low weight, we use the frequency-filling approach in Section~\ref{sec:freqfill}. To determine the filled region, we estimate the $v$-weighted sampling density by computing the Fourier transform $\hat{Q}$ of the associated point-spread function $Q$ defined in Eq.~\ref{eq:Q} (using $v$ instead of $w$). Here $\hat{Q}$ is computed on a uniform grid $\mathcal G$ via an FFT. Then, we construct $\mathcal V_{\rm model} = \{\mathbf q \in \mathcal G: |\hat{Q}(\mathbf q)| < 0.5 \}$, which estimates all $\mathbf q$ on the grid that are more than half a Nyquist length away from any reliable Fourier information involved in the DWBA summation for the measured intensity. 

In the following subsections, we formulate a series of new Bregman projection operators in the $w$-weighted $\ell^2$ norm, where $w$ is computed via Section~\ref{sec:DirectInversion} on the frequency-filled $\mathbf q$ points. In particular, this weighting mitigates any bias in the iteration steps caused by the clustering of points in the nonuniform sampling, and it allows us to leverage the properties in Section~\ref{sec:DirectInversion} to simplify some of the calculations needed to compute the projections. Furthermore, by Eqs.~\ref{eq:PC} and \ref{eq:isometry}, projecting a real-space function to a constraint set using the $\ell^2$ norm is equivalent to projecting the NDFT of that function to the corresponding constraint set in Fourier space using the $w$-weighted $\ell^2$ norm, which makes the nonuniform Fourier formulation of the projections consistent with uniform-grid real-space formulations described in Section~\ref{sec:IPA}.

\subsection{DWBA Magnitude Projection}\label{sec:PD}
Now we seek to formulate a DWBA magnitude projection $P_D$ that, for each $\mathbf q$ and $j$, maps the set of four Fourier values $\hat{\rho}_{j,s}(\mathbf q)$ for $s=1,\hdots,4$ to the closest values (in the $w$-weighted norm) consistent with the CSSI intensity measurement $I_j(\mathbf q)$. This projection can be expressed as the solution to the constrained minimization
\begin{equation}\label{eq:projmin}
\begin{aligned}
P_{D}\hat{\rho} = &\argmin_h ||h-\hat{\rho}||_w\\
&\text{subject to  } \left|\sum_{s=1}^4 R_{j,s}(\qz) h_{j,s}(\mathbf q) \right| = \sqrt{I_{j}(\mathbf q)},\\ &\qquad \text{for all } j = 1,\hdots,J \text{ and } \mathbf q \text{ with } m(\mathbf q) = 1.
\end{aligned}
\end{equation}
For any $\mathbf q$ where $I_j(\mathbf q)$ is not available, i.e., where the mask $m_j(\mathbf q) = 0$ or at any frequency-filled points, the projection allows the associated Fourier values to float: $(P_D\hat{\rho})_{j,s}(\mathbf q) = \hat{\rho}_{j,s}(\mathbf q)$.

Recall from Section~\ref{sec:CartAccel} that $\hat{\rho}$ contains duplicate points when defined on a DWBA grid. Note that the constraint in Eq.~\ref{eq:projmin} does not enforce the projected values $P_{\rm D}\hat{\rho}$ to be the same on duplicate points. This feature allows the minimization in Eq.~\ref{eq:projmin} to be decoupled across different $\mathbf q$ and $j$, thus simplifying the projection calculation. Specifically, for all $\mathbf q$ and $j$ where $\sum_{s} R_{j,s}(\qz)\hat{\rho}_{j,s}(\mathbf q) \neq 0$, the solution to Eq.~\ref{eq:projmin} has the following analytic expression (derived in Section S2 of the Supplemental Material):
\begin{equation}\label{eq:PD}
P_{\rm D}\hat{\rho}_{j,s} = \hat{\rho}_{j,s} + \frac{\overline{R_{j,s}}\!\sum_{s'}\! R_{j,s'} \hat{\rho}_{j,s'}}{w_{j,s}\!\sum_{s'}\! w_{j,s'}^{-1}|R_{j,s'}|^2}\hspace*{-.28em}\left(\hspace*{-.28em} \frac{\sqrt{I_{j}}}{|\!\sum_{s'} R_{j,s'}\hat{\rho}_{j,s'}\!|}-1\hspace*{-.30em}\right)\hspace*{-.25em},
\end{equation}
where $\mathbf q$ dependencies have been suppressed. For $\mathbf q$ and $j$ where $\sum_{s} R_{j,s}(\qz)\hat{\rho}_{j,s}(\mathbf q) = 0$, Eq.~\ref{eq:projmin} has multiple solutions. At these coordinates, we choose the convention $P_{\rm D}\hat{\rho}_{j,s} = \hat{\rho}_{j,s}+ \overline{R_{j,s}}\sqrt{I_j}/(w_{j,s}\sum_{s'} w_{j,s'}^{-1}|R_{j,s'}|^2)$.

Consistency between duplicate points is enforced as part of a separate linear projection, introduced in the next subsection.

\subsection{Fourier Column-Space Projection}
Here we formulate a Bregman projection operator in Fourier space, using the $w$-weighted norm, to simultaneously enforce equality of values at duplicate Fourier coordinates and consistency with a specified real-space support region $S$. This operation is equivalent to projecting a function defined on a set of nonuniform Fourier coordinates to the closest function that is the NDFT of a real-space function supported in $S$ (the set of all such functions is the column space of the NDFT restricted to $S$). Here we assume that either the nonuniform Fourier coordinates have sufficient coverage of Fourier space to stably solve Eq.~\ref{eq:delta} or that any undersampled regions have been frequency-filled as described in Section~\ref{sec:freqfill}.

Note that the set of all functions that are the NDFT of a function supported in $S$ can be expressed in Fourier space as the column space of $FP_S$, defined as
\begin{equation}
\text{Col}(FP_S) = \{h: \text{ there exists } g, \text{ such that } h = FP_Sg  \}.
\end{equation}

The Bregman projection operator that maps a function defined on a set of nonuniform Fourier coordinates to the closest function in $\text{Col}(FP_S)$ using the $w$-weighted norm is given by
\begin{equation}\label{eq:PF}
    P_F\hat{\rho} = \argmin_{h\in\text{Col}(FP_S)}||h-\hat{\rho}||_w.
\end{equation}
To compute Eq.~\ref{eq:PF}, we first note that, by definition, $h \in {\rm Col}(FP_S)$ means that $h = FP_Sg$ for some function $g$. Therefore, we have that
\begin{equation}\label{eq:PFW}
||h-\hat{\rho}||_w = ||\sqrt{W}(h - \hat{\rho}) ||_2 = ||\sqrt{W}FP_Sg - \sqrt{W}\hat{\rho} ||_2.
\end{equation}
If $h$ solves Eq.~\ref{eq:PF}, then $g$ minimizes Eq.~\ref{eq:PFW}, which is equivalent to satisfying the associated normal equations
\begin{equation}\label{eq:PFnormal}
P_SF^*WFP_Sg = P_SF^*W\hat{\rho}.
\end{equation}
Here we used the fact that $P_S$ is Hermitian, i.e., $P_S^* = P_S$. 

Now take $w$ to be the optimal weights described in Section~\ref{sec:DirectInversion}, which causes $F^*WF$ to become the identity matrix. Furthermore, since $P_S$ is a projection matrix we have that $P_SP_S = P_S$, and so we can simplify Eq.~\ref{eq:PFnormal} to
\begin{equation}
P_Sg = P_SF^*W\hat{\rho}.
\end{equation}
Therefore, the minimizer in Eq.~\ref{eq:PF} is 
\begin{equation}
\begin{aligned}    
h &= FP_Sg\\ 
&= FP_SF^*W\hat{\rho},
\end{aligned}
\end{equation}
allowing us to express the column-space projection operator as 
\begin{equation} \label{eq:PF2}
    P_F\hat{\rho} = FP_SF^*W\hat{\rho}.
\end{equation}

Note that the above projection is valid for an arbitrary set of Fourier coordinates, as long as the Fourier volume is sufficiently sampled or the frequency-filling technique has been applied to ensure that the linear system in Eq.~\ref{eq:delta} needed for the optimal weight calculation is full rank. However, the calculation of $F$ and $F^*W$ in $P_F$ can be accelerated if the Fourier coordinates are located on the structured CSSI grid described in Section~\ref{sec:acceleration}. 

\subsection{Nonuniform Iterative Phasing for 3D CSSI Reconstruction}\label{sec:CSSIIPA}
We now describe how to combine the projections in the previous subsections in an iterative phasing scheme to reconstruct a 3D density from its CSSI rotation-series data measured at one or more incident angles. These schemes are based on generalizations of the ER and HIO schemes discussed in Section~\ref{sec:IPA}. 

We start with an initial random density $\rho^{(0)}$ and compute its Fourier transform values $\hat{\rho}^{(0)}$ on the nonuniform Fourier coordinates described in Section~\ref{sec:prelim}. We then iteratively apply the following schemes to produce the Fourier iterate $\hat{\rho}^{(n)}$ after $n$ iterations. The corresponding density $\rho^{(n)}$ can be retrieved as needed via the direct nonuniform Fourier inversion in Section~\ref{sec:acceleration}. 

The first iterative scheme we discuss is a local optimizer based on reformulating the ER scheme in Fourier space, which simply consists of alternating between the DWBA and column-space projections and is given by
\begin{equation}
\hat{\rho}^{(n+1)} = P_F P_{\rm D}\hat{\rho}^{(n)}.
\end{equation}
Similar to the original ER method discussed in Section~\ref{sec:IPA}, the above scheme is a local optimizer, which excels at reducing overall error but quickly gets trapped in local minima.

To escape local minima, we formulate an extension of the coordinate-free HIO representation in Eq.~\ref{eq:hio_coord_free} in Fourier space, which is given by 
\begin{equation}\label{eq:HIO2}
\begin{aligned}
\hat{\rho}^{(n+1)} &= P_F{P_{\rm D}\hat{\rho}^{(n)}} + \hat{\eta}^{(n+1)},\\
\hat{\eta}^{(n+1)} &= \hat{\eta}^{(n)} - \beta m({P_{\rm D}\hat{\rho}^{(n)}} - P_F{P_{\rm D}\hat{\rho}^{(n)}}),
\end{aligned}
\end{equation}
where $\beta \in (0,1]$ is a constant. Here, the negative feedback update term is initialized as $\hat{\eta}^{(0)} = \hat{\rho}^{(0)}-P_F\hat{\rho}^{(0)}$. 

Note that this Fourier formulation of HIO allows the mask $m$ (described in Section~\ref{sec:prelim}) to be used to turn the negative feedback off for any $\mathbf q$ not sampled by the intensity data. This masking procedure helps stabilize the iterative scheme, since without it the magnitude of $\hat{\eta}^{(n)}(\mathbf q)$ could grow without bound for any $\mathbf q$ not sampled by the data.

Another benefit of this Fourier-space formulation of HIO is that it does not require faithfully representing $\hat{\eta}$ in real space, i.e., such that $F\eta = \hat{\eta}$. Due to the use of duplicate Fourier coordinates that can take inconsistent values under the DWBA projector, such a real-space representation is not possible in our formulation. Moreover, even if no duplicate points were present or values were consistent among them, $\hat{\eta}$ is generally not bandlimited, meaning that representing $\eta$ on a real-space grid would require at least as many grid points as there are unique Fourier coordinates, which could be as many as $(2J+1)(2N_x-1)(2N_y-1)(2N_z-1)$ for the DWBA grid. Since Eq.~\ref{eq:HIO2} updates $\hat{\eta}$ in Fourier space and since the NDFT does not require any zero padding to compute an oversampled $\hat{\rho}$, we only need to compute and invert the NDFT for $\rho$ on an $N_x \times N_y \times N_z$ real-space grid, thus reducing the overall memory and computational requirements.

\begin{figure}[h!] 
 \centering
\includegraphics[width = .475\textwidth]{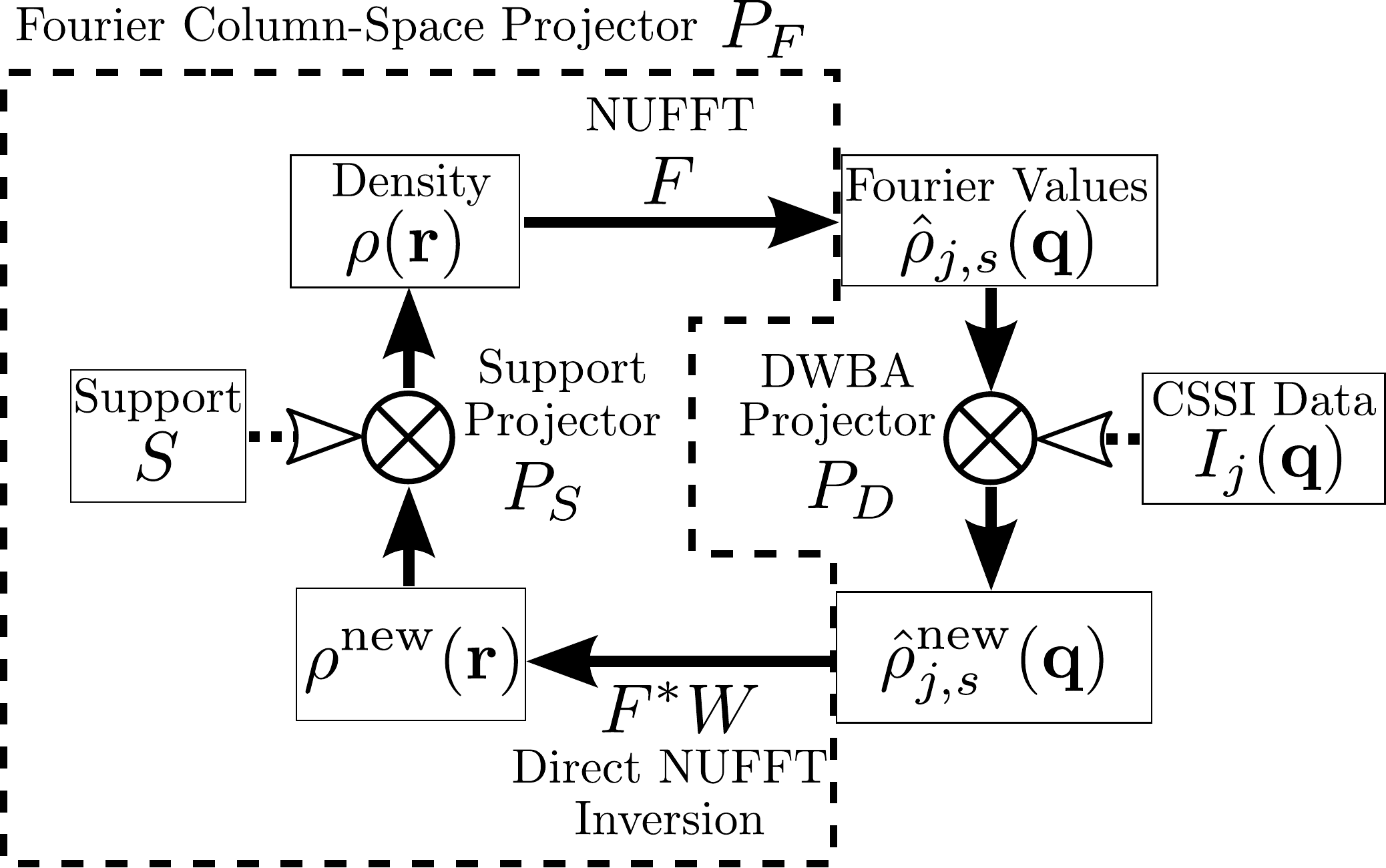}
\caption{Flow diagram for the nonuniform iterative phasing CSSI reconstruction algorithm. Solid arrows represent the order of operations, open-dashed arrows represent dependencies, and crossed circles represent operations that combine multiple quantities. The three operations on the left (direct NUFFT inversion, support projection, and NUFFT) make up the column space projection $P_F$.}
\label{fig:mtip}
\end{figure}

In practice, we run several cycles, each consisting of several iterations of our nonuniform Fourier HIO scheme, followed by several iterations of our nonuniform Fourier ER scheme, and then Shrinkwrap to update the support estimate $S$ that is used in the column-space projection $P_F$. See Fig.~\ref{fig:mtip} for a summary of the operations in each iteration.

The presented nonuniform iterative phasing framework can be extended to reconstruction problems for other experimental modalities by replacing $P_D$ with the Bregman projection that enforces the data constraint for that modality. For example, this framework could be especially useful in cases where intensity data is sparsely sampled or most naturally represented on polar or spherical-polar grids, e.g., \cite{donatelli2015iterative, donatelli2017reconstruction}. In addition, other popular iterative phasing schemes \cite{elser2003phase, elser2007searching, luke2004relaxed, bauschke2003hybrid, elser2018benchmark} can be generalized to this framework.

\section{Experimental Parameter Requirements and Uniqueness Properties}\label{sec:uniqueness}

In this section, we derive conditions on the experimental parameters needed for 3D \textit{ab initio} CSSI reconstruction. We then discuss general uniqueness properties of the CSSI reconstruction problem and how they differ from the classical phase problem.

Here we assume that the CSSI diffraction patterns are oversampled by at least a factor of two, and we seek conditions on the experimental parameters to ensure that the associated diffraction volume is oversampled by at least the same amount. Whereas oversampling is necessary for the classical phase problem \cite{hayes1982reconstruction}, CSSI reconstruction may be possible without it if additional data is included from sufficiently many incident angles, as the phases combine differently in the DWBA for each incident angle. However, from a practical standpoint, sampling at a rate that is too low washes out the signal, as each pixel measures an average of the intensity over its surface. Ideally, the diffraction patterns should be oversampled by a modestly large factor (e.g., eight) to mitigate this averaging effect. Although it may be possible to model the intensity averaging within each pixel, we do not address this here. 

\subsection{Incident-Angle Requirements}\label{sec:incident}

Here we investigate how the solvability of the CSSI reconstruction problem is related to the set of incident angles used in the CSSI data measurements. Specifically, we demonstrate how this solvability is closely linked to the magnitude of the incident Fresnel reflection coefficient $|R_i|$, which depends on the substrate's index of refraction $n_r$ and the incident angle $\alpha_i$ via Eqs.~\ref{eq:k}-\ref{eq:r}. See Fig.~\ref{fig:Ri} for an example of the reflection coefficient profile. We now analyze the solvability across three incident-angle regimes characterized by the value of $|R_i|$. 

\medskip\noindent\emph{Case 1: small $\alpha_i$, $|R_i| \approx 1$.}
Recall from Eqs.~\ref{eq:g}-\ref{eq:gintoI} that the CSSI intensity can be represented as 
\begin{equation}
    I(\qp,\qz) = |\hat{g}(\qp,\qz) + R_f(\qz)\hat{g}(\qp,-\qz+\qzi)|^2,
\end{equation}
where 
\begin{equation}
    g(\rp,\rz) = \left(1 + R_i e^{2\pi i \qzi \rz}\right)\rho(\rp,\rz).
\end{equation}
Define the interference factor $\mathfrak s(\rz,\qzi)$ as the above term in parentheses, i.e.,
\begin{equation}
    \mathfrak s(r_z,\qzi) = 1 + R_i e^{2\pi i \qzi \rz}.
\end{equation}
Note that $\mathfrak s(r_z,\qzi) \approx 0$ near all $r_z$ satisfying
\begin{equation}\label{eq:rz}
    \rz = \frac{-\arg(R_i)+(2n+1)\pi}{2\pi\qzi},\quad n \in \mathbb Z,
\end{equation}
where $\arg$ refers to the complex argument. For all $\rz$ close to satisfying Eq.~\ref{eq:rz}, $g(\rp,\rz) \approx 0$, implying that $g$, and consequently the intensity $I$, are essentially invariant to any changes in $\rho$ at these coordinates. Therefore, the density at these locations cannot be reliably reconstructed from the intensity measured from a single incident angle in this range. Note that this does not cause an issue when the vertical extent of the specimen does not overlap with the troughs in the interference function, i.e., when $|\mathfrak s(\rz,\qzi)| \gg 0$ for all $\rz < D_z$, where $D_z$ is the specimen height.

The above reconstruction breakdown can be avoided by including data from multiple incident angles $\alpha_{i,j}$ if, for each point in the vertical range of the specimen, at least one of the corresponding interference functions is sufficiently large, i.e., when
\begin{equation}\label{eq:multiple}
\max_j|\mathfrak s(\rz,\qzij)| \gg 0, \quad \text{ for all } \rz < D_z.
\end{equation}
Eq.~\ref{eq:multiple} can typically be satisfied with just two different incident angles, ideally chosen to maximize the maximum interference function magnitude at each point. However, more incident angles could be used to further increase the maximum interference function magnitude, which may help improve reconstruction stability.

\medskip\noindent\emph{Case 2: large $\alpha_i$, $|R_i| \approx 0$.}
In this case, the second and fourth terms of the DWBA in Eq.~\ref{eq:CSSI} drop out. Since data for $\alpha_f < 0$ is not collected in a reflection geometry, the first term (i.e., the Born term) only samples the Fourier transform $\hat{\rho}$ for $\qz = (\sin(\alpha_f) + \sin(\alpha_i))/\lambda \geq \qzi/2$. While the third term samples $\hat{\rho}$ for $\qz = (-\sin(\alpha_f)+\sin(\alpha_i))/\lambda \leq \qzi/2$, note that its corresponding reflection coefficient $R_f(\alpha_f) \rightarrow R_i(\alpha_i)\approx 0$ as $-\sin(\alpha_f)+\sin(\alpha_i) \rightarrow 0$. Therefore, the third DWBA term also drops out for $-\sin(\alpha_f)+\sin(\alpha_i) \approx 0$, so it is unable to sample $\hat{\rho}$ for small $\qz$. Consequently, for $\qz$ below a certain threshold, $\hat{\rho}(\qp,\qz)$  cannot be sampled by any of the four terms in the DWBA, making \textit{ab initio} reconstruction impossible at these incident angles unless additional constraints can be applied to help compensate for the missing Fourier information.

\medskip\noindent\emph{Case 3: moderate $\alpha_i$, $0\ll|R_i|\ll1$.}
In this case, $|\mathfrak s(\rz,\qzi)| \gg 0$ for all $\rz$ and each Fourier transform value $\hat{\rho}(\qp,\qz)$ is sampled by at least one term in the DWBA for the entire $\qz$ range. This suggests that \textit{ab initio} CSSI reconstruction should be possible with data collected at just a single incident angle, as long as the other sampling conditions discussed in this section are satisfied.

\medskip\noindent\emph{Case Summary:} The relationship between the solvability of the CSSI reconstruction problem and the incident angle can be summarized as follows:
\begin{itemize}
    \item For very small $\alpha_i$, where $|R_i| \approx 1$, CSSI data from two or more incident angles may be needed for accurate \textit{ab initio} reconstruction. 
    \item For moderate $\alpha_i$, where $0\ll|R_i|\ll1$, CSSI data from a single incident angle should be sufficient for \textit{ab initio} reconstruction. 
    \item For larger $\alpha_i$, where $|R_i|\approx 0$, \textit{ab initio} reconstruction is generally infeasible. 
\end{itemize}
There are no strict cutoff points for the above ranges. However, the stability, convergence, and robustness to noise of the reconstruction may gradually degrade as $|R_i|$ approaches 0 or 1 when the data is collected at a single incident angle. Incident angles slightly larger than the critical angle typically yield $|R_i|$ values in the middle of this range, maximizing both the data's information content and the fidelity of the reconstruction. Combining data from incident angles across these three ranges could also be beneficial in some cases, e.g., it may allow sampling a larger region of Fourier space than may be possible with angles from a single range.

\subsection{Exit-Angle Requirements}\label{sec:exit}

We now discuss the conditions on the range $[\alpha_f^{\min},\alpha_f^{\max}]$ of measured exit angles required for \textit{ab initio} reconstruction. In this angular range, the CSSI intensity is sampled for $\qz \in [q_z^{\min},q_z^{\max}]$, where 
\begin{equation}
\begin{aligned}
q_z^{\min} &= \frac{1}{\lambda}\left(\sin(\alpha_f^{\min})+\sin(\alpha_i)\right), \\
q_z^{\max} &= \frac{1}{\lambda}\left(\sin(\alpha_f^{\max})+\sin(\alpha_i)\right).
\end{aligned}
\end{equation}
 The four terms in the DWBA for these intensities sample the Fourier values $\hat{\rho}(\qp,\qz)$ for $\qz$ in the intervals $[q_z^{\min}, q_z^{\max}]$, $[q_z^{\min}- \qzi, q_z^{\max} -\qzi]$, $[-q_z^{\max} +\qzi,-q_z^{\min}+\qzi]$, and $[-q_z^{\max},-q_z^{\min}]$, respectively.

 To reconstruct the density $\rho$ at maximal resolution, its Fourier transform must be sampled for $\qz$ over the range $[-q_z^{\max}, q_z^{\max}]$. This sampling can be achieved if $\alpha_f^{\min}$ and $\alpha_f^{\max}$ are chosen such that the above four intervals sampled by the DWBA cover this range. If the density is real-valued, the Fourier values for the third and fourth intervals can be obtained from the first and second via Friedel symmetry. However, for a complex-valued density, the reflection coefficients $R_f(\qz)$ would need to remain sufficiently large over $[q_z^{\min}, q_z^{\max}]$ to reliably sample the Fourier values in the third and fourth intervals. 

 We now consider two cases based on how close $\alpha_f^{\min}$ is to the horizon.

\medskip\noindent\emph{Case 1: $\alpha_f^{\min}\approx 0$.}
Suppose CSSI data near $\alpha_f^{\min}\approx 0$ can be measured, i.e., $q_z^{\min}\leq \sin(\alpha_i)/\lambda + \epsilon$ where $\epsilon$ is a fraction of a Nyquist pixel, e.g., $\epsilon \leq 1/(8D_z)$. In this case, by Eqs.~\ref{eq:ki}-\ref{eq:kf} and the definition $\qzi = 2\sin(\alpha_i)/\lambda$, we have that the upper bound of the third interval satisfies
\begin{equation}
\begin{aligned}
-q_z^{\min}+\qzi &\approx \frac{1}{\lambda}\left(-\sin(\alpha_i)+2\sin(\alpha_i)\right)\\
&= \frac{1}{\lambda}\sin(\alpha_i)\\
&\approx q_z^{\min}.
\end{aligned}
\end{equation}
Therefore, the first and third DWBA terms sample the Fourier transform for $\qz$ in the interval $[-q_z^{\max}+\qzi, q_z^{\max}]$, and, by symmetry, the second and fourth terms sample it for the interval $[-q_z^{\max}, q_z^{\max}-\qzi]$. 

To ensure that the two previous intervals fully cover $[-q_z^{\max},q_z^{\max}]$, it is required that 
\begin{equation}
\begin{aligned}
0 &\geq -q_z^{\max}+\qzi\\
&= \frac{1}{\lambda}\left(-\sin(\alpha_f^{\max}) + \sin(\alpha_i)\right),
\end{aligned}
\end{equation}
 or equivalently 
\begin{equation}\label{eq:alphafmax1}
    \alpha_f^{\max} \geq \alpha_i.
\end{equation}

\medskip\noindent\emph{Case 2: $\alpha_f^{\min}\gg 0$.}
Now suppose that the data cannot be measured for $\alpha_f^{\min} \approx 0$, e.g., due to physical limitations in the experimental setup. In this situation, the first and third DWBA terms cannot sample $[-q_z^{\min}+\qzi,q_z^{\min}]$, and the second and fourth terms cannot sample $[-q_z^{\min},q_z^{\min}-\qzi]$. 

Note that if $q_z^{\max} - \qzi \geq q_z^{\min}$, then the second DWBA term samples $[-q_z^{\min}+\qzi,q_z^{\min}]$, and the third term samples $[-q_z^{\min},q_z^{\min}-\qzi]$. This condition is equivalent to the maximum exit angle satisfying
\begin{equation}\label{eq:alphafmax2}
    \alpha_f^{\max} \geq \arcsin\left(\sin(\alpha_f^{\min})+2\sin(\alpha_i)\right).
\end{equation}
However, note that Eq.~\ref{eq:alphafmax2} requires a significantly larger maximum exit angle than Eq.~\ref{eq:alphafmax1}.

Alternatively, an $\alpha_f^{\max}$ satisfying Eq.~\ref{eq:alphafmax1} is sufficient in this case if CSSI data from a second incident angle can be included in the reconstruction to sample the intervals missed by the DWBA terms for the first incident angle. In particular, (using subscripts 1 and 2 to distinguish the terms for each incident angle) this can be achieved if $q_{z,1}^{\min} \leq -q_{z,2}^{\min}+q_{z,2}^i$ or $q_{z,2}^{\min} \leq -q_{z,1}^{\min}+q_{z,1}^i$. If $\alpha_f^{\min} = \alpha_{f,1}^{\min} = \alpha_{f,2}^{\min}$, then this condition simplifies to
\begin{equation}\label{eq:alphamin}
    \alpha_f^{\min}  \leq \arcsin\left(\frac{|\sin(\alpha_{i,1}) - \sin(\alpha_{i,2})|}{2}\right).
\end{equation}
Eq.~\ref{eq:alphamin} implies that a larger minimum exit angle is sufficient if the two incident angles are sufficiently far apart.

\subsection{In-Plane Rotation-Angle Sampling Requirements}\label{sec:Inplane}

Due to the experimental geometry and curvature of the Ewald sphere, the CSSI rotation-series data typically needs to be collected with more than 180 degrees of in-plane specimen rotation to maximally sample the 3D CSSI diffraction volume. In this subsection, we formulate sufficient conditions on the rotational sampling to ensure that the measurable regions of the diffraction volume are oversampled by at least a factor of two.

Suppose that the data is collected for in-plane specimen rotations $\phi \in [0,\phi^{\max}]$, horizontal scattering angles $\theta_f \in[-\theta_f^{\max},\theta_f^{\max}]$, and exit angles $\alpha_f \in [\alpha_f^{\min}, \alpha_f^{\max}]$, where all the minimum/maximum angles are nonnegative and less than $\pi/2$. At the reference orientation $\phi = 0$, a pixel of the diffraction pattern measured at angles $\alpha_f$ and $\theta_f$ samples the intensity at $(\qp,\qz)$ where the components of $\qp = (q_x,q_y)$ are 
\begin{equation}
\begin{aligned}
    q_x &= \frac{1}{\lambda}\sin(\theta_{f})\cos(\alpha_{f}),\\
    q_y &= \frac{1}{\lambda}\big(\cos(\theta_{f})\cos(\alpha_{f}) - \cos(\alpha_i)\big),
\end{aligned}
\end{equation}
 while the pixel at $\alpha_f$ and $-\theta_f$ samples it at the horizontal coordinate $(-q_x, q_y)$. At a specimen rotation of $\phi = \pi$ these pixels sample the intensity at horizontal coordinates $(-q_x,-q_y)$ and $(q_x,-q_y)$ respectively. 

\begin{figure}[h] 
 \centering
\includegraphics[width = 1\columnwidth]{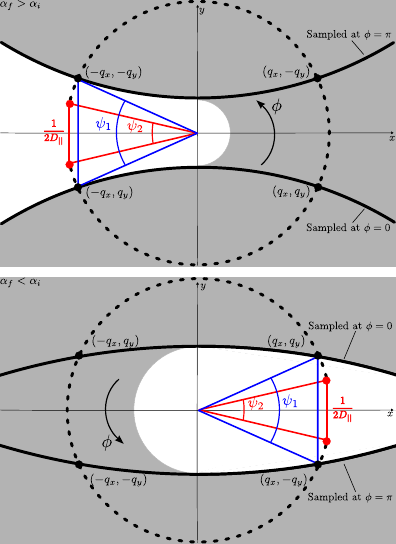}
\caption{Illustrations of how the CSSI rotation series samples the diffraction volume along a horizontal cross-section for $\alpha_f > \alpha_i$ (top) and $\alpha_f < \alpha_i$ (bottom). The diffraction patterns sample the bottom and top circular arcs at $\phi =0$ and $\phi=\pi$, or vice-versa. The data collected between these orientations samples the gray region but does not sample the white regions. Each pair of mirror-opposite pixels samples the volume at horizontal coordinates $(q_x,q_y)$ and $(-q_x,q_y)$ for $\phi = 0$ and at $(-q_x,-q_y)$ and $(q_x, -q_y)$ for $\phi = \pi$. To oversample the diffraction volume by at least a factor of two, the distance between the red endpoints of the dashed circular arc sampled by these pairs of pixels as the specimen is rotated should be less than half a Nyquist length, i.e., $1/(2D_{||})$. This requires an additional rotation beyond $\pi$ of at least $\psi_1 - \psi_2$, where $\psi_1 = 2\arctan(|q_y/q_x|)$ and $\psi_2 = 2\arcsin(1/(4D_{||}|\qp|))$.}\label{fig:phi_max_1}
\end{figure}

As shown in Fig.~\ref{fig:phi_max_1}, if rotation-series data is collected for $0\leq \phi \leq \pi$, then these two pixels do not sample the diffraction volume along the circular arc of radius $|\qp|$ between $(-q_x,-q_y)$ and $(-q_x,q_y)$ if $q_y < 0$, and it does not sample the arc between $(q_x,-q_y)$ and $(q_x, q_y)$ if $q_y > 0$. The first case occurs when $\alpha_f \geq \alpha_i$ and $\theta_f \geq 0$ or when $\alpha_f < \alpha_i$ and $\theta_f \geq \arccos(\cos(\alpha_i)/\cos(\alpha_f))$. The second case occurs when $\alpha_f < \alpha_i$ and $0\leq\theta_f\leq \arccos(\cos(\alpha_i)/\cos(\alpha_f)) $. The angle of the missing arc is $2\arctan(|q_y/q_x|)$. Therefore, for these two pixels to sample a complete circle of the diffraction volume via Eq.~\ref{eq:q}, the specimen would need to be rotated by up to $\pi+2\arctan(|q_y/q_x|)$. 

 Note that when $\theta_f = 0$, the above condition would require a $2\pi$ (360-degree) rotation.
 However, since our goal is only to oversample the diffraction volume by a factor of two, we can relax this condition to require only that the chord length of the missing arc is less than or equal to half a Nyquist length, i.e., for all $(q_x,q_y)$ sampled by the diffraction pattern at the reference orientation and satisfying $|q_y| > 1/(4D_{||})$,  
 \begin{equation}\label{eq:phimaxorig}
 \phi^{\max} \geq \pi+2\arctan\left(\left|\frac{q_y}{q_x}\right|\right) - 2\arcsin\left(\frac{1}{4D_{||}|\qp|}\right).
 \end{equation}
 For $|q_y| \leq 1/(4D_{||})$, $\phi^{\max} \geq \pi$ is sufficient. Eq.~\ref{eq:phimaxorig} avoids a 360-degree rotation requirement as long as the incident/exit angles are sufficiently small.
 
 As shown in Section S3 of the Supplemental Material, the right-hand side of Eq.~\ref{eq:phimaxorig} is maximized over the feasible region $|q_y| > 1/(4D_{||})$ when $\theta_f = 0$ and either $\alpha_f = \alpha_f^{\min}$ or $\alpha_f = \alpha_f^{\max}$, or when $\theta_f = \theta_f^{\max}$ and $\alpha_f = \alpha_f^{\max}$. For the $\theta_f=\theta_f^{\max}$ case, the bound can be further simplified by dropping the $\arcsin$ component, which is very small when $\theta_f^{\max}$ is large enough so that several Nyquist pixels are sampled in the $\qp$ dimensions. 
 
 Therefore, to maximize the region of the diffraction volume that is oversampled by at least a factor of two, the maximum specimen rotation angle should satisfy the following inequalities:
 \begin{equation}\label{eq:phimax}
 \begin{aligned}
     &\phi^{\max} \geq \pi + 2\arctan\left(\hspace{-.2em}\frac{|\cos(\alpha_i)-\cos(\theta_f^{\max})\cos(\alpha_f^{\max})|}{\sin(\theta_f^{\max})\cos(\alpha_f^{\max})}\hspace{-.2em}\right),\\
     &\phi^{\max} \geq 2\pi - 2\arcsin\left(\min\left(\frac{1}{4D_{||}q_{||,0}^{\max}},1\right)\right), \text{ where}\\
     & q_{||,0}^{\max}\hspace*{-.095em}\!=\!\frac{\!\max\!\left(\cos(\alpha_f^{\min})-\cos(\alpha_i), \cos(\alpha_i)-\cos(\alpha_f^{\max})\right)}{\lambda}.
\end{aligned}
 \end{equation}
Typically, the first inequality is dominant when smaller incident/exit angles are used, and the second inequality is dominant for larger angles. 

In addition, to ensure that the diffraction volume is sampled at a sufficient rate, the in-plane rotation angle should sample the interval $[0,\phi^{\max}]$ at least at twice the Nyquist angular rate, i.e., with an angular spacing of 
\begin{equation}\label{eq:deltaphi}
    \Delta\phi \leq \frac{1}{2D_{||}q_{||}^{\max}},
\end{equation}
where $q_{||}^{\max}$ is the magnitude of the largest horizontal component $\qp$ where the data samples the intensity. This magnitude can be calculated in terms of the minimum/maximum scattering and exit angles as
\begin{equation}
\begin{aligned}
    &q_{||}^{\max}\!=\!\max(q_{||,\text{t}}^{\max}, q_{||,\text{b}}^{\max}), \text{ where}\\
    &q_{||,\text{t}}^{\max} \!=\!\frac{\!\sqrt{\!\cos^2(\hspace{-.095em}\alpha_f^{\max}\hspace{-.095em})\!+\!\cos^2(\hspace{-.095em}\alpha_i\hspace{-.095em})\!\shortminus\!2{\cos(\hspace{-.095em}\theta_f^{\max}\hspace{-.095em})}{\cos(\hspace{-.095em}\alpha_f^{\max}\hspace{-.095em})}{\cos(\hspace{-.095em}\alpha_i\hspace{-.095em})}}}{\lambda}\!,\\
    &q_{||,\text{b}}^{\max} \!=\!\frac{\!\sqrt{\!\cos^2(\hspace{-.095em}\alpha_f^{\min}\hspace{-.095em})\!+\!\cos^2(\hspace{-.095em}\alpha_i\hspace{-.095em})\!\shortminus\!2{\cos(\hspace{-.095em}\theta_f^{\max}\hspace{-.095em})}{\cos(\hspace{-.095em}\alpha_f^{\min}\hspace{-.095em})}{\cos(\hspace{-.095em}\alpha_i\hspace{-.095em})}}}{\lambda}\!.\\
\end{aligned}
\end{equation}

\subsection{Missing Candlestick Region}\label{sec:candlestick}
Due to the experimental geometry of CSSI, the DWBA cannot sample certain regions of the specimen's 3D Fourier volume. This limitation arises from the curvature of the Ewald sphere combined with the non-orthogonal alignment of the rotation axis relative to the incident beam. In more traditional diffraction and microscopy experiments, this phenomenon is commonly referred to as the ``missing cone'' or ``missing apple core" problem \cite{vertu2009diffraction, lim2015comparative}, due to the shape of the unsampled region. 

\begin{figure}[h] 
 \centering
\includegraphics[width = \columnwidth]{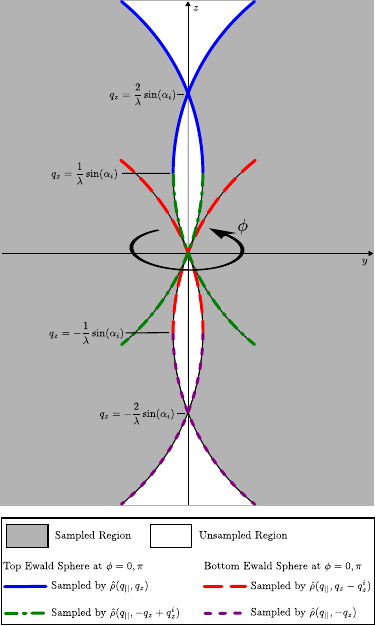}
\caption{Illustration of the missing candlestick region. The four terms in the DWBA sample the 3D Fourier volume along two Ewald spheres of radius $1/\lambda$, whose centers have vertical components $\pm\sin(\alpha_i)/\lambda$. The top Ewald sphere is sampled by the first (blue) and third (green) terms of the DWBA, and the bottom Ewald sphere is sampled by the second (red) and fourth (purple) terms. The rotation series samples the intensity along these Ewald spheres as they are rotated by $\phi$, resulting in the gray sampled region, while leaving the white candlestick-shaped region unsampled.}\label{fig:candlestick}
\end{figure}

In contrast, for CSSI, the DWBA terms sample the Fourier volume along two different rotating Ewald spheres. This results in a candlestick-shaped unsampled region, given by the set of coordinates satisfying
\begin{equation}\label{eq:dhg}
\begin{aligned}
    &\left(|\qp| + \frac{1}{\lambda}\cos(\alpha_i)\right)^2\hspace*{-.5em} + \left(|\qz| - \frac{1}{\lambda}\sin(\alpha_i)\right)^2\hspace*{-.5em} < \left(\frac{1}{\lambda}\right)^2\hspace*{-.5em},\\
    & |\qz| < \frac{2}{\lambda}\sin(\alpha_i),\\
    &\text{ or}\\
    &\left(|\qp| - \frac{1}{\lambda}\cos(\alpha_i)\right)^2\hspace*{-.5em} + \left(|\qz| - \frac{1}{\lambda}\sin(\alpha_i)\right)^2\hspace*{-.5em} > \left(\frac{1}{\lambda}\right)^2\hspace*{-.5em},\\
    & |\qz| \geq \frac{2}{\lambda}\sin(\alpha_i).
    \end{aligned}
\end{equation}
See Section S4 in the Supplemental Material for a proof of Eq.~\ref{eq:dhg} and see Fig.~\ref{fig:candlestick} for a visualization of this region.

For $\alpha_f < \alpha_i$, the largest $|\qp|$ on the boundary of Eq.~\ref{eq:dhg} occurs when $\alpha_f = 0$, meaning that the diffraction volume is unable to sample the disk defined by
\begin{equation}\label{eq:missreg1}
    |\qp| < \frac{1}{\lambda}(1-\cos(\alpha_i)),\quad |\qz| = \frac{1}{\lambda}\sin(\alpha_i). 
\end{equation}
To ensure the diffraction volume is oversampled by a factor of two near the missing region in Eq.~\ref{eq:missreg1}, its width should be less than half a Nyquist length, i.e., 
\begin{equation}\label{eq:miss1}
    \frac{1}{\lambda}(1-\cos(\alpha_i)) < \frac{1}{4D_{||}}.
\end{equation}

For $\alpha_f > \alpha_i$, the largest $|\qp|$ on the boundary of the missing region occurs when $\alpha_f = \alpha_f^{\max}$, yielding the unsampled disk
\begin{equation}\label{eq:missreg2}
\begin{aligned}
    |\qp| &< \frac{1}{\lambda}\left(\cos(\alpha_i)-\cos(\alpha_f^{\max})\right),\\
    |\qz| &= \frac{1}{\lambda}\left(\sin(\alpha_f^{\max})+\sin(\alpha_i)\right). 
    \end{aligned}
\end{equation}
Oversampling the diffraction volume by a factor of two in the region near Eq.~\ref{eq:missreg2} requires that
\begin{equation}\label{eq:miss2}
    \frac{1}{\lambda}\left(\cos(\alpha_i)-\cos(\alpha_f^{\max})\right) < \frac{1}{4D_{||}}.
\end{equation}

To maximize the information content of the data, the experimental parameters should be chosen to satisfy Eqs.~\ref{eq:miss1} and \ref{eq:miss2}. In particular, Eq.~\ref{eq:miss1} is crucial for the stability of the reconstruction because the region in Eq.~\ref{eq:missreg1} contains low-frequency information, which is needed to determine the overall shape and support of the reconstruction. In contrast, violation of Eq.~\ref{eq:miss2} is more tolerable, because the region in Eq.~\ref{eq:missreg2} only contains high-frequency information, whose loss may limit the resolution of the reconstruction but might not hinder the recovery of its overall shape.
 
A mild violation of these conditions likely results in only a small loss of resolution in the reconstruction. However, larger violations, especially in the low-frequency region, may prevent a successful reconstruction, unless additional constraints are enforced to help compensate for the unsampled information. This issue is more pronounced for larger specimens since they have more Nyquist voxels in the unsampled region. In practice, the volume of the missing candlestick can be reduced by using a shorter X-ray beam wavelength along with smaller incident and exit angles.

\subsection{Uniqueness Properties: CSSI Intensity Invariances}\label{sec:invariance}

Similar to the classical phase problem, the CSSI intensity is invariant to various operations on the specimen density. This means that the reconstruction is only able to determine the correct solution up to these invariant operations. In particular, the CSSI reconstruction problem shares many of the uniqueness properties of the classical phase problem, discussed in Section~\ref{sec:IPA}, but there are a few key differences. Here we discuss these invariant operations and uniqueness properties. 

To simplify the following expressions, here we represent the CSSI intensity as
\begin{equation}
    I(\qp,\qz) = \left|\sum_{s=1}^4 R_s(\qz)\hat{\rho}(\qp,q_{z,s})\right|^2,
\end{equation}
where $R_s(\qz)$ and $q_{z,s}$ are the reflection coefficient and vertical argument of the $s$-th term in Eq.~\ref{eq:CSSI}.

\medskip\noindent\emph{Invariance to Phase-Factor/Sign.} Similar to the classical phase problem, the CSSI intensity is invariant to multiplying the density by a constant phase factor $e^{i\omega}$, as this term can be factored out of the four-term DWBA summation and has unit magnitude. If the density is real-valued, then this implies that the sign of the reconstructed density is ambiguous. However, this sign ambiguity can typically be overcome by simply choosing the sign that makes the density mostly positive.

\medskip\noindent\emph{Invariance to Horizontal Translation.} While the intensity in the classical phase problem is invariant to any 3D translation of the solution, the CSSI intensity is only invariant for translations in the parallel component $r_{||}$. This can be seen by noting that, due to the Fourier shift theorem, a translation $\rho(\mathbf r + \boldsymbol \tau)$ of the density by $\boldsymbol\tau = (\tau_{||},\tau_z)$ has the phase-shifted Fourier transform $\hat{\rho}(\qp,\qz)e^{2\pi i (\tau_{||},\tau_z) \cdot \mathbf (\qp,\qz)}$. When $\tau_z = 0$, this phase shift can be factored out of the DWBA, meaning that the CSSI intensity  of the translated density is
\begin{equation}
\begin{aligned}
   I_{\tau_{||}}(\qp,\qz) &=\left|\sum_{s=1}^4 R_s(\qz)\hat{\rho}(\qp,q_{z,s})e^{2\pi i \tau_{||} \cdot \qp}\right|^2\\
    &= \left|\sum_{s=1}^4 R_s(\qz)\hat{\rho}(\qp,q_{z,s})\right|^2\left|e^{2\pi i \tau_{||} \cdot \qp}\right|^2\\
    &= \left|\sum_{s=1}^4 R_s(\qz)\hat{\rho}(\qp,q_{z,s})\right|^2\\
    &= I(\qp,\qz),
\end{aligned}
\end{equation}
i.e., the CSSI intensity is invariant to parallel density translations.

However, when $\tau_z \neq 0$, the phase factor generally cannot be factored out since the $z$-term in the phase shift is different for each of the four DWBA terms, i.e., in this case, the intensity of a generic translated density is
\begin{equation}
\begin{aligned}
   I_{\tau_{z}}(\qp,\qz) &=\left|\sum_{s=1}^4 R_s(\qz)\hat{\rho}(\qp,q_{z,s})e^{2\pi i \tau_z q_{z,s}}\right|^2\\
    &\neq I(\qp,\qz),
\end{aligned}
\end{equation}
i.e., the CSSI intensity is not generally invariant to vertical density translations.

\medskip\noindent\emph{Non-Invariance to Reflection Through the Origin.} While the intensity in the classical phase problem is invariant to reflecting the density through the origin, this is not true for CSSI. By the Fourier conjugation property, the reflected density $\rho(-\mathbf r)$ has the complex-conjugated Fourier transform $\overline{\hat{\rho}(\qp,\qz)}$. Since $R_2=R_3=R_4 = 0$ in the classical phase problem, the associated intensity is invariant to this operation, as $|\hat{\rho}| = |\overline{\hat{\rho}}|$. However, for CSSI, the reflection coefficients are non-negligible, meaning that the CSSI intensity of a generic reflected density is 
\begin{equation}
\begin{aligned}
   I_{\rm ref}(\qp,\qz) &=\left|\sum_{s=1}^4 R_s(\qz)\overline{\hat{\rho}(\qp,q_{z,s})}\right|^2\\
    &\neq I(\qp,\qz),
\end{aligned}
\end{equation}
since, for $s\neq1$, $R_s$ generally has a substantial imaginary part, implying that $\overline{R_s} \neq R_s$.

\medskip\noindent\emph{Invariance to Horizontally Homometric Structures.} Another difference from the classical phase problem is that not all homometric structures lead to nonuniqueness for the CSSI reconstruction problem. For the classical phase problem, if $\rho = \rho_1\ast \rho_2$ is a homometric solution, then so is $\rho_h = \rho_1\ast \rho_{2r}$, where $\rho_{2r}(\mathbf r) = \rho_2(-\mathbf r)$ since, by the convolution theorem, $|\hat{\rho}_h| = |\hat{\rho}_1\hat{\rho}_{2r}| = |\hat{\rho}_1\, \overline{\hat{\rho}_2}| = |\hat{\rho}_1| |\overline{\hat{\rho}_2}| = |\hat{\rho}_1\hat{\rho}_2| = |\hat{\rho}|$. However, for the CSSI reconstruction problem, replacing $\hat{\rho}_1\hat{\rho}_2$ with $\hat{\rho}_1\hat{\rho}_{2r}$ in Eq.~\ref{eq:CSSI} typically leads to different CSSI intensity values, since, for a generic $\rho_1$, $\hat{\rho}_{2r}$ cannot be factored out unless $\hat{\rho}_2$ is constant in $\qz$. This condition would imply that $\rho_2$ is of the form $\rho_2(r_{||},r_z)=h(r_{||})\delta(r_z)$, where $h$ is a 2D function and $\delta$ is the 1D delta function. In this case, $\hat{\rho}_2(\qp,\qz) = \hat{h}(\qp)$, since $\hat{\delta}(\qz) = 1$, meaning that if $I$ is the CSSI intensity for $\rho = \rho_1\ast \rho_2$, then the intensity for $\rho_h$ is 

\begin{align}
    I_h(\qp,\qz) &= \left|\sum_{s=1}^4 R_s(\qz)\hat{\rho}_1(\qp,q_{z,s})\hat{\rho}_{2r}(\qp,q_{z,s})\right|^2\notag\\
    &= \left|\sum_{s=1}^4 R_s(\qz)\hat{\rho}_1(\qp,q_{z,s})\overline{\hat{h}(\qp)}\right|^2\notag\\
    &= \left|\sum_{s=1}^4 R_s(\qz)\hat{\rho}_1(\qp,q_{z,s})\right|^2\left|\hat{h}(\qp)\right|^2\\
    &= \left|\sum_{s=1}^4 R_s(\qz)\hat{\rho}_1(\qp,q_{z,s})\hat{h}(\qp)\right|^2\notag\\
    &= \left|\sum_{s=1}^4 R_s(\qz)\hat{\rho}_1(\qp,q_{z,s})\hat{\rho}_2(\qp,\qz)\right|^2\notag\\
    &= I(\qp,\qz).\notag
    \end{align}

In summary, if $\rho_1$ and $\rho_2$ are noncentrosymmetric and $\rho = \rho_1 \ast \rho_2$ is a homometric solution to the CSSI reconstruction problem, then, in general, $\rho_h = \rho_1 \ast \rho_{2r}$ is also a solution if and only if $\rho_2$ is of the form $\rho_2(r_{||},r_z) = h(r_{||})\delta(r_z)$, for some 2D function $h$.

\medskip\noindent\emph{Partial Invariance to 180-Degree Rotation for Unitary Reflection Coefficients.} Unlike the classical phase problem, the CSSI intensity $I(\qp,\qz)$ is approximately invariant to in-plane rotation of the specimen by 180 degrees when $|R_i| \approx 1$ and $|R_f(\qz)|\approx 1$. This can be seen by noting that Friedel symmetry gives $\overline{\ph(\qp,-\qz)}$ for the Fourier transform of the rotated density, so in this case the CSSI intensity of the rotated specimen is
\begin{equation}
    \begin{aligned}
    I_{\rm rot}(\qp,\qz)\! &=\! \big|\overline{\ph(\qp,-\qz)} + R_i\overline{\ph(\qp,-\qz\!+\!\qzi)}\\
    &\hspace{1em}+ R_f(\qz)\overline{\ph(\qp,\qz\! -\!\qzi)} +  R_iR_f(\qz)\overline{\ph(\qp,\qz)}\big|^2\\
    &= \big|\ph(\qp,-\qz) + \overline{R_i}\ph(\qp,-\qz\!+\!\qzi)\\
    &\hspace{1em}+ \overline{R_f(\qz)}\ph(\qp,\qz\! -\!\qzi) +  \overline{R_i R_f(\qz)}\ph(\qp,\qz)\big|^2\\
    &\approx \big|\overline{R_i R_f(\qz)}\big(R_iR_f(\qz)\ph(\qp,-\qz)\\
    &\hspace{1em}+ R_f(\qz)\ph(\qp,-\qz\!+\!\qzi)+ R_i\ph(\qp,\qz\! -\!\qzi)\\
    &\hspace{1em}+  \ph(\qp,\qz)\big)\big|^2\\
    &\approx \big|\ph(\qp,\qz) + R_i\ph(\qp,\qz\! -\!\qzi)\\
    &\hspace{1em}+ R_f(\qz)\ph(\qp,-\qz\!+\!\qzi)\\
    &\hspace{1em}+ R_iR_f(\qz)\ph(\qp,-\qz)\big|^2\\
    &= I(\qp,\qz)
    \end{aligned}
\end{equation}

Note that $|R_i|\approx 1$ and $|R_f(\qz)|\approx 1$ occur only for small angles below the critical angle (see Fig.~\ref{fig:Ri}). Therefore, if the data is collected only for small incident and exit angles, then both orientations of the specimen may be consistent with the entire dataset, leading to a 50/50 chance of either orientation being reconstructed.

When data is collected at small incident angles and large maximum exit angles, then $|R_f(\qz)|\approx 1$ will hold only for a subset of the data, typically the low-resolution parts. In this case, data at small $\qz$ (lower resolution) may be approximately invariant to the rotation, while data at larger $\qz$ (higher resolution) may not be. This can cause the reconstruction procedure to become trapped in local minima that capture the correct low-resolution structure of the rotated specimen but miss high-resolution details, making the reconstruction consistent with the low-resolution data but inconsistent with the high-resolution data. Performing multiple reconstructions with different starting conditions and selecting the one with the best fit to the data can help mitigate this problem. See Section S6 in the Supplemental Material for an example of the partial invariance described above.

\section{Results}\label{sec:results}
In this section, we test the algorithmic components of our CSSI reconstruction methodology described in the previous sections. 

Our numerical experiments use the following 3D test objects, which are first generated with a uniform density on an oversampled grid and then downsampled to a $121 \times 121 \times 31$ grid via inverse-distance weighting within each downsampled voxel. The first test object is a porous medium structure (PM) constructed within an 800 nm $\times$ 800 nm $\times$ 200 nm volume on a twofold-oversampled grid by removing balls of random radii and centers from a thin cylinder, as shown in Fig.~\ref{fig:pm_grid_recs}. The second test object is a 3D conical extension of a Siemens star (SS) constructed within a 200 nm $\times$ 200 nm $\times$ 50 nm volume on a tenfold-oversampled grid, as shown in Fig.~\ref{fig:pss_grid_recs}.

All CSSI simulations are performed using a silicon substrate and an X-ray wavelength of $\lambda = 1.7$ \AA, yielding a complex refractive index of $n_r = 1.0 - 9.25111908 \times 10^{-6} + i 2.54238785\times 10^{-7}$ and a critical angle $\alpha_c \approx 0.246\degree$. All calculations are performed using double-precision arithmetic, and NUFFTs are computed using the FINUFFT library with an error tolerance of $10^{-11}$. To verify the consistency of our physical units and DWBA formulation, we validated our NUFFT-based forward model against the BornAgain software \cite{pospelov2020bornagain} (see Section S7 in the Supplemental Material for an example). The presented runtimes are measured using a single core of a 3.6-GHz i9-9900K processor.

For each numerical experiment, we quantify the reconstruction error as follows. If necessary, we first align each reconstruction $\rho_{\rm rec}$ with the corresponding ground truth object $\rho_{\rm truth}$ to within a resolution of one-tenth of a voxel via trigonometric interpolation. The error between the aligned structure and ground truth is then reported as a relative $\ell^2$ error $||\rho_{\rm rec}-\rho_{\rm truth}||_2/||\rho_{\rm truth}||_2$ and/or a Fourier shell correlation (FSC), defined as
\begin{equation}
    {\rm FSC}(q) = \frac{\sum_{\mathbf q \in \Omega(q)} \hat{\rho}_{\rm rec}(\mathbf q)\overline{\hat{\rho}}_{\rm truth}(\mathbf q)}{\sqrt{\sum_{\mathbf q\in \Omega(q)}|\hat{\rho}_{\rm rec}(\mathbf q)|^2\sum_{\mathbf q\in \Omega(q)}|\hat{\rho}_{\rm truth}(\mathbf q)|^2}},
\end{equation}
where $\Omega(q) = \{\mathbf q \in \mathcal G: q- 1/(2N_z)\leq|\mathbf q|\leq q+1/(2N_z)\}$ and $\mathcal G$ is an $N_x\times N_y\times N_z$ uniform Fourier grid.

\subsection{Direct Nonuniform Fourier Inversion}
Here we evaluate the accuracy and speed of the direct real-valued nonuniform Fourier inversion, introduced in Section~\ref{sec:DirectInversion}, for different sets of nonuniform Fourier sampling sets $\{\mathbf q_k\}_k$ and pre-weight factors $v$. Our Fourier sampling sets consist of:
\begin{enumerate}[1)]
    \item Clipped cylindrical grid, constructed with $1621$ angles uniformly spaced in $[0, 2\pi)$, $258$ radii uniformly spaced in $[0,\tfrac{\sqrt{2}}{2})$, and $15$ heights uniformly spaced in $[0,\tfrac{1}{2})$, with all points outside of the box $[-\tfrac{1}{2},\tfrac{1}{2}] \times [-\tfrac{1}{2},\tfrac{1}{2}] \times [-\tfrac{1}{2},\tfrac{1}{2}]$ removed.
    \item  DWBA grid, described in Section~\ref{sec:CartAccel}, for $J=1$ incident angle $\alpha_i = 0.26$   and $(2\cdot121-1) \times (2\cdot121-1) \times (8\cdot31)$ grid points.
    \item $8\cdot 121 \cdot 121 \cdot 31$ random 3D coordinates whose components are sampled from a uniform distribution over the interval $[-\tfrac{1}{2},\tfrac{1}{2}]$. 
\end{enumerate}
The Fourier data for each object is simulated on each of the above sampling sets via an NUFFT. 

For each of these Fourier sampling sets, we compute the weights $w$ using two choices of pre-weight factors. The first set of weights is computed using $v(\mathbf q) = 1$. A second set of weights is computed using $v(\mathbf q) = |\qp|$ for the clipped cylindrical grid, $v_s(\mathbf q) = \sqrt{|R_s(\qz)|}$ for the DWBA grid, and $v(\mathbf q)$ randomly sampled from a uniform distribution on $[0,1]$ for the random coordinate set. 
A relative error tolerance of $10^{-10}$ is used for the stopping criterion of conjugate gradient in the weight calculation. For each sampling set and choice of $v$, we use the above weights to apply the direct nonuniform Fourier inversion to the Fourier data for each test object via Eq.~\ref{eq:real_inverse}, where an NUFFT is used to compute the $F^*$ operation. 

Computing the weights using $v(\mathbf q)=1$ required 353, 9, and 218 iterations of conjugate gradient to reach the stopping criterion for the first, second, and third sampling sets, respectively. For the second set of pre-weight factors, 36, 15, and 296 iterations were required. Each conjugate gradient iteration took 3.0 seconds, and after the weights were computed, direct inversion took 7.1, 12.4, and 5.0 seconds for the three sampling sets, respectively. For the DWBA grid, using the accelerations in Eq.~\ref{eq:DINUFTaccel} reduced the computation time to 0.5 seconds, yielding a 25-fold speedup, and resulted in an almost identical accuracy, i.e., within the $10^{-11}$ error tolerance used in FINUFFT. 

\begin{figure}[h!] 
\centering
\includegraphics[width = \columnwidth]{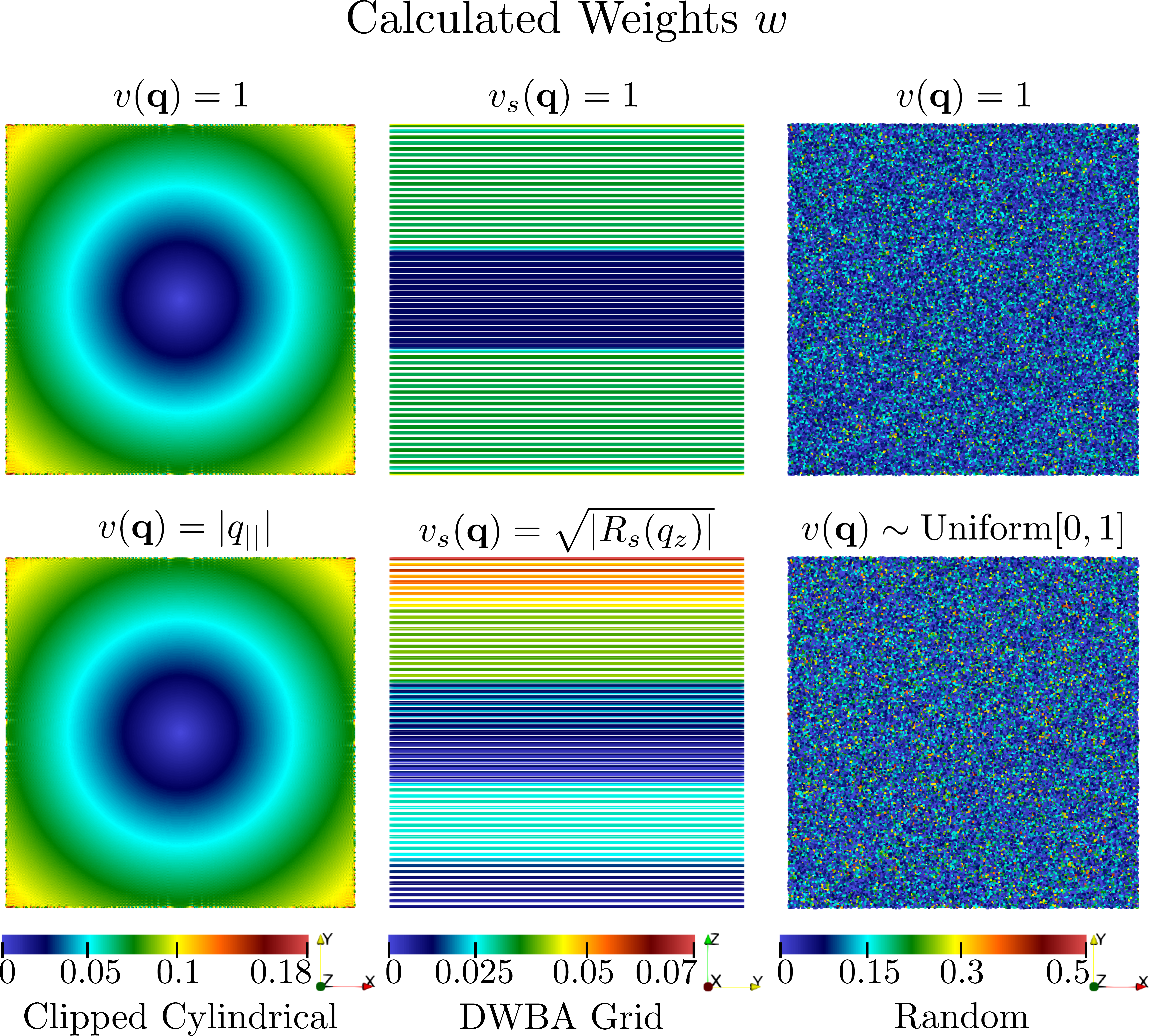}
\caption{2D slices of the calculated weights $w$ used in the direct nonuniform Fourier inversion with different pre-weight factors $v(\mathbf q)$ for a clipped cylindrical sampling, DWBA grid, and random sampling. The slices are in the $xy$ plane for the clipped cylindrical and random samplings and are in the $yz$ plane for the DWBA grid sampling. Color bars and orientation axes for each sampling set are shown at the bottom.}\label{fig:weights}
\end{figure}

The relative $\ell^2$ errors between the reconstructed and ground-truth densities for each sampling set and pre-weight factor are provided in Tables~\ref{table:DINUFT1}-\ref{table:DINUFT2}, and the calculated weights $w$ are shown in Fig.~\ref{fig:weights}. In each case, the error is on the order of the tolerance for the conjugate gradient stopping criterion. 

As shown in Fig.~\ref{fig:weights}, the computed weights depend on both the sampling distribution and the chosen pre-weight factor, and they approximate the pre-weighted sampling density. For the clipped cylindrical grid, using $v(\mathbf q) = |\qp|$ as a pre-weight factor has little effect on the calculated weights but significantly improves convergence. This improvement in convergence demonstrates how the weight calculation can be accelerated by choosing a pre-weight factor that approximates the inverse of the nonuniform sampling density, as is the case here. This example exhibits oscillations in the weights at the boundary due to discontinuities in the periodic extension of the sampling density. For the DWBA grid, using $v_s(\mathbf q) = \sqrt{|R_s(\mathbf q)|}$ significantly affects the weights, resulting in higher values for points associated with larger reflection coefficients, which contribute more significantly to the DWBA summation and thus are more important in the reconstruction. For the random coordinate set, using a random $v(\mathbf q)$ simply yields a slightly different set of weights, but it demonstrates that this procedure is robust to completely unstructured sampling sets and pre-weight factors.

\begin{table}[h]
\begin{center}
\begin{tabular}{l@{\hspace{1em}} c c c} 
\midrule
&\multicolumn{3}{c}{Fourier Sampling Sets}\\
Object  & Cylindrical & DWBA Grid & Random \\
\midrule
PM & $8.5 \times 10^{-10}$ & $1.5 \times 10^{-11}$ & $5.1 \times 10^{-11}$ \\
SS & $9.1 \times 10^{-10}$ & $1.5 \times 10^{-11}$ & $4.7 \times 10^{-11}$  \\
\midrule
\end{tabular}
\end{center}
\vspace*{-1em}
\caption{Relative $\ell^2$ errors of direct nonuniform Fourier inversion with $v(\mathbf q) = 1$ for the porous medium (PM) and conical Siemens star (SS) objects, using three different Fourier sampling sets.}\label{table:DINUFT1}
\end{table}

\begin{table}[h]
\begin{center}
\begin{tabular}{l@{\hspace{1em}} c c c} 
\midrule
&\multicolumn{3}{c}{Fourier Sampling Sets}\\
Object  & Cylindrical & DWBA Grid & Random \\ 
\midrule
PM & $3.3 \times 10^{-10}$ & $2.1 \times 10^{-11}$ & $5.3 \times 10^{-11}$ \\
SS & $1.8 \times 10^{-10}$ & $2.2 \times 10^{-11}$ & $5.0 \times 10^{-11}$  \\
\midrule
\end{tabular}
\end{center}
\vspace*{-1em}
\caption{Relative $\ell^2$ errors of direct nonuniform Fourier inversion for the porous medium (PM) and conical Siemens star (SS) objects, with $v(\mathbf q) = |\qp|$ for the clipped cylindrical grid, $v_s(\mathbf q) = \sqrt{|R_s(\qz)|}$ for the DWBA grid, and $v(\mathbf q)\sim {\rm Uniform}[0,1]$ for the random point set.}\label{table:DINUFT2}
\end{table}

\subsection{Reconstruction from Complete CSSI Diffraction Volumes}\label{sec:results_volume}
Here we evaluate the performance of our nonuniform iterative phasing algorithm, introduced in Section~\ref{sec:NIP}, by applying it to reconstruct test objects from complete CSSI data generated on the staggered CSSI intensity grids described in Section~\ref{sec:Cartesian}. This allows us to study the fundamental properties of the reconstruction algorithm across the different incident-angle ranges discussed in Section~\ref{sec:incident}. See Section S5 in the Supplemental Material for additional reconstruction results demonstrating the utility of using a staggered intensity grid over a standard uniform intensity grid.

For both the porous medium and conical Siemens star test objects, we perform reconstructions from CSSI data simulated on staggered CSSI intensity grids with the five sets of incident angles indicated in Table~\ref{table:recparams}. In each case, the maximum exit angle is determined by the incident angle and the maximum $q_z$ value represented on the nonuniform grid via $\alpha_f^{\max} = \arcsin(\lambda q_z^{\max} - \sin(\alpha_i))$, which is different for the two test objects since they are generated with different sizes. Data is only simulated for $q_z$ that correspond to a nonnegative exit angle, i.e., where $q_z \geq \sin(\alpha_i)/\lambda$. Any grid points outside this region are masked out and allowed to float as described in Section~\ref{sec:prelim}. 

\begin{figure}[h!] 
 \centering
 \includegraphics[width = \columnwidth]{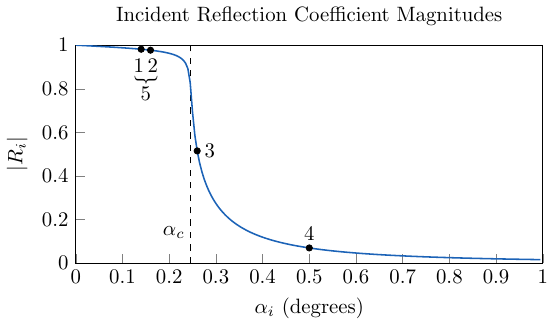}
\caption{Plot of the incident reflection coefficient magnitudes $|R_i|$ versus the incident angle $\alpha_i$, with the critical angle $\alpha_c$ and the incident angles used in the test cases 1--5 indicated. Test case 5 combines data from test cases 1 and 2.}\label{fig:inc_ref_coeff}
\end{figure}

\begin{figure}[h!] 
 \centering
 \includegraphics[width = \columnwidth]{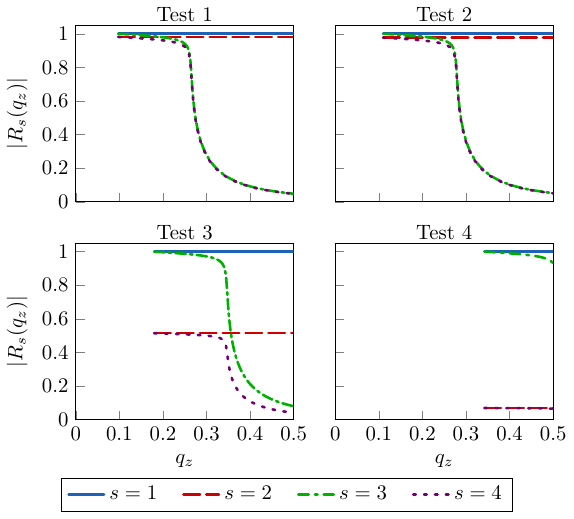}
\caption{DWBA reflection coefficient magnitudes $|R_s(\qz)|$, $s=1,2,3,4$, over the $\qz$ range sampled by the CSSI intensity in the five porous medium test cases. Test case 5 combines data from test cases 1 and 2.  }\label{fig:PM_ref_coeff}
\end{figure}

\begin{figure}[h!] 
 \centering
 \includegraphics[width = \columnwidth]{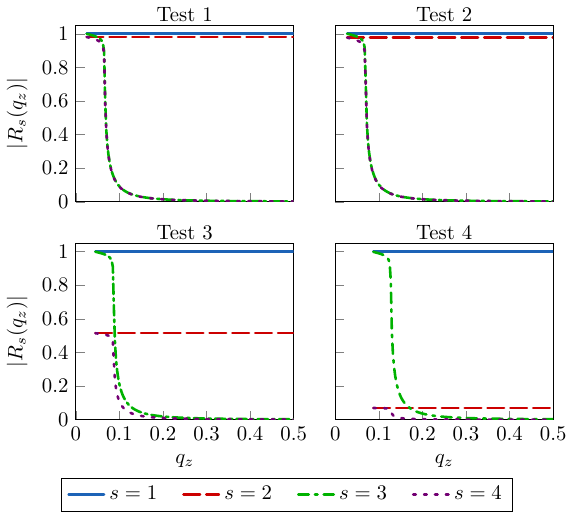}
\caption{DWBA reflection coefficient magnitudes $|R_s(\qz)|$, $s=1,2,3,4$, over the $\qz$ range sampled by the CSSI intensity in the five conical Siemens star test cases. Test case 5 combines data from test cases 1 and 2.  }\label{fig:SS_ref_coeff}
\end{figure}

Fig.~\ref{fig:inc_ref_coeff} compares the magnitudes of the incident reflection coefficients for the five test cases and the critical angle. Figs.~\ref{fig:PM_ref_coeff}-\ref{fig:SS_ref_coeff} display the profiles of the four DWBA reflection coefficients $|R_s(\qz)|$, $s=1,2,3,4$, in each test case. As can be seen from these figures, all four reflection coefficients make a significant contribution to the data, which thus requires the full DWBA to accurately model. Although some of the terms for the conical Siemens star tests drop off at higher $\qz$, all four terms contribute significantly to the lower $\qz$ data and thus are important in determining the overall shape of the reconstruction. For the porous medium tests, the reflection coefficients maintain a significant magnitude throughout most of the simulated $\qz$ range.

\begin{table}[h]
\begin{center}
\begin{tabular}{c@{\hspace{1em}} c@{\hspace{1em}} c c@{\hspace{.5em}} c@{\hspace{.5em}} c} 
\midrule            
Test  & $J$ & $\alpha_{i}$ & $|R_i(\alpha_{i})|$ & $\alpha_{f}^{\max}$(PM) & $\alpha_{f}^{\max}$(SS) \\
\midrule
1 & $1$ & $0.14\degree$  & $0.98$ & $0.59\degree$ & $2.78\degree$ \\
2 & $1$ & $0.16\degree$ & $0.98$ & $0.57\degree$ & $2.76\degree$\\
3 & $1$ & $0.26\degree$  & $0.51$& $0.47\degree$ & $2.66\degree$\\
4 & $1$ & $0.50\degree$  & $0.07$& $0.23\degree$ & $2.42\degree$\\
\multirow{2}{*}{5} & \multirow{2}{*}{$2$} & $0.14\degree$ & $0.98$ & $0.59\degree$ & $2.78\degree$\\ 
                       &                        & $0.16\degree$  & $0.98$ & $0.57\degree$ & $2.76\degree$\\
\midrule
\end{tabular}
\end{center}
\vspace*{-1em}
\caption{Incident angles, incident reflection coefficient magnitudes, and exit angles used in the numerical tests for the reconstructions of the porous medium (PM) and conical Siemens star (SS) from complete CSSI diffraction volume data.}\label{table:recparams}
\end{table}

We run our nonuniform iterative phasing algorithm on each dataset with $\beta = 0.9$ and $N_{\rm cycle}$ cycles consisting of 60 iterations of the nonuniform HIO scheme, 40 iterations of the nonuniform ER scheme, and an application of Shrinkwrap using the parameters described in Section~\ref{sec:IPA} with $\tau$ set to $5\%$ of the maximum density. After the last cycle, we refine the solution with 200 more iterations of the nonuniform ER scheme. We use $N_{\rm cycle} = 10$ for the porous medium reconstructions, and $N_{\rm cycle} = 20$ for the conical Siemens star reconstructions, which took longer to converge due to large variations in the density near the center of the object. We recenter the reconstruction after each cycle to prevent it from potentially being cut off by the zero-padding/restriction operations in the nonuniform Fourier operations. Since the CSSI intensity is invariant to negating the density, we ensure that a mostly positive solution is reconstructed by multiplying the estimated density by the sign of the average density before the Shrinkwrap step. When run on a single CPU core, each cycle takes approximately 3.3 minutes for $J=1$ incident angle and 5.9 minutes for $J=2$ incident angles.

The reconstructions are initialized with $\rho^{(0)}$ set to random values uniformly sampled on the interval $[0,0.1]$, and with the support $S$ initialized to a cylinder, with a diameter equal to the grid width and a height one pixel less than the grid height. 

All presented reconstructions use the same random seed to generate the starting density, except for porous medium test case 2. In this case, $|R_i|\approx 1$ and $|R_f(\qz)| \approx 1$ for a significant portion of the simulated $\qz$ range, making the reconstruction susceptible to becoming trapped in local minima induced by the partial 180-degree rotation invariance for unitary reflection coefficients, described in Section~\ref{sec:invariance}. Therefore, for this case, we show the results of rerunning the reconstruction with a second random seed, which successfully avoids the local minima and leads to a better fit to the data. In particular, the $\ell^2$ error between the input intensity and the intensity simulated from the reconstruction is 29 times smaller for the second reconstruction compared to the first. See Section S6 in the Supplemental Material for a comparison of these two reconstructions. 

\begin{figure}[h!] 
 \centering
 \includegraphics[width = \columnwidth]{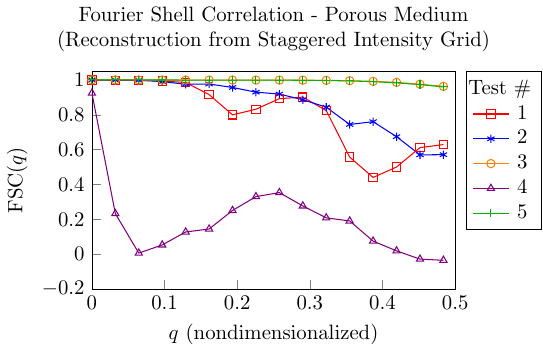}
\caption{Fourier shell correlation (FSC) for the reconstructions of the porous medium object from staggered CSSI intensity grid data for the five test cases. }\label{fig:FSC_PM}
\end{figure}

\begin{figure}[h!] 
 \centering
 \includegraphics[width = \columnwidth]{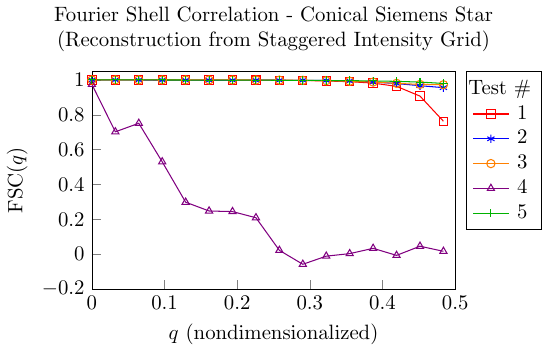}
\caption{Fourier shell correlation (FSC) for the reconstructions of the conical Siemens star object from staggered CSSI intensity grid data for the five test cases.  }\label{fig:FSC_SS}
\end{figure}

\begin{figure*}[h!] 
 \centering
\includegraphics[width = .89\textwidth]{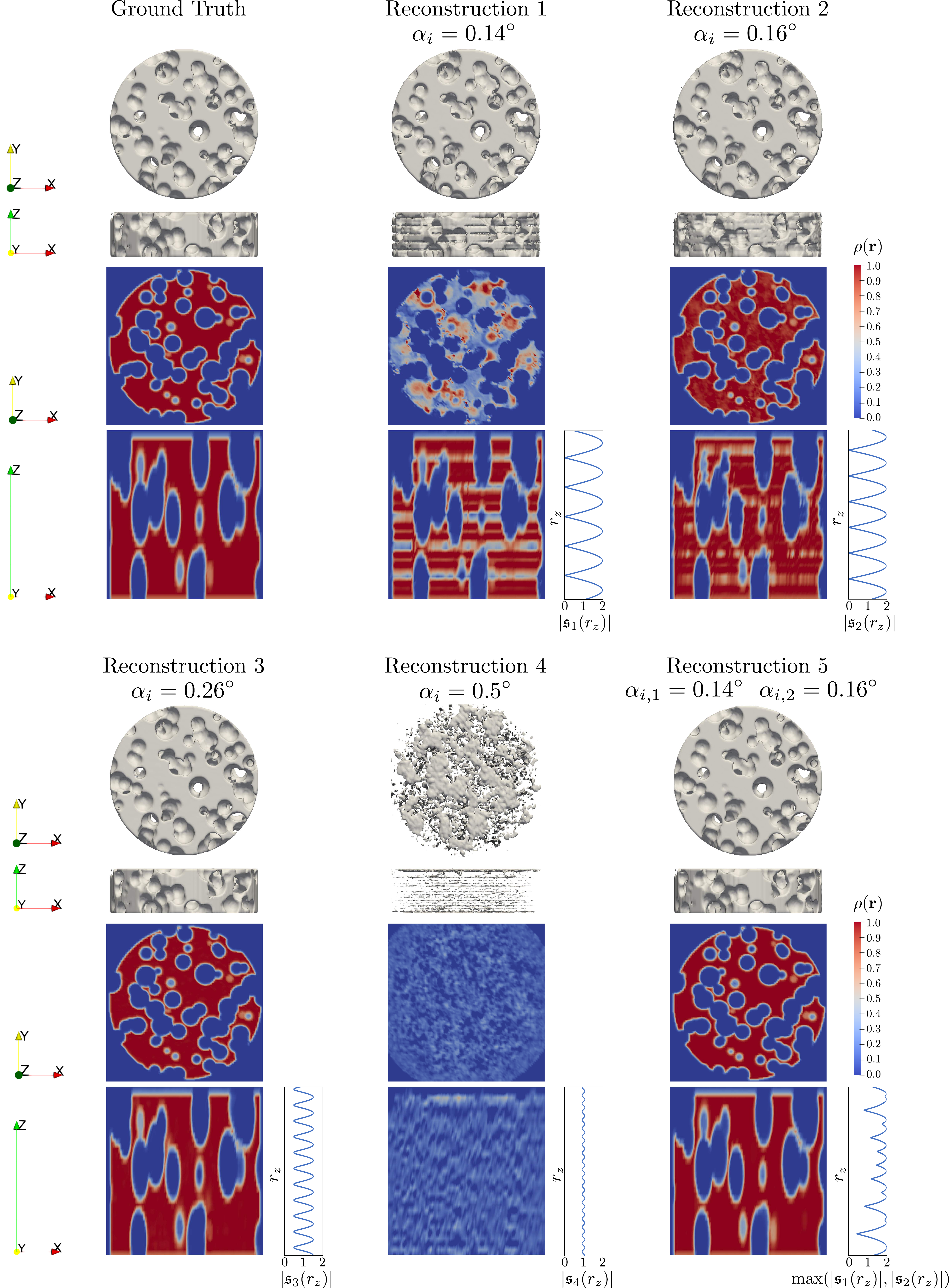}
\vspace*{-1em}
\caption{Ground truth and reconstructions of the porous medium from staggered CSSI intensity grid data. For each example, the top two rows display different orientations of the density isosurfaces at $\rho = 0.5$ for test cases 1, 2, 3, and 5 and $\rho = 0.25$ for test case 4. The bottom two rows show horizontal and vertical slices of each density through the middle of the volume. In the bottom row, each slice is stretched in the $z$ direction by a factor of four and compared against the corresponding interference function to illustrate the alignment of reconstruction deformities with the zeros of the interference function. }\label{fig:pm_grid_recs}
\end{figure*}

\begin{figure*}[h!] 
 \centering
\includegraphics[width = .89\textwidth]{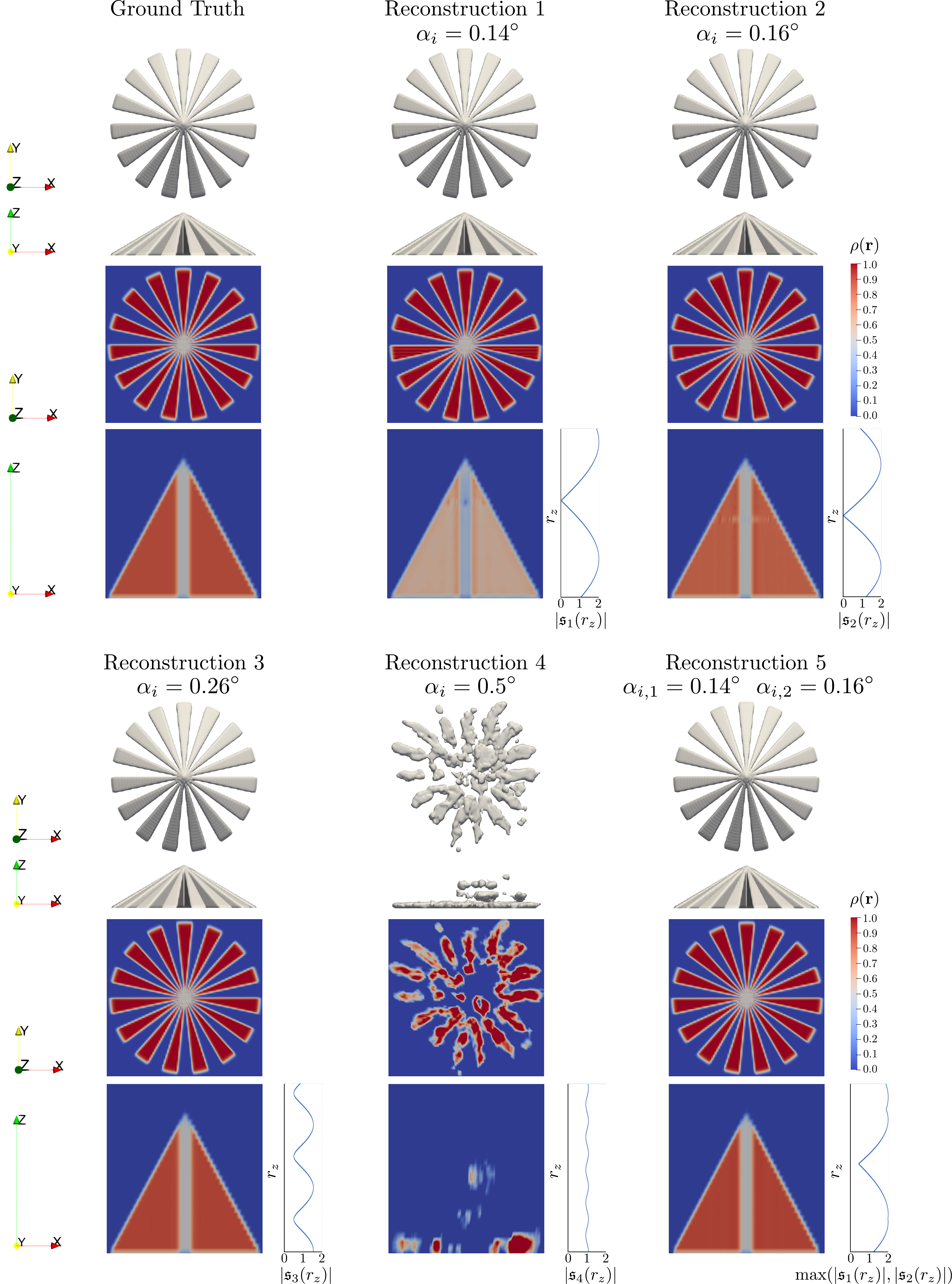}
\caption{Ground truth and reconstructions of the conical Siemens star from staggered CSSI intensity grid data. For each example, the top two rows display different orientations of the density isosurfaces at $\rho = 0.25$. The bottom two rows show horizontal and vertical slices of each density through the bottom and middle of the volume, respectively. In the bottom row, each slice is stretched in the $z$ direction by a factor of four and compared against the corresponding interference function to illustrate the alignment of reconstruction deformities with the zeros of the interference function.}\label{fig:pss_grid_recs}
\end{figure*}

The reconstruction and interference function for each test case are visualized in Figs.~\ref{fig:pm_grid_recs}-\ref{fig:pss_grid_recs}. The relative $\ell^2$ errors and FSC curves for each reconstruction are reported in Table~\ref{table:rel_error} and Figs.~\ref{fig:FSC_PM}-\ref{fig:FSC_SS}, respectively.

In test cases 1 and 2, where the incident angle is below the critical angle and $|R_i| \approx 1$, the reconstructions of both test objects recover most of the correct overall shape, except at $r_z$ where the associated interference function $\mathfrak s(r_z)$ is close to 0. However, in these two tests, the interference functions for the porous-medium reconstructions have a large number of approximate zeros, leading to more instabilities and resolution loss in the reconstruction compared to the conical Siemens star. 

In test case 3, where the incident angle is slightly above the critical angle and $|R_i|\approx 0.5$, the interference function remains large throughout the entire $r_z$ range, leading to accurate reconstructions with data from just a single incident angle. 

In test case 4, where the incident angle is much larger than the critical angle and $|R_i| \approx 0$, the reconstructions perform poorly, as predicted, since this causes several of the DWBA terms to vanish, preventing full sampling of the density's Fourier transform. Here, the maximum exit angle for the porous medium example also fails to satisfy Eq.~\ref{eq:alphafmax1}. Although the maximum exit angle for the conical Siemens star example satisfies this condition, the small $|R_i|$ still leads to reconstruction instability.

In test case 5, where the datasets from tests 1 and 2 are combined, the reconstructions for both objects are accurate throughout the $r_z$ range since, for each $r_z$, at least one of the two interference functions is sufficiently large.

In summary, these reconstruction results validate the theory described in Sections~\ref{sec:incident}-\ref{sec:exit} and demonstrate that our proposed nonuniform iterative phasing algorithm is capable of accurately reconstructing 3D structure from CSSI intensities when the incident- and exit-angle conditions for this theory are satisfied, i.e., for test cases 3 and 5. In these cases, the reconstructions exhibit a low relative error of about 3\% and yield FSC values close to 1 throughout the simulated $q$ range. Even when the data is generated from a single incident angle far below the critical angle, the algorithm is still able to reconstruct many of the correct features in regions away from the approximate zeros of the corresponding interference function.

\begin{table}[h!]
\begin{center}
\begin{tabular}{l@{\hspace{1em}} c c c c c} 
\midrule
& \multicolumn{5}{c}{Test Cases} \\
Object  & 1 & 2 & 3 & 4 & 5 \\ 
\midrule
PM & 0.29 & 0.18 & 0.03 & 0.90 & 0.03  \\
SS & 0.06 & 0.05 & 0.03 & 0.84 & 0.03  \\
\midrule
\end{tabular}
\end{center}
\vspace*{-1em}
\caption{Relative $\ell^2$ errors of the reconstructions of the porous medium (PM) and conical Siemens star (SS) objects from staggered CSSI intensity grid data for the five test cases.}\label{table:rel_error}
\end{table}

\subsection{Reconstruction from CSSI Rotation-Series Data}\label{sec:rot_series}
Here we present 3D reconstructions from rotation series of 2D CSSI diffraction images. This is accomplished by first applying the techniques in Section~\ref{sec:reduction} to reduce and resample the data onto a staggered CSSI intensity grid and then applying the nonuniform iterative phasing algorithm in Section~\ref{sec:CSSIIPA} to reconstruct the target objects from the resampled data on this grid.

\begin{figure*}[h!] 
 \centering
\includegraphics[width = 1\textwidth]{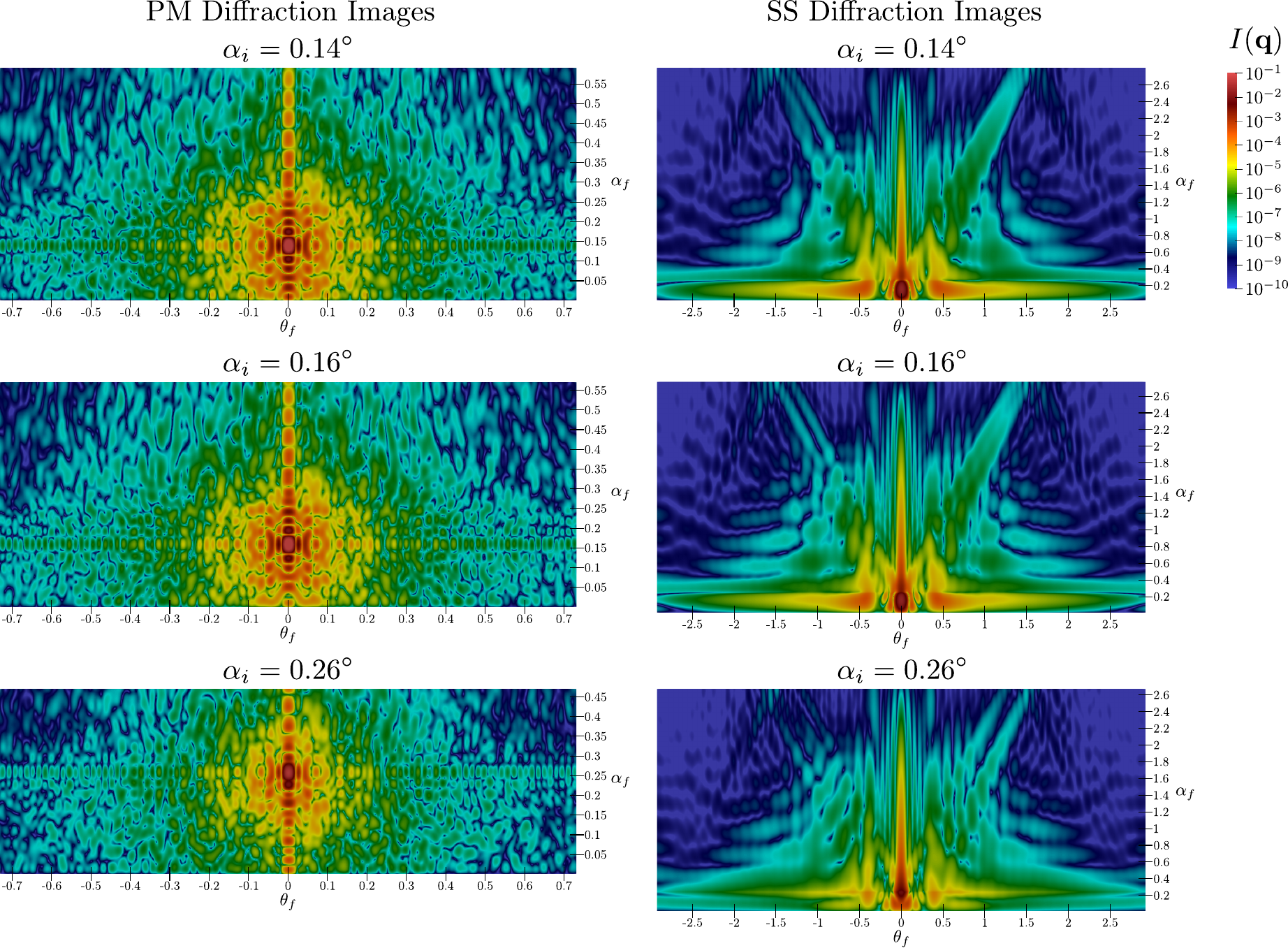}
\caption{Examples of the simulated CSSI diffraction images generated from the porous medium (PM) and conical Siemens star (SS) for each simulated incident angle. Displayed images are generated from the reference orientation $\phi=0$.}\label{fig:diff_images}
\end{figure*}

Rotation-series data is simulated for both the porous medium and conical Siemens star using experimental parameters to mirror test cases 3 and 5 in Section~\ref{sec:results_volume}. More specifically, we generate three rotation-series datasets with incident angles of $0.14\degree$, $0.16\degree$, and $0.26\degree$. The maximum exit and scattering angles are selected to be the largest possible values that prevent aliasing in the diffraction patterns, while the specimen-rotation angles are chosen to enforce the conditions in Eqs.~\ref{eq:phimax}-\ref{eq:deltaphi}. The data is simulated using a pixel size of 50 \si{\um} and sample-to-detector distances of 2000 mm for the porous medium and 500 mm for the conical Siemens star, which oversamples the intensity by a factor of 8-10 in each case. The bottom of the detector was placed 50 \si{\um} above the substrate plane, yielding minimum exit angles of $0.0014\degree$ and $0.0057\degree$ for the respective detector setups. These parameters are summarized in Table~\ref{table:rot_series_params}, and examples of the simulated images are displayed in Fig.~\ref{fig:diff_images}.

\begin{table}[h!]
\begin{center}
\begin{tabular}{l@{\hspace{1em}} c c c c c c} 
\midrule            
Object  & $\alpha_i$ & $\alpha_f^{\max}$ & $\theta_f^{\max}$ & $\phi^{\max}$ & \#Images& \#Pixels\\ 
\midrule
PM  & $0.14\degree$ & $0.59\degree$ & $0.73\degree$ & $181\degree$  & 384 & 1021x412\\
PM  & $0.16\degree$ & $0.57\degree$ & $0.73\degree$ & $181\degree$  & 384 & 1021x398\\
PM  & $0.26\degree$ & $0.47\degree$ & $0.73\degree$ & $181\degree$  & 384 & 1021x328\\
SS  & $0.14\degree$ & $2.78\degree$ & $2.92\degree$ & $339\degree$  & 718 & 1021x486\\
SS  & $0.16\degree$ & $2.76\degree$ & $2.92\degree$ & $339\degree$  & 717 & 1021x482\\
SS  & $0.26\degree$ & $2.66\degree$ & $2.92\degree$ & $337\degree$  & 713 & 1021x465\\
\midrule
\end{tabular}
\end{center}
\vspace*{-1em}
\caption{Experimental parameters for the CSSI rotation series simulations of the porous medium (PM) and conical Siemens star (SS) test objects.}\label{table:rot_series_params}
\end{table}

\begin{figure*}[h!] 
 \centering
\includegraphics[width = .99\textwidth]{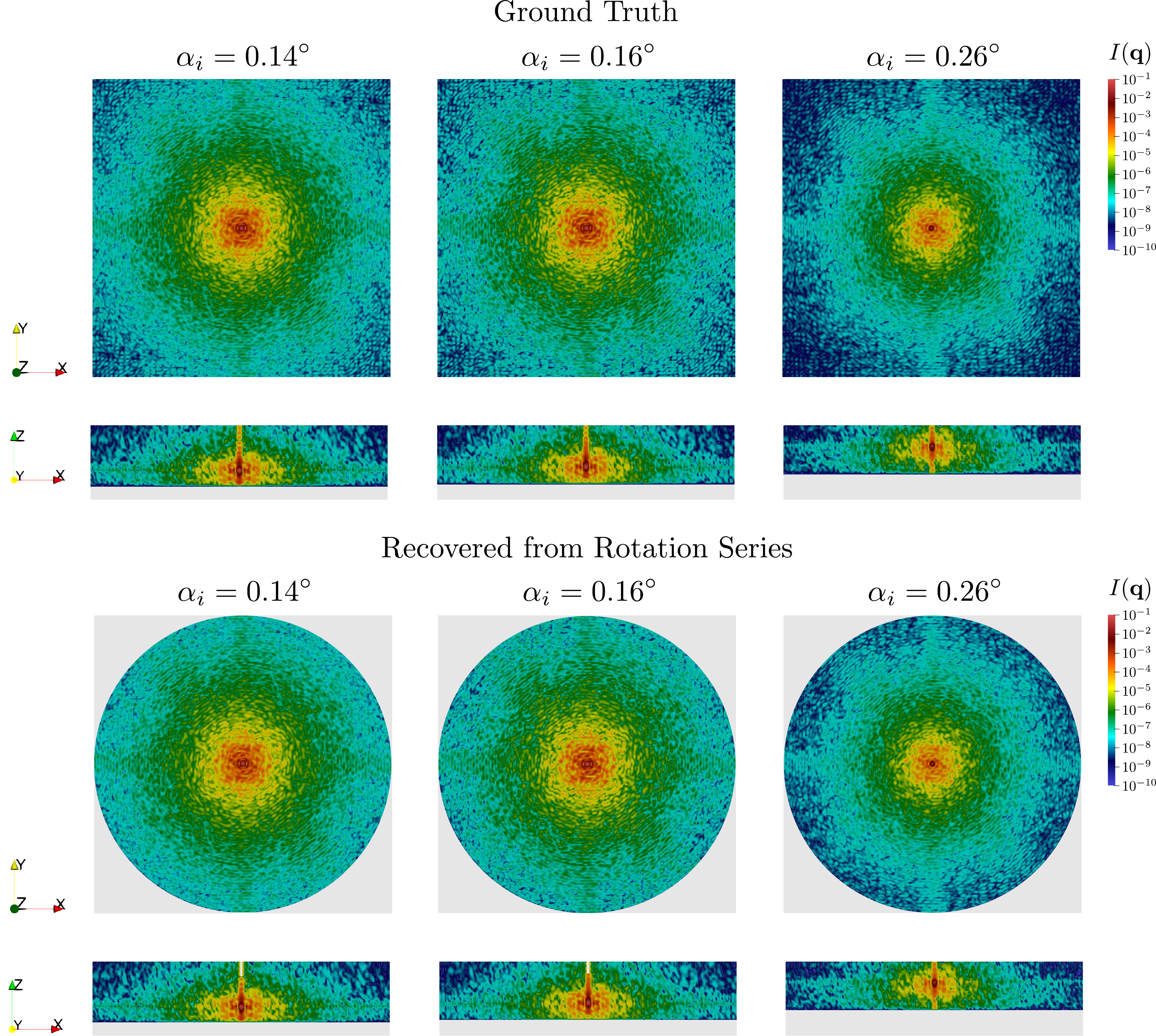}
\caption{2D slices of the ground-truth porous-medium diffraction volumes compared with the diffraction volumes recovered from the porous-medium rotation-series data, for each simulated incident angle. Each pixel corresponds to one Nyquist pixel. The recovered volumes are obtained by first reducing the rotation-series data to their compressed real-space representations and then transforming these representations to CSSI intensities defined on staggered intensity grids. Horizontal slices are shown for $\qz=2\sin(\alpha_i)/\lambda$ and vertical slices are shown for $q_y = 0$. Gray regions indicate unsampled areas that are masked out in the subsequent reconstruction.}\label{fig:rot_reductions_PM}
\end{figure*}

\begin{figure*}[h!] 
 \centering
\includegraphics[width = .99\textwidth]{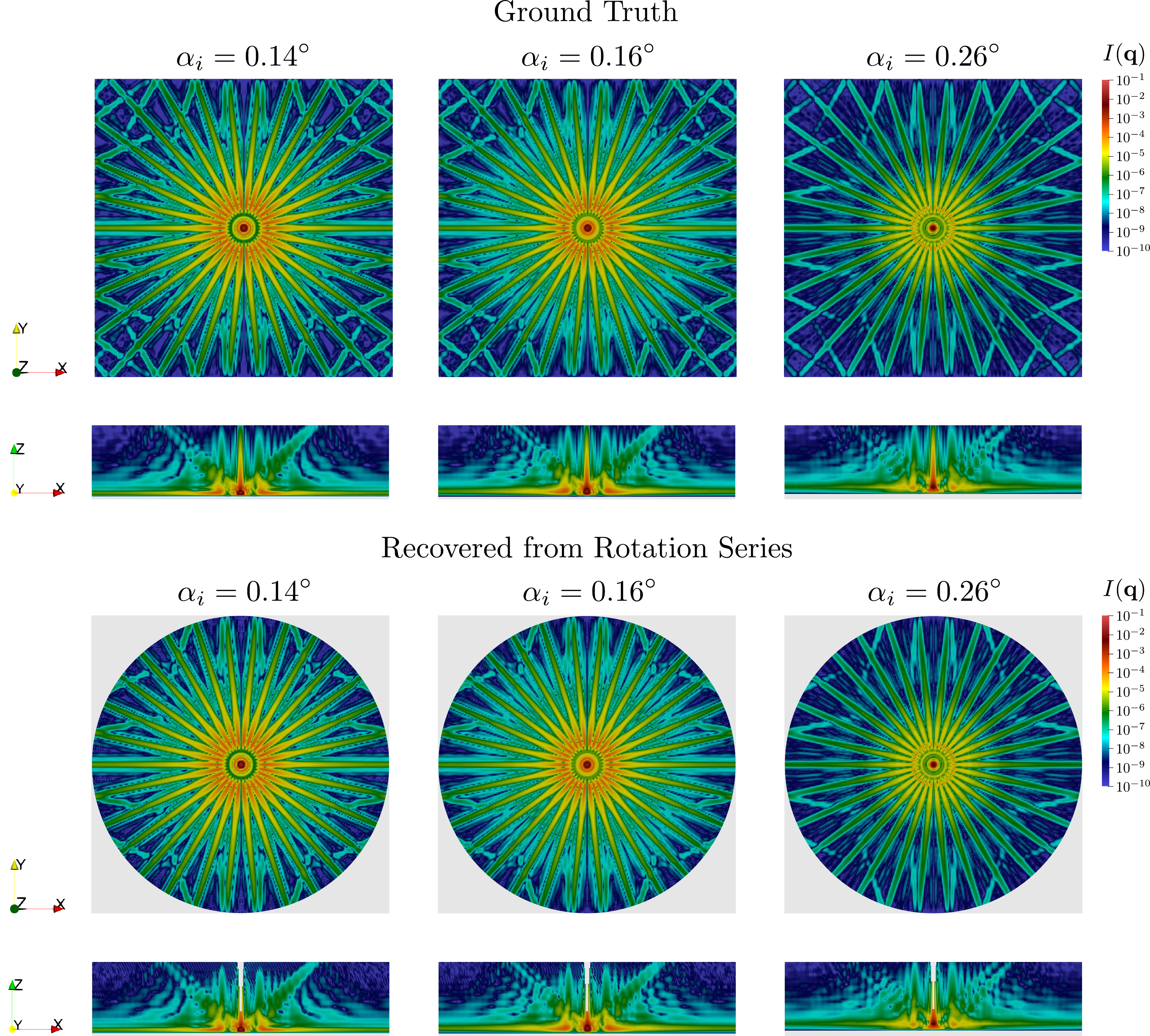}
\caption{ 2D slices of the ground-truth conical-Siemens-star diffraction volumes compared with the diffraction volumes recovered from the conical-Siemens-star rotation-series data, for each simulated incident angle. Each pixel corresponds to one Nyquist pixel. The recovered volumes are obtained by first reducing the rotation-series data to their compressed real-space representations and then transforming these representations to CSSI intensities defined on staggered intensity grids. Horizontal slices are shown for $\qz=2\sin(\alpha_i)/\lambda$ and vertical slices are shown for $q_y = 0$. Gray regions indicate unsampled areas that are masked out in the subsequent reconstruction.}\label{fig:rot_reductions_SS}
\end{figure*}

We begin by applying the methods in Section~\ref{sec:reduction} to reduce/resample the rotation-series datasets onto staggered intensity grids. For each dataset, we solve Eq.~\ref{eq:Anormal} to obtain the reduced real-space representations $f_1$ and $f_2$. This solution is obtained by applying 2000 conjugate-gradient iterations using the generalized Toeplitz acceleration in Eq.~\ref{eq:GMLToeplitz}. We then input these real-space representations into Eq.~\ref{eq:linred} to compute CSSI diffraction volumes defined on staggered intensity grids. For each staggered grid, a mask is applied to all points more than half a Nyquist length away from the set of $\mathbf q$ sampled by the corresponding rotation series. These reduced diffraction volumes are displayed in Figs.~\ref{fig:rot_reductions_PM}-\ref{fig:rot_reductions_SS}, and their corresponding errors are reported in Table~\ref{table:intensity_error_rot_series}.

\begin{table}[h!]
\begin{center}
\begin{tabular}{l@{\hspace{1em}} c c c} 
\midrule
& \multicolumn{3}{c}{Incident Angle} \\
Object  & $0.14\degree$ & $0.16\degree$ & $0.26\degree$\\ 
\midrule
PM\rule{0pt}{2.4ex} & $1.5\times 10^{-5}$ & $9.2\times 10^{-6}$ & $1.4\times 10^{-4}$  \\
SS & $7.5\times 10^{-5}$ & $5.1\times 10^{-5}$ & $6.0\times 10^{-5}$ \\
\midrule
\end{tabular}
\end{center}
\vspace*{-1em}
\caption{Relative $\ell^2$ errors of the staggered-grid CSSI diffraction volumes reduced from simulated rotation series images for the porous medium (PM) and conical Siemens star (SS) objects, using the parameters in Table~\ref{table:rot_series_params}. Only voxels in the sampled regions of the volume (i.e., outside of the masked regions) are included in the error calculation.}\label{table:intensity_error_rot_series}
\end{table}

Next, we apply the nonuniform iterative phasing algorithm in Section~\ref{sec:NIP} to reconstruct the densities of the target objects from the resampled intensity data. In particular, we replicate test case 3 by performing the reconstruction using the intensity from the $\alpha_i = 0.26\degree$ dataset, and test case 5 by using the combined intensity data from the $\alpha_{i} = 0.14\degree$ and $\alpha_{i} = 0.16\degree$ datasets. We use the same reconstruction parameters described in Section~\ref{sec:results_volume}. 

\begin{figure}[h!] 
 \centering
 \includegraphics[width = \columnwidth]{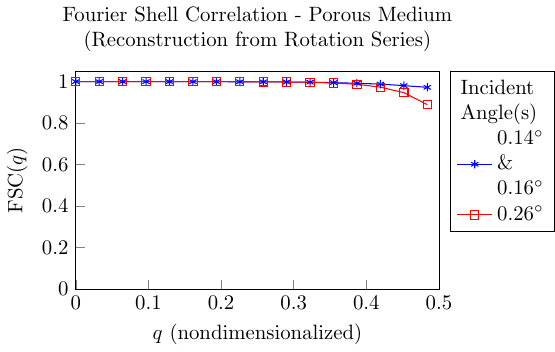}
\caption{Fourier shell correlation (FSC) for the reconstructions of the porous medium from the rotation-series data. }\label{fig:FSC_PM_rot}
\end{figure}

\begin{figure}[h!] 
\centering
\includegraphics[width = \columnwidth]{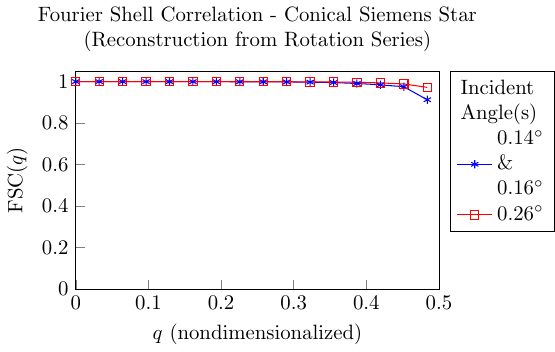}
\caption{Fourier shell correlation (FSC) for the reconstructions of the conical Siemens star from the rotation-series data. }\label{fig:FSC_SS_rot}
\end{figure}

\begin{figure*}[h!] 
 \centering
\includegraphics[width = \textwidth]{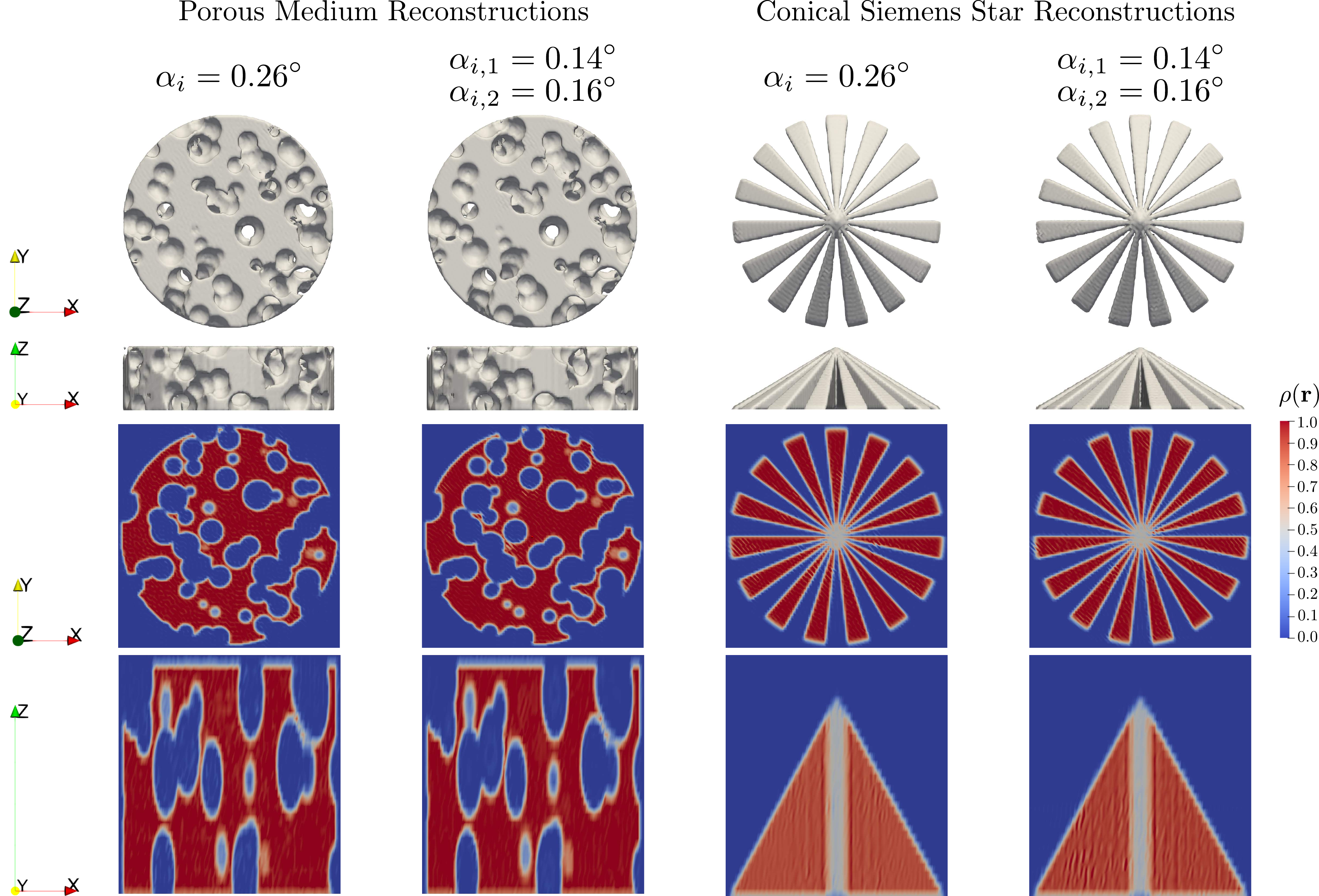}
\caption{Reconstructions of the porous medium and conical Siemens star from the reduced rotation-series data. For each object, a reconstruction is performed using data from a single incident angle of $\alpha_i = 0.26\degree$, and another is performed using data from two incident angles of $\alpha_{i,1}=0.14\degree$ and $\alpha_{i,2}=0.16\degree$.  For each example, the top two rows display different orientations of the density isosurfaces at $\rho = 0.5$ for the porous medium and $\rho = 0.25$ for the conical Siemens star. The bottom two rows
show horizontal and vertical slices of each density. In the bottom
row, each slice is stretched in the $z$ direction by a factor of four.}\label{fig:rot_reconstructions}
\end{figure*}

The reconstructions are visualized in Fig.~\ref{fig:rot_reconstructions}, their relative $\ell^2$ errors are listed in Table~\ref{table:rel_error_rot_series}, and their corresponding FSC curves are shown in Figs.~\ref{fig:FSC_PM_rot}-\ref{fig:FSC_SS_rot}. As the figures and table show, the reconstructions from the rotation-series data are highly accurate across the simulated resolution range. However, compared to test cases 3 and 5 in Section~\ref{sec:results_volume}, there is a slight decline in reconstruction quality. This is due to the inability of the rotation series to sample the diffraction volume near the corners of the $xy$ plane and near the $z$ axis for large $\qz$ values, resulting in the missing-candlestick effect described in Section~\ref{sec:candlestick}. These inaccessible regions are shown in gray in Figs.~\ref{fig:rot_reductions_PM}-\ref{fig:rot_reductions_SS}.

\begin{table}[h!]
\begin{center}
\begin{tabular}{l@{\hspace{1em}} c c} 
\midrule
& \multicolumn{2}{c}{\hspace{1em}Incident Angle(s)} \\
Object  & $0.14\degree \&\, 0.16\degree$ &$0.26\degree$  \\ 
\midrule
PM & 0.05 & 0.05  \\
SS & 0.08 & 0.06  \\
\midrule
\end{tabular}
\end{center}
\vspace*{-1em}
\caption{Relative $\ell^2$ errors of the reconstructions of the porous medium (PM) and conical Siemens star (SS) objects from simulated rotation-series data for the test cases described in Table~\ref{table:rot_series_params}.}\label{table:rel_error_rot_series}
\end{table}

\subsection{Reconstruction from Noisy CSSI Data}
\begin{figure*}[h!] 
 \centering
\includegraphics[width = .99\textwidth]{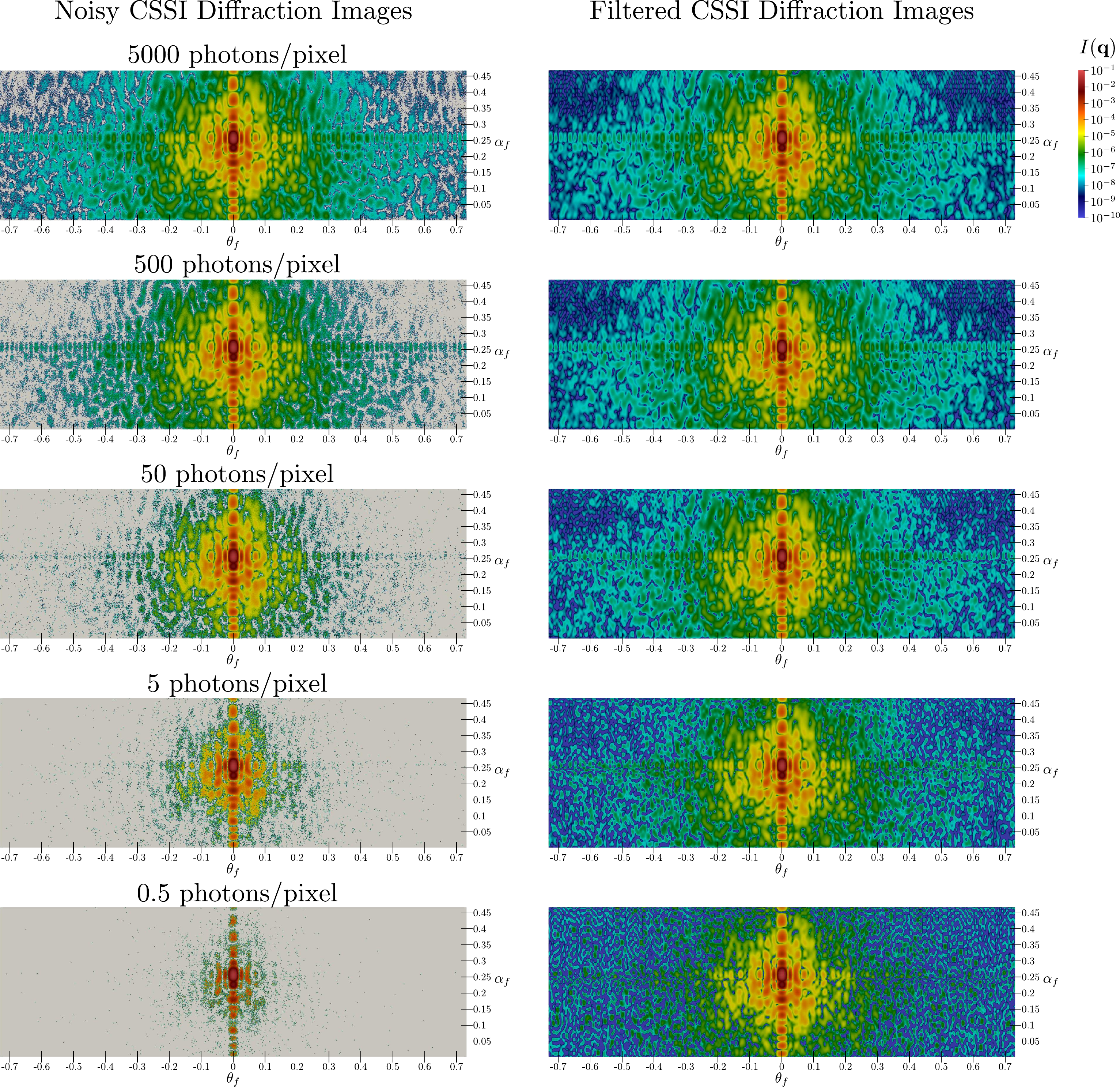}
\caption{Examples of noisy CSSI diffraction images simulated from the porous medium with $\alpha_i = 0.26\degree$ and their corresponding filtered images, obtained by first reducing the rotation-series data to their compressed real-space representations and then transforming these representations to CSSI intensities defined on the original pixel coordinates. Gray regions correspond to pixels where no photons were measured.}\label{fig:noisy_filtered_PM}
\end{figure*}

\begin{figure*}[h!] 
 \centering
\includegraphics[width = .99\textwidth]{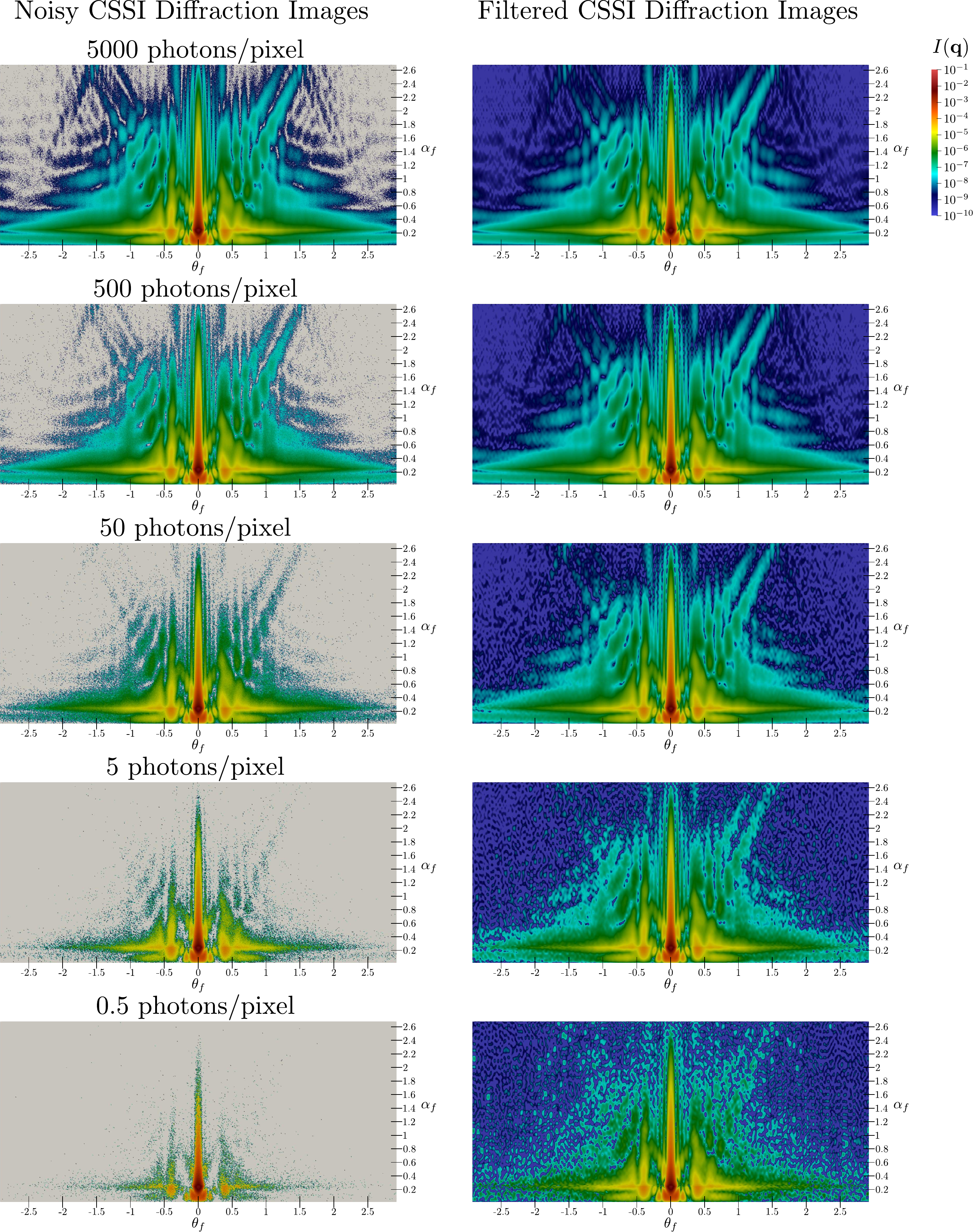}
\caption{Examples of noisy CSSI diffraction images simulated from the conical Siemens star with $\alpha_i = 0.26\degree$ at different noise levels (quantified as average number of photons per pixel), and their corresponding filtered images obtained by first reducing the rotation-series data to their compressed real-space representations and then transforming these representations to CSSI intensities defined on the original pixel coordinates. Gray regions correspond to pixels where no photons were measured.}\label{fig:noisy_filtered_SS}
\end{figure*}

Here we test the robustness of our reduction and reconstruction frameworks to different noise levels.  We simulate rotation-series data for both of our test objects with five different levels of shot noise, using the parameters in Table~\ref{table:rot_series_params} for $\alpha_i = 0.26\degree$. The noisy data is generated by first multiplying the original clean data by a scale factor, and then replacing each pixel's intensity with a value randomly sampled from a Poisson distribution, where the mean is given by the scaled pixel intensity. The scale factors are chosen to produce datasets with average photon counts of about 5000, 500, 50, 5, and 0.5 photons per pixel. The noisy images are then divided by their corresponding scale factor to ensure that all results are on a consistent scale. Examples of the noisy images are illustrated in Figs.~\ref{fig:noisy_filtered_PM}-\ref{fig:noisy_filtered_SS}.

\begin{figure*}[h!] 
 \centering
\includegraphics[width = .99\textwidth]{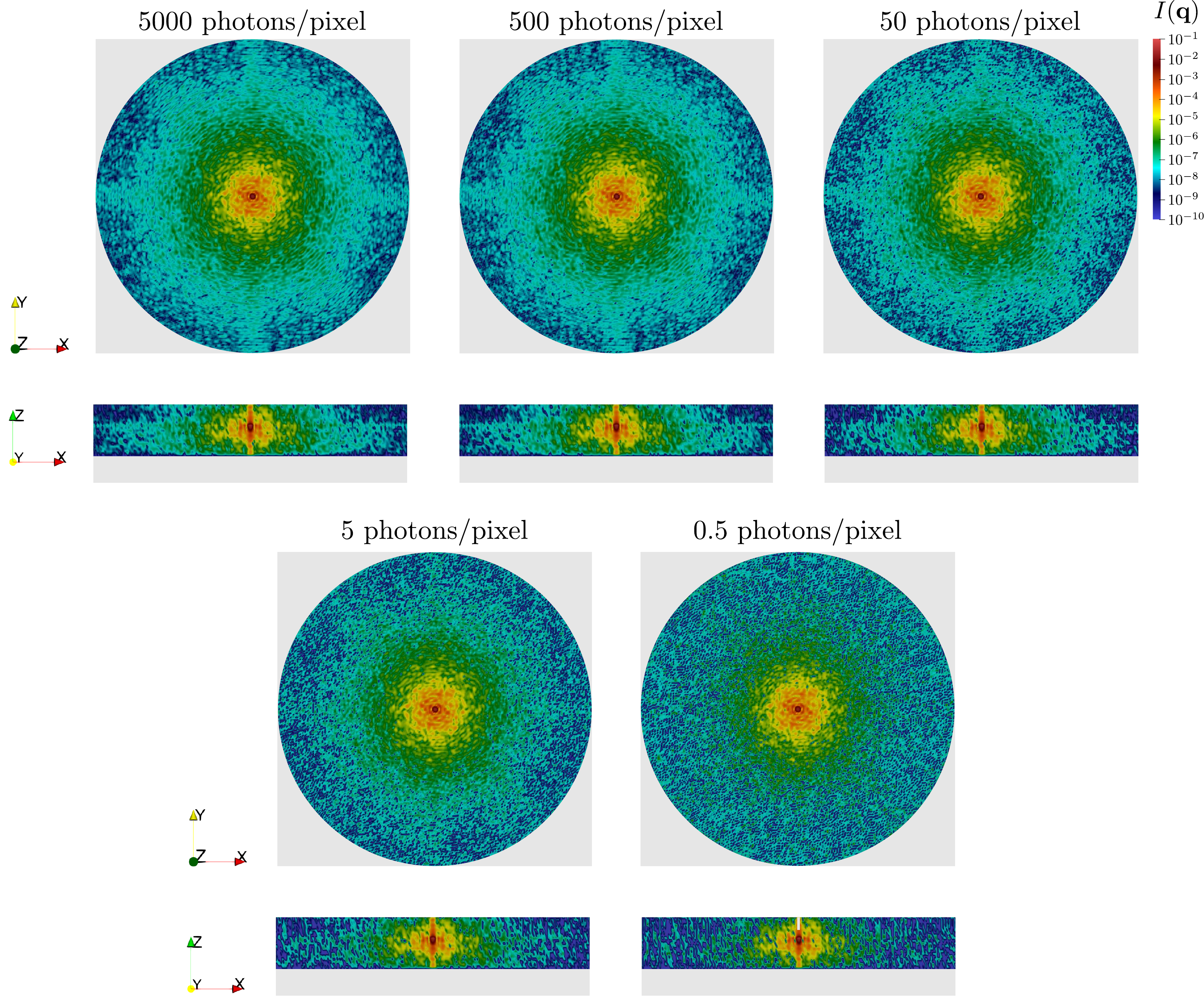}
\caption{2D slices of the CSSI diffraction volumes recovered from the porous-medium rotation-series data simulated with $\alpha_i = 0.26\degree$ at different noise levels (quantified as average number of photons per pixel). Each pixel corresponds to one Nyquist pixel. Horizontal slices are shown for $\qz=2\sin(\alpha_i)/\lambda$ and vertical slices are shown for $q_y = 0$. Gray regions indicate unsampled areas that are masked out in the subsequent reconstruction.}\label{fig:reduced_noisy_PM}
\end{figure*}

\begin{figure*}[h!] 
 \centering
\includegraphics[width = .99\textwidth]{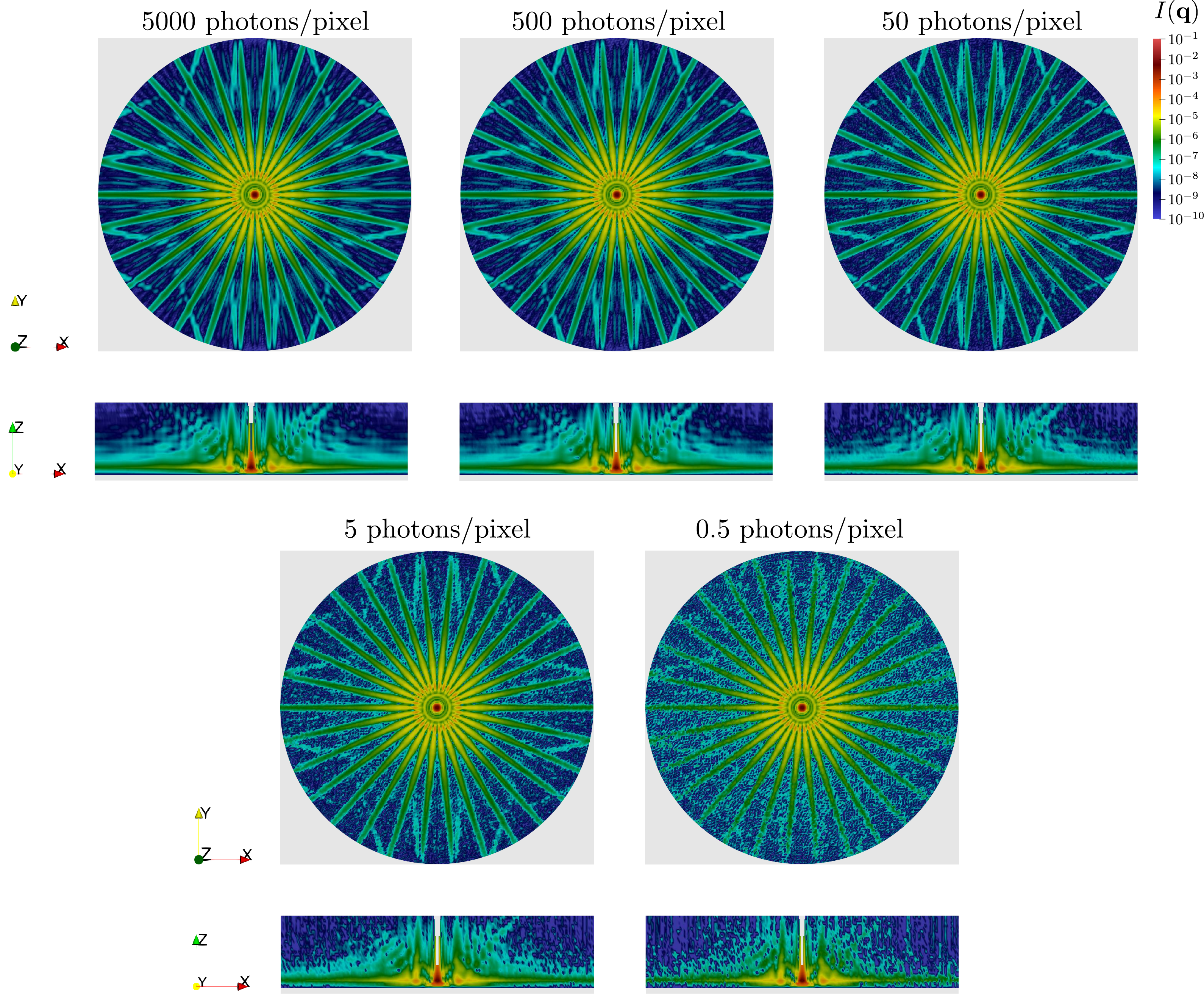}
\caption{2D slices of the CSSI diffraction volumes recovered from the conical-Siemens-star rotation-series data simulated with $\alpha_i = 0.26\degree$ at different noise levels (quantified as average number of photons per pixel). Each pixel corresponds to one Nyquist pixel. Horizontal slices are shown for $\qz=2\sin(\alpha_i)/\lambda$ and vertical slices are shown for $q_y = 0$. Gray regions indicate unsampled areas that are masked out in the subsequent reconstruction.}\label{fig:reduced_noisy_SS}
\end{figure*}

We first repeat the data reduction/resampling and reconstruction procedures in Section~\ref{sec:rot_series} for the simulated noisy data. In addition, to illustrate the robustness of the reduction/resampling procedure to noise, we also resample the reduced data onto the original pixel coordinates to obtain filtered versions of the diffraction images. We show examples of the noisy and filtered images in Figs.~\ref{fig:noisy_filtered_PM}-\ref{fig:noisy_filtered_SS} and the recovered diffraction volumes in Figs.~\ref{fig:reduced_noisy_PM}-\ref{fig:reduced_noisy_SS}. The relative $\ell^2$ errors of the noisy and filtered images and the recovered diffraction volumes are reported in Tables~\ref{table:rot_series_error_noisy} and  \ref{table:diff_volume_noisy}, respectively. 

As shown by the filtered images and recovered diffraction volumes, the reduction/resampling procedure can recover accurate pixel intensities even when the images are very sparse. Compared to the noisy images, the relative errors in the recovered volumes and filtered images are between one and two orders of magnitude lower, which is on the same order as the ratios of the number of measurements to linear degrees of freedom (see Section~\ref{sec:linear_reduction}), which are about 36 for the porous medium data and 96 for the conical Siemens star data. In regions with insufficient signal, this procedure produces small oscillations in the recovered intensity, but these oscillations remain bounded as the signal decreases. 

\begin{table}[h!]
\begin{center}
\begin{tabular}{@{\hspace{-.1em}}l@{\hspace{.5em}} c c c c c} 
\midrule
              & \multicolumn{5}{c}{Noise level (photons/pixel)} \\
Images & 5000 & 500 & 50 & 5 & 0.5 \\ 
 \midrule
PM(\hspace*{-.08em}orig\hspace*{-.08em})\hspace*{-.2em}\rule{0pt}{2.4ex} & \hspace*{-.08em}$5.6\!\times\!10^{-4}$\hspace*{-.08em} & \hspace*{-.08em}$1.8\!\times\!10^{-3}$\hspace*{-.08em} & \hspace*{-.08em}$5.6\!\times\!10^{-3}$\hspace*{-.08em} & \hspace*{-.08em}$1.7\!\times\!10^{-2}$\hspace*{-.08em} & \hspace*{-.08em}$5.6\!\times\!10^{-2}$\hspace*{-.08em}  \\
PM(filt) & \hspace*{-.08em}$1.6\!\times\!10^{-4}$\hspace*{-.08em} & \hspace*{-.08em}$1.7\!\times\!10^{-4}$\hspace*{-.08em} & \hspace*{-.08em}$1.9\!\times\!10^{-4}$\hspace*{-.08em} & \hspace*{-.08em}$3.9\!\times\!10^{-4}$\hspace*{-.08em} & \hspace*{-.08em}$1.1\!\times\!10^{-3}$\hspace*{-.08em}\\
SS(orig) & \hspace*{-.08em}$6.7\!\times\!10^{-4}$\hspace*{-.08em} & \hspace*{-.08em}$2.1\!\times\!10^{-3}$\hspace*{-.08em} & \hspace*{-.08em}$6.7\!\times\!10^{-3}$\hspace*{-.08em} & \hspace*{-.08em}$2.1\!\times\!10^{-2}$\hspace*{-.08em} & \hspace*{-.08em}$6.7\!\times\!10^{-2}$\hspace*{-.08em}  \\
SS(filt) & \hspace*{-.08em}$7.2\!\times\!10^{-5}$\hspace*{-.08em} & \hspace*{-.08em}$8.4\!\times\!10^{-5}$\hspace*{-.08em} & \hspace*{-.08em}$1.6\!\times\!10^{-4}$\hspace*{-.08em} & \hspace*{-.08em}$4.1\!\times\!10^{-4}$\hspace*{-.08em} & \hspace*{-.08em}$1.3\!\times\!10^{-3}$\hspace*{-.08em}  \\
 \midrule
\end{tabular}
\end{center}
\vspace*{-1em}
\caption{
Relative $\ell^2$ errors of the original (orig) and filtered (filt) CSSI rotation series images simulated for the porous medium (PM) and conical Siemens star (SS) objects at different noise levels (quantified as average number of photons per pixel).}\label{table:rot_series_error_noisy}
\end{table}

\begin{table}[h!]
\begin{center}
\begin{tabular}{l@{\hspace{.1em}} c c c c c} 
\midrule
& \multicolumn{5}{c}{Noise level (photons/pixel)} \\
Object  & 5000 & 500 & 50 & 5 & 0.5 \\ 
\midrule
PM & $1.4\!\times\! 10^{-4}$ & $1.9\!\times\! 10^{-4}$ & $4.4\!\times\! 10^{-4}$ & $1.3\!\times\! 10^{-3}$ & $4.0\!\times\! 10^{-3}$  \\
SS & $8.5\!\times\! 10^{-5}$ & $2.0\!\times\! 10^{-4}$ & $6.0\!\times\! 10^{-4}$ & $1.9\!\times\! 10^{-3}$& $5.7\!\times\! 10^{-3}$  \\
\midrule
\end{tabular}
\end{center}
\vspace*{-1em}
\caption{Relative $\ell^2$ errors of the staggered-grid CSSI diffraction volumes reduced from rotation series images simulated at different noise levels (quantified as the average number of photons per pixel) for the porous medium (PM) and conical Siemens star (SS) objects. Only voxels in the sampled regions of the volume (i.e., outside of the masked regions) are included in the error calculation.}\label{table:diff_volume_noisy}
\end{table}

\begin{figure}[h!] 
\centering
\includegraphics[width = \columnwidth]{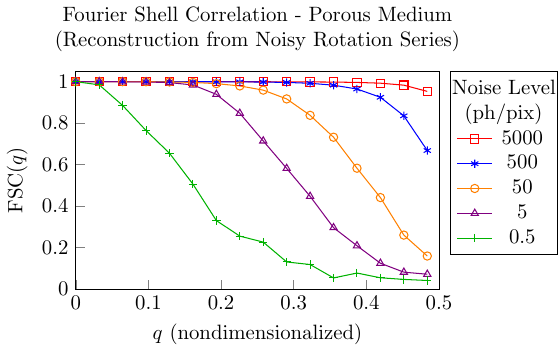}
\caption{Fourier shell correlation (FSC) for the reconstructions of the porous medium from rotation-series data simulated with five different noise levels (quantified as average number of photons per pixel). }\label{fig:FSC_PM_noise}
\end{figure}

\begin{figure}[h!] 
\centering
\includegraphics[width = \columnwidth]{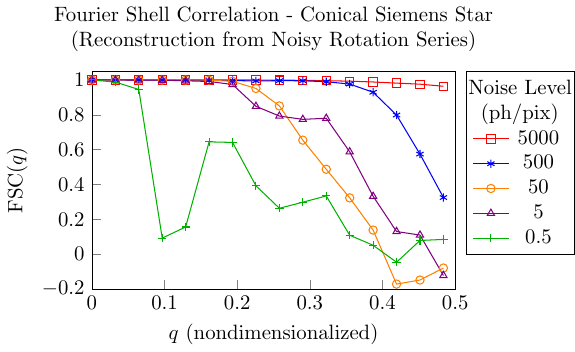}
\caption{Fourier shell correlation (FSC) for the reconstructions of the conical Siemens star from rotation-series data simulated with five different noise levels (quantified as average number of photons per pixel). }\label{fig:FSC_SS_noise}
\end{figure}

\begin{figure*}[h!] 
 \centering
\includegraphics[width = .89\textwidth]{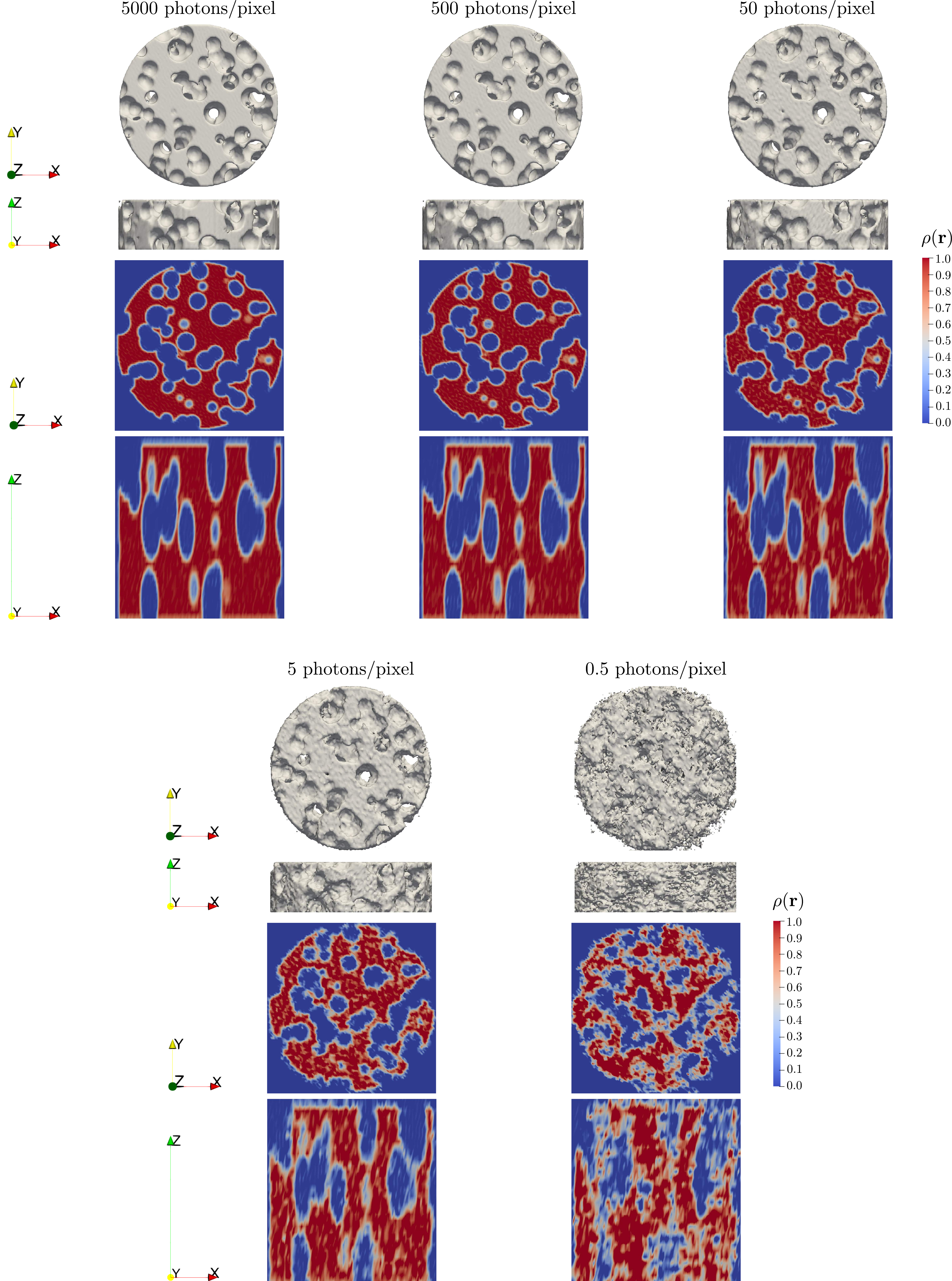}
\caption{Reconstructions of the porous medium from its rotation-series data simulated with $\alpha_i = 0.26\degree$ at different noise levels (quantified as average number of photons per pixel). For each example, the top two rows display different orientations of the density isosurfaces at $\rho = 0.5$. The bottom two rows show horizontal and vertical slices of each density through the middle of the volume. In the bottom row, each slice is stretched in the $z$ direction by a factor of four.}\label{fig:noisy_reconstructions_PM}
\end{figure*}

\begin{figure*}[h!] 
 \centering
\includegraphics[width = .89\textwidth]{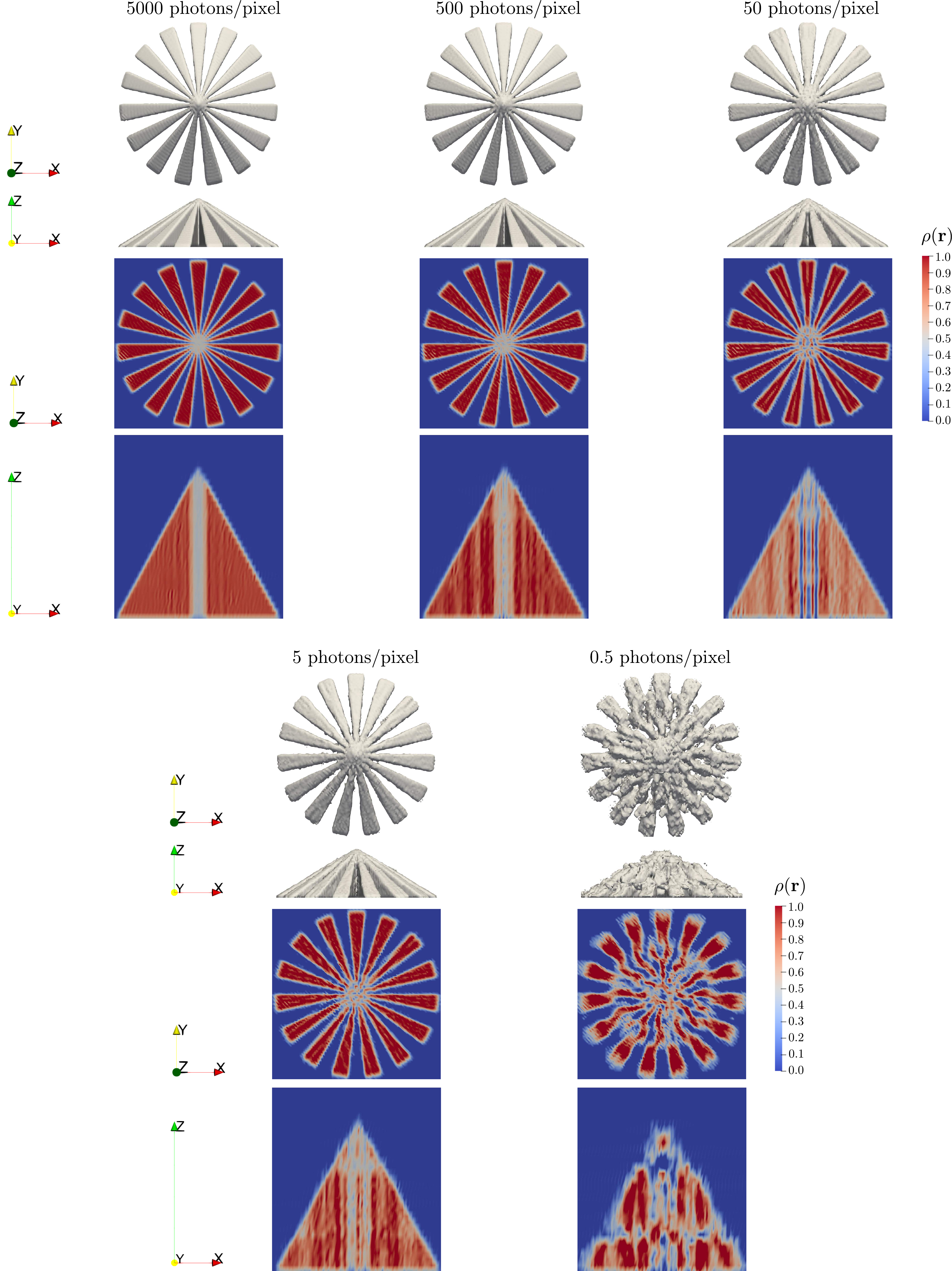}
\caption{Reconstructions of the conical Siemens star from its rotation-series data simulated with $\alpha_i = 0.26\degree$ at different noise levels (quantified as average number of photons per pixel). For each example, the top two rows display different orientations of the density isosurfaces at $\rho = 0.25$. The bottom two rows show horizontal and vertical slices of each density through the bottom and middle of the volume, respectively. In the bottom row, each slice is stretched in the $z$ direction by a factor of four.}\label{fig:noisy_reconstructions_SS}
\end{figure*}

Next, we apply the nonuniform iterative phasing algorithm to the staggered-grid intensity volumes resampled from the noisy rotation-series images, using the same reconstruction parameters as in Section~\ref{sec:results_volume}. The resulting reconstructions are displayed in Figs.~\ref{fig:noisy_reconstructions_PM}-\ref{fig:noisy_reconstructions_SS}, the relative $\ell^2$ reconstruction errors are reported in Table~\ref{table:rel_error_noisy}, and the corresponding FSC curves are shown in Figs.~\ref{fig:FSC_PM_noise}-\ref{fig:FSC_SS_noise}.

The reconstruction results indicate the robustness of the end-to-end nonuniform iterative reconstruction framework to noise. In particular, the reconstructions accurately recover the overall shape of the target objects, even from very sparse data, with a gradual loss of detail as noise levels increase. Notably, the reconstructions only begin to significantly deteriorate at the extreme sparsity level of 0.5 photons/pixel. Interestingly, when the average photon count is 50 photons/pixel or less, the conical Siemens star reconstructions exhibit artifacts along horizontal planes at heights near the minima of the interference function, and these artifacts become more pronounced as the photon counts decrease. This observation further suggests that the stability of the CSSI reconstruction is closely linked to the shape of the interference function.

\begin{table}[h!]
\begin{center}
\begin{tabular}{l@{\hspace{.6em}} c@{\hspace{.6em}} c@{\hspace{.6em}} c@{\hspace{.6em}} c c} 
\midrule
& \multicolumn{5}{c}{Noise level (photons/pixel)} \\
Object  & 5000 & 500 & 50 & 5 & 0.5 \\ 
\midrule
PM & 0.05 & 0.08 & 0.16 & 0.26 & 0.57  \\
SS & 0.07 & 0.13 & 0.26 & 0.26 & 0.78  \\
\midrule
\end{tabular}
\end{center}
\vspace*{-1em}
\caption{Relative $\ell^2$ errors of the reconstructions of the porous medium and conical Siemens star objects from rotation-series data simulated at different noise levels (quantified as average number of photons per pixel).}\label{table:rel_error_noisy}
\end{table}

\section{Conclusion}
We have presented a new nonuniform iterative reconstruction framework capable of \textit{ab initio} 3D structure reconstruction from CSSI rotation-series data, even when all terms in the DWBA are significant, and requiring as few as one or two incident beam angles. In addition, we have derived several experimental parameter/sampling requirements and uniqueness properties for CSSI reconstruction, which should be taken into account during experimental design and data collection to maximize the information content of the data and resolution of the reconstruction.  

Applying the presented reconstruction techniques to experimental data will likely require additional processing steps to ensure sufficient data quality. This may involve masking out streaks, handling detector gaps, performing alignment and jitter corrections, and carrying out other calibrations. Many of these procedures can potentially be integrated directly into the iterative phasing framework, similar to \cite{kurta2017correlations, donatelli2017reconstruction, pande2018ab, chen2017phase}, if the algorithm is applied to the original CSSI images rather than the resampled data on the staggered grid. Alternatively, these procedures could be performed as part of the linear reduction step, similar to joint iterative reconstruction and alignment in tomography \cite{pande2022joint, yang2005unified, ramos2017automated, yu2018automatic}. Ongoing improvements in X-ray light source coherence, brightness, and instrumentation, as well as dedicated beamlines such as the CSSI-feature beamline at the Advanced Photon Source \cite{jiang2025design}, are expected to significantly enhance CSSI data quality, making high-resolution 3D structure determination from CSSI measurements feasible.

We are also investigating extensions of the proposed framework to handle more general samples and experimental conditions. These include modeling additional dynamical scattering events for thicker or more strongly scattering materials \cite{venkatakrishnan2016multi, chu2023probing, myint2023multislice}, combining the rotation-series approach with ptychographic scanning \cite{myint2024three, sung2025automatic} for specimens larger than the beam width, and accounting for surface roughness and other effects arising from substrate imperfections \cite{lerondel1999fresnel}.

Finally, the nonuniform iterative reconstruction framework introduced here provides a general approach to extend conventional iterative phasing techniques to more general phase problems that inherently involve nonuniform Fourier sampling or are more conveniently expressed using polar or spherical-polar grids, e.g., \cite{donatelli2015iterative, donatelli2017reconstruction}. A detailed analysis of this framework and its extensions to other experiments will be investigated in an upcoming manuscript. 

\begin{acknowledgments}
This work was supported by the U.S. Department of Energy (DOE), Office of Science, Office of Advanced Scientific Computing Research (ASCR) Early Career Research Program and Applied Mathematics Competitive Portfolios Program, and by the Center for Advanced Mathematics for Energy Research Applications (CAMERA), jointly funded by ASCR and the Office of Basic Energy Sciences (BES), under Contract No.\@ DE-AC02-05CH11231.
Additional support was provided by the U.S. DOE, Office of Science, \mbox{ASCR} and \mbox{BES} awards ``\mbox{ILLUMINE} - Intelligent Learning for Light Source and Neutron Source User Measurements Including Navigation and Experiment Steering" and ``X-ray \& Neutron Scientific Center for Optimization, Prediction, \& Experimentation (\mbox{XSCOPE})", by the BES Early Career Research Program, and used resources of the Advanced Photon Source (\mbox{APS}), a U.S. DOE, Office of Science, BES user facility operated by Argonne National Laboratory, under Contract No.\@ DE-AC02-06CH11357.

\end{acknowledgments}

\bibliography{references}

\end{document}

\title{Supplemental Material: Nonuniform Iterative Phasing Framework and Sampling Requirements for 3D Dynamical Inversion from Coherent Surface Scattering Imaging}
\author{Jeffrey J. Donatelli}
\affiliation{Mathematics Department, Lawrence Berkeley National Laboratory, Berkeley, CA USA 94720}
\affiliation{Center for Advanced Mathematics for Energy Research Applications, Lawrence Berkeley National Laboratory, Berkeley, CA USA 94720}
\author{Miaoqi Chu}
\affiliation{X-ray Science Division, Advanced Photon Source, Argonne National Laboratory, Lemont, IL USA 60439}
\author{Zixi Hu}
\affiliation{Mathematics Department, Lawrence Berkeley National Laboratory, Berkeley, CA USA 94720}
\affiliation{Center for Advanced Mathematics for Energy Research Applications, Lawrence Berkeley National Laboratory, Berkeley, CA USA 94720}
\author{Zhang Jiang}
\affiliation{X-ray Science Division, Advanced Photon Source, Argonne National Laboratory, Lemont, IL USA 60439}
\author{Nicholas Schwarz}
\affiliation{X-ray Science Division, Advanced Photon Source, Argonne National Laboratory, Lemont, IL USA 60439}
\author{Jin Wang}
\affiliation{X-ray Science Division, Advanced Photon Source, Argonne National Laboratory, Lemont, IL USA 60439}

\author{James A. Sethian}
\affiliation{Mathematics Department, Lawrence Berkeley National Laboratory, Berkeley, CA USA 94720}
\affiliation{Center for Advanced Mathematics for Energy Research Applications, Lawrence Berkeley National Laboratory, Berkeley, CA USA 94720}
\affiliation{Department of Mathematics, University of California, Berkeley, CA USA 94720}
\maketitle

\section{S1.\hspace{.5em} Derivation of Generalized Multilevel Toeplitz Acceleration for CSSI Data Reduction}
Here we derive the generalized multilevel Toeplitz representation of the normal equations for CSSI linear reduction, introduced in Section~\ref{sec:acceleration} of the main text. This representation allows the corresponding matrix-vector operations to be expressed very efficiently in terms of convolutions, reflections, and multiplication by complex exponentials. 

The forward operator $A$ mapping $\mathbf f = \left[\begin{smallmatrix} f_1 \\ f_2 \end{smallmatrix}\right]$ to the CSSI data $I$ is defined as 
\begin{equation}
\begin{aligned}
(A\mathbf f)[k] &= \fho(\qpk,\qzk) + |\tilde{R}_f(\qzk)|^2\fho(\qpk,-\qzk+\qzi) + \overline{\tilde{R}_f(\qzk)}\fht(\qpk,\qzk-\qzit)\\
&+ \tilde{R}_f(\qzk)\fht(\qpk,-\qzk+\qzit)\\
&= \sum_{\mathbf m} f_1[\mathbf m]\ef{(\qpk,\qzk)\cdot(\mpp,\mz)} + |\tilde{R}_f(\qzk)|^2\sum_{\mathbf m} f_1[\mathbf m]\ef{(\qpk,-\qzk+\qzi)\cdot(\mpp,\mz)}\\
&+\overline{\tilde{R}_f(\qzk)}\sum_{\mathbf m} f_2[\mathbf m]\ef{(\qpk,\qzk-\qzit)\cdot(\mpp,\mz)}\\ &+\tilde{R}_f(\qzk)\sum_{\mathbf m} f_2[\mathbf m]\ef{(\qpk,-\qzk+\qzit)\cdot(\mpp,\mz)}.
\end{aligned}
\end{equation}

Now express the above equations as $A\mathbf f = A_1 f_1 + A_2 f_2$, where
\begin{equation}
\begin{aligned}
(A_1 f_1)[k] &= \sum_{\mathbf m} \left(\ef{(\qpk,\qzk)\cdot(\mpp,\mz)} + |\tilde{R}_f(\qzk)|^2\ef{(\qpk,-\qzk+\qzi)\cdot(\mpp,\mz)}\right)f_1[\mathbf m],\\
(A_2 f_2)[k] &= \sum_{\mathbf m}\left(\overline{\tilde{R}_f(\qzk)} \ef{(\qpk,\qzk-\qzit)\cdot(\mpp,\mz)} +\tilde{R}_f(\qzk)\ef{(\qpk,-\qzk+\qzit)\cdot(\mpp,\mz)}\right)f_2[\mathbf m].
\end{aligned}
\end{equation}
Their corresponding adjoints are given by
\begin{equation}
\begin{aligned}
(A_1^* b)[\npp,\nz] &= \sum_k \left(\eb{(\qpk,\qzk)\cdot(\npp,\nz)} + |\tilde{R}_f(\qzk)|^2\eb{(\qpk,-\qzk+\qzi)\cdot(\npp,\nz)}\right)b[k],\\
(A_2^* b)[\npp,\nz] &= \sum_k\left(\tilde{R}_f(\qzk) \eb{(\qpk,\qzk-\qzit)\cdot(\npp,\nz)} +\overline{\tilde{R}_f(\qzk)}\eb{(\qpk,-\qzk+\qzit)\cdot(\npp,\nz)}\right)b[k].
\end{aligned}
\end{equation}

To solve the weighted symmetrized normal equations $\mathcal SA^*WA\mathcal S\mathbf f = \mathcal A^*I$, we require efficient computation of $A^*WA$, which we express in terms of two components:
\begin{equation}
A^*WA = \begin{bmatrix}A_1^*WA \\ A_2^*WA \end{bmatrix}.
\end{equation}
These components are given by
\begin{equation}
\begin{aligned}
(A_1^*WA\mathbf f) [\npp,\nz] &= (A_1^*WA_1 f_1) [\npp,\nz] + (A_1^*WA_2 f_2) [\npp,\nz]\\
&= \sum_k\sum_{\mathbf m} \Big(\eb{(\qpk,\qzk)\cdot(\npp-\mpp,\nz-\mz)}\\
&+ |\tilde{R}_f(\qzk)|^4 \eb{(\qpk,-\qzk+\qzi)\cdot(\npp-\mpp,\nz-\mz)}\\
&+|\tilde{R}_f(\qzk)|^2\eb{(\qpk,\qzk-\qzi)\cdot(\npp-\mpp,\nz+\mz)}\eb{\qzi\nz}\\
&+|\tilde{R}_f(\qzk)|^2\eb{(\qpk,-\qzk)\cdot(\npp-\mpp,\nz+\mz)}\eb{\qzi\nz}\Big)f_1[\mathbf m]w[k]\\
&+\sum_k\sum_{\mathbf m}\Big(\overline{\tilde{R}_f(\qzk)}\eb{(\qpk,\qzk-\qzit)\cdot(\npp-\mpp,\nz-\mz)}\ebh{\qzi\nz}\\
&+{\tilde{R}_f(\qzk)}\eb{(\qpk,\qzk-\qzit)\cdot(\npp-\mpp,\nz+\mz)}\ebh{\qzi\nz}\\
&+\overline{\tilde{R}_f(\qzk)}|\tilde{R}_f(\qzk)|^2\eb{(\qpk,-\qzk+\qzit)\cdot(\npp-\mpp,\nz+\mz)}\ebh{\qzi\nz}\\
&+{\tilde{R}_f(\qzk)}|\tilde{R}_f(\qzk)|^2\eb{(\qpk,-\qzk+\qzit)\cdot(\npp-\mpp,\nz-\mz)}\ebh{\qzi\nz}\Big)f_2[\mathbf m]w[k],
\end{aligned}
\end{equation}
\begin{equation}
\begin{aligned}
(A_2^*WA\mathbf f) [\npp,\nz] &= (A_2^*WA_1 f_1) [\npp,\nz] + (A_2^*WA_2 f_2) [\npp,\nz]\\
&= \sum_k\sum_{\mathbf m}\Big({\tilde{R}_f(\qzk)}\eb{(\qpk,\qzk)\cdot(\npp-\mpp,\nz-\mz)}\efh{\qzi\nz}\\
&+\overline{\tilde{R}_f(\qzk)} \eb{(\qpk,-\qzk)\cdot(\npp-\mpp,\nz+\mz)}\ebh{\qzi\nz}\\
&+{\tilde{R}_f(\qzk)}|\tilde{R}_f(\qzk)|^2\eb{(\qpk,\qzk-\qzi)\cdot(\npp-\mpp,\nz+\mz)}\ebh{\qzi\nz}\\
&+\overline{\tilde{R}_f(\qzk)}|\tilde{R}_f(\qzk)|^2\eb{(\qpk,-\qzk+\qzi)\cdot(\npp-\mpp,\nz-\mz)}\efh{\qzi\nz}\Big)f_1[\mathbf m]w[k]\\
&+\sum_k\sum_{\mathbf m}\Big(|\tilde{R}_f(\qzk)|^2 \eb{(\qpk,\qzk-\qzit)\cdot(\npp-\mpp,\nz-\mz)}\\
&+ |\tilde{R}_f(\qzk)|^2 \eb{(\qpk,-\qzk+\qzit)\cdot(\npp-\mpp,\nz-\mz)}\\
&+ \tilde{R}_f(\qzk)^2 \eb{(\qpk,\qzk-\qzit)\cdot(\npp-\mpp,\nz+\mz)}\\
&+ \overline{\tilde{R}_f(\qzk)}^2 \eb{(\qpk,-\qzk+\qzit)\cdot(\npp-\mpp,\nz+\mz)}\Big)f_2[\mathbf m]w[k].
\end{aligned}
\end{equation}

We now group terms that have the same $f_1$ or $f_2$ factor and the same $\nz\pm m_z$ term in the exponent. Then, we factor out exponentials from each group to make the $\mathbf q$ arguments of the exponentials within the $k$ summations consistent across all groups:

\begin{align}\label{eq:A1sWfgroup}
(A_1^*WA\mathbf f) [\npp,\nz] = \ebh{\qzi \nz}\!&\sum_{\mathbf m}\!\sum_k\!\Big(\eb{(\qpk,\qzk-\qzit)\cdot(\npp-\mpp,\nz-\mz)}\notag\\
&\hspace*{1.3em}+ |\tilde{R}_f(\qzk)|^4 \eb{(\qpk,-\qzk+\qzit)\cdot(\npp-\mpp,\nz-\mz)}\!\Big)w[k]\efh{\qzi\mz}f_1[\mathbf m]\notag\\
+\,\ebh{\qzi\nz}\!&\sum_{\mathbf m}\! \sum_k\!\Big(|\tilde{R}_f(\qzk)|^2\eb{(\qpk,\qzk-\qzit)\cdot(\npp-\mpp,\nz+\mz)}\notag\\
&\hspace*{1.3em}+ |\tilde{R}_f(\qzk)|^2\eb{(\qpk,-\qzk+\qzit)\cdot(\npp-\mpp,\nz+\mz)}\!\Big)w[k]\efh{\qzi \mz}f_1[\mathbf m]\notag\\
+\,\ebh{\qzi\nz}\!&\sum_{\mathbf m}\! \sum_k\!\Big(\overline{\tilde{R}_f(\qzk)}\eb{(\qpk,\qzk-\qzit)\cdot(\npp-\mpp,\nz-\mz)}\notag\\
&\hspace*{1.3em}+\tilde{R}_f(\qzk)|\tilde{R}_f(\qzk)|^2\eb{(\qpk,-\qzk+\qzit)\cdot(\npp-\mpp,\nz-\mz)}\Big)w[k]f_2[\mathbf m]\notag\\
+\,\ebh{\qzi\nz}\!&\sum_{\mathbf m}\!\sum_k\!\Big({\tilde{R}_f(\qzk)}\eb{(\qpk,\qzk-\qzit)\cdot(\npp-\mpp,\nz+\mz)}\notag\\
&\hspace*{1.3em}+\overline{\tilde{R}_f(\qzk)}|\tilde{R}_f(\qzk)|^2\eb{(\qpk,-\qzk+\qzit)\cdot(\npp-\mpp,\nz+\mz)}\Big)w[k]f_2[\mathbf m],
\end{align}

\begin{align}\label{eq:A2sWfgroup}
(A_2^* WA\mathbf f)[\npp,\nz] = &\sum_{\mathbf m}\!\sum_k\Big(\tilde{R}_f(\qzk)\eb{(\qpk,\qzk-\qzit)\cdot(\npp-\mpp,\nz-\mz)}\notag\\
&\hspace*{1.3em}+\overline{\tilde{R}_f(\qzk)}|\tilde{R}_f(\qzk)|^2\eb{(\qpk,-\qzk+\qzit)\cdot(\npp-\mpp,\nz-\mz)}\!\Big)w[k]\efh{\qzi \mz}f_1[\mathbf m]\notag\\
+&\sum_{\mathbf m}\!\sum_k\Big(\tilde{R}_f(\qzk)|\tilde{R}_f(\qzk)|^2\eb{(\qpk,\qzk-\qzit)\cdot(\npp-\mpp,\nz+\mz)}\notag\\
&\hspace*{1.3em}+\overline{\tilde{R}_f(\qzk)} \eb{(\qpk,-\qzk+\qzit)\cdot(\npp-\mpp,\nz+\mz)}\Big)w[k]\efh{\qzi\mz}f_1[\mathbf m]\notag\\
+&\sum_{\mathbf m}\!\sum_k\Big(|\tilde{R}_f(\qzk)|^2\eb{(\qpk,\qzk-\qzit)\cdot(\npp-\mpp,\nz-\mz)}\notag\\
&\hspace*{3em}+ |\tilde{R}_f(\qzk)|^2\eb{(\qpk,-\qzk+\qzit)\cdot(\npp-\mpp,\nz-\mz)}\Big)w[k]f_2[\mathbf m]\notag\\
+ &\sum_{\mathbf m}\!\sum_k\Big(\tilde{R}_f(\qzk)^2 \eb{(\qpk,\qzk-\qzit)\cdot(\npp-\mpp,\nz+\mz)}\notag\\
&\hspace*{3em} \overline{\tilde{R}_f(\qzk)}^2\eb{(\qpk,-\qzk+\qzit)\cdot(\npp-\mpp,\nz+\mz)}\Big)w[k]f_2[\mathbf m].
\end{align}

We now define the point-spread functions $Q_l$ for $l=1,\hdots,6$ as the 6 distinct summations over $k$ in Eqs.~\ref{eq:A1sWfgroup}-\ref{eq:A2sWfgroup}:
\begin{equation}
\begin{aligned}
    Q_{l}[\mathbf n] = \sum_k\Big(&C_{l,1}[k]\eb{(\qpk,\qzk-\qzi/2)\cdot(\npp,\nz)}\\
    + &C_{l,2}[k]\eb{(\qpk,-\qzk+\qzi/2)\cdot(\npp,\nz)}\Big)w[k],
\end{aligned}
\end{equation}

\begin{equation}\label{eq:RjlS}
C_{l,t}[k] = \tilde{R}_f^{a_{l,t}}(\qzk)\overline{\tilde{R}_f^{b_{l,t}}(\qzk)},
\end{equation}
where the exponents $a_{l,t}$ and $b_{l,t}$ are given in Table~\ref{table:exponentsS}.
\begin{table}[h]
\begin{center}
\begin{tabular}{l c c c c c c} 
\midrule
&\multicolumn{6}{c}{$l$}\\ 
 Exponents & $1$ & $2$ & $3$ & $4$ & $5$ & $6$\\
 \midrule
$(a_{l,1},b_{l,1})$ & $(0,0)$ & $(1,1)$ & $(0,1)$ & $(1,0)$ & $(2,1)$ & $(2,0)$ \\ 
 $(a_{l,2},b_{l,2})$ & $(2,2)$ & $(1,1)$ & $(2,1)$ & $(1,2)$ & $(0,1)$ & $(0,2)$\\
 \midrule
\end{tabular}
\end{center}
\vspace*{-1em}
\caption{Reflection coefficient exponents used in Eq.~\ref{eq:RjlS}.}\label{table:exponentsS}
\end{table}

Replacing the summations over $k$ in Eqs.~\ref{eq:A1sWfgroup}-\ref{eq:A2sWfgroup} with these point-spread functions gives the following:

\begin{equation}\label{eq:A1sWAq}
\begin{aligned}
(A_1^*WA\mathbf f) [\npp,\nz] = \ebh{\qzi\nz}\Big(&\sum_{\mathbf m} Q_{1}[\npp-\mpp,\nz-\mz]\efh{\qzi \mz}f_1[\mathbf m]\\
+&\sum_{\mathbf m} Q_{2}[\npp-\mpp,\nz+\mz]\efh{\qzi \mz}f_1[\mathbf m]\\
+&\sum_{\mathbf m} Q_{3}[\npp-\mpp,\nz-\mz]f_2[\mathbf m] \\
+&\sum_{\mathbf m} Q_{4}[\npp-\mpp,\nz+\mz]f_2[\mathbf m]\Big),
\end{aligned}
\end{equation}

\begin{equation}\label{eq:A2sWAq}
\begin{aligned}
(A_2^*WA\mathbf f) [\npp,\nz] = \Big(&\sum_{\mathbf m} Q_{4}[\npp-\mpp,\nz-\mz]\efh{\qzi \mz}f_1[\mathbf m]\\
+&\sum_{\mathbf m} Q_{5}[\npp-\mpp,\nz+\mz]\efh{\qzi \mz}f_1[\mathbf m]\\
+&\sum_{\mathbf m} Q_{2}[\npp-\mpp,\nz-\mz]f_2[\mathbf m] \\
+&\sum_{\mathbf m} Q_{6}[\npp-\mpp,\nz+\mz]f_2[\mathbf m]\Big).
\end{aligned}
\end{equation}

To transform the second and fourth summations in Eqs.~\ref{eq:A1sWAq}-\ref{eq:A2sWAq} into convolutions, we first apply the identity $\sum_{\mathbf m} h[\mpp,\mz] = \sum_{\mathbf m} h[\mpp,-\mz]$, which is just an index reordering:

\begin{equation}\label{eq:A1sWAq2}
\begin{aligned}
(A_1^*WA\mathbf f) [\npp,\nz] = \ebh{\qzi\nz}\Big(&\sum_{\mathbf m} Q_{1}[\npp-\mpp,\nz-\mz]\efh{\qzi \mz}f_1[\mpp,\mz]\\
+&\sum_{\mathbf m} Q_{2}[\npp-\mpp,\nz-\mz]\ebh{\qzi \mz}f_1[\mpp,-\mz]\\
+&\sum_{\mathbf m} Q_{3}[\npp-\mpp,\nz-\mz]f_2[\mpp,\mz] \\
+&\sum_{\mathbf m} Q_{4}[\npp-\mpp,\nz-\mz]f_2[\mpp,-\mz]\Big),
\end{aligned}
\end{equation}

\begin{equation}\label{eq:A2sWAq2}
\begin{aligned}
(A_2^*WA\mathbf f) [\npp,\nz] = \Big(&\sum_{\mathbf m} Q_{4}[\npp-\mpp,\nz-\mz]\efh{\qzi \mz}f_1[\mpp,\mz]\\
+&\sum_{\mathbf m} Q_{5}[\npp-\mpp,\nz-\mz]\ebh{\qzi \mz}f_1[\mpp,-\mz]\\
+&\sum_{\mathbf m} Q_{2}[\npp-\mpp,\nz-\mz]f_2[\mpp,\mz] \\
+&\sum_{\mathbf m} Q_{6}[\npp-\mpp,\nz-\mz]f_2[\mpp,-\mz]\Big).
\end{aligned}
\end{equation}
By defining $\mathcal R_z f_t[\mpp,\mz] = f_t[\mpp,-\mz]$ to be reflection though the $xy$ plane, and defining $Ef_t[\mpp, \mz] = f_t[\mpp,\mz]\efh{\qzi\mz}$ and $\overline{E}f_t[\mpp,\mz] = f_t[\mpp,\mz]\ebh{\qzi\mz}$ to be multiplication by the complex exponentials, we can express Eqs.~\ref{eq:A1sWAq2}-\ref{eq:A2sWAq2} as 
\begin{equation}
\begin{aligned}
(A_1^*WA\mathbf f) [\npp,\nz] = \overline{E}\Big(&\sum_{\mathbf m} Q_{1}[\npp-\mpp,\nz-\mz]Ef_1[\mathbf m]\\
+&\sum_{\mathbf m} Q_{2}[\npp-\mpp,\nz-\mz]\mathcal R_zEf_1[\mathbf m]\\
+&\sum_{\mathbf m} Q_{3}[\npp-\mpp,\nz-\mz]f_2[\mathbf m] \\
+&\sum_{\mathbf m} Q_{4}[\npp-\mpp,\nz-\mz]\mathcal R_zf_2[\mathbf m]\Big),
\end{aligned}
\end{equation}

\begin{equation}
\begin{aligned}
(A_2^*WA\mathbf f) [\npp,\nz] = &\sum_{\mathbf m} Q_{4}[\npp-\mpp,\nz-\mz]Ef_1[\mathbf m]\\
+&\sum_{\mathbf m} Q_{5}[\npp-\mpp,\nz-\mz]\mathcal R_zEf_1[\mathbf m]\\
+&\sum_{\mathbf m} Q_{2}[\npp-\mpp,\nz-\mz]f_2[\mathbf m] \\
+&\sum_{\mathbf m} Q_{6}[\npp-\mpp,\nz-\mz]\mathcal R_zf_2[\mathbf m].
\end{aligned}
\end{equation}

The equations can be simplified by expressing the summations as convolutions:
\begin{equation}
\begin{aligned}
(A_1^*WA\mathbf f) &= \overline{E}\big((Ef_1) \ast Q_{1} + (\mathcal R_z E f_1) \ast Q_{2} + f_2 \ast Q_{3}+(\mathcal R_z f_2) \ast Q_{4}\big),\\
(A_2^*WA\mathbf f) &=   (Ef_1) \ast Q_{4} + (\mathcal R_z Ef_1) \ast Q_{5} + f_2 \ast Q_{2} + (\mathcal R_z f_2) \ast Q_{6},
\end{aligned}
\end{equation}
which gives Eqs.~\ref{eq:Qij}-\ref{eq:GMLToeplitz} in the main text.

\section{S2.\hspace{.5em} Derivation of the DWBA Magnitude Projection }
Here we derive the analytic expression for the DWBA magnitude projector $P_{D}$ in Eq.~\ref{eq:PD} in the main text. Recall that this projection is defined for the set of sampled $\mathbf q$ as
\begin{equation}\label{eq:Sprojmin}
P_{D}\hat{\rho} = \argmin_h ||h-\hat{\rho}||_w \text{ subject to  } \left|\sum_{s=1}^4 R_{j,s}(\qz) h_{j,s}(\mathbf q) \right| = \sqrt{I_{j}(\mathbf q)}, \text{ for all $j$ and $\mathbf q$}.
\end{equation}
Since the constrained minimization in Eq.~\ref{eq:Sprojmin} is decoupled across $\mathbf q$ and $j$, we can solve for each component separately. To simply notation, here we fix a $\mathbf q$ and $j$ and drop dependencies on those variables (e.g., $h_s$ refers to $h_{j,s}(\mathbf q)$). For the moment, assume that $\sum_{s} R_{j,s}(\qz)\hat{\rho}_{j,s}(\mathbf q) \neq 0$. 

Now define the following matrix and vectors:
\begin{equation}
W = \begin{bmatrix} w_1 & 0 & 0 & 0 \\ 0 & w_2 & 0 & 0 \\ 0 & 0 & w_3 & 0 \\ 0 & 0 & 0 & w_4
\end{bmatrix},\ \ R = \begin{bmatrix} R_{1} \\ R_{2} \\ R_{3} \\ R_{4} \end{bmatrix},\ \ h = \begin{bmatrix} h_{1} \\ h_{2} \\ h_{3} \\ h_{4} \end{bmatrix},\ \ \hat{\rho} = \begin{bmatrix} \hat{\rho}_{1} \\ \hat{\rho}_{2} \\ \hat{\rho}_{3} \\ \hat{\rho}_{4} \end{bmatrix}.
\end{equation}

The above notation allows us to express the constraint on $h$ in Eq.~\ref{eq:Sprojmin} as
\begin{equation}\label{eq:constraint}
R^Th = \sqrt{I}e^{i\phi}, \text{ for some } \phi \in \mathbb R.
\end{equation}
Therefore, we can express the constrained minimization in Eq.~\ref{eq:Sprojmin} as 
\begin{equation}\label{eq:constrained_min}
\argmin_h ||h-\hat{\rho}||_w \text{ subject to } R^Th = \sqrt{I}e^{i\phi}\text{ for some } \phi \in \mathbb R.
\end{equation}

We first fix $\phi$ and seek the corresponding minimizer for $h$. To find the minimum $w$-weighted-norm solution, we first rewrite the constraint in Eq.~\ref{eq:constraint} as 
\begin{equation}\label{eq:weightedeq}
R^TW^{-\frac{1}{2}}W^\frac{1}{2}(h-\hat{\rho}) = \sqrt{I}e^{i\phi} - R^T\hat{\rho}.
\end{equation}

Now define $u = W^\frac{1}{2}(h-\hat{\rho})$ and substitute this into Eq.~\ref{eq:weightedeq}, which yields
\begin{equation}\label{eq:u}
R^TW^{-\frac{1}{2}}u = \sqrt{I}e^{i\phi} - R^T\hat{\rho}.
\end{equation}
The minimum-$\ell^2$-norm solution for $u$ in Eq.~\ref{eq:u} can be expressed in terms of the pseudoinverse $(R^TW^{-\frac{1}{2}})^{\dagger}$ as
\begin{equation}\label{eq:usolution}
u = (R^TW^{-\frac{1}{2}})^{\dagger}(\sqrt{I}e^{i\phi} - R^T\hat{\rho}).
\end{equation}

Since $||u ||_2 = ||W^\frac{1}{2}(h-\hat{\rho}))||_2 = ||h - \hat{\rho}||_w$, the minimum $w$-weighted-norm solution of Eq.~\ref{eq:weightedeq} for $h$ is then given by $h = \hat{\rho} + W^{-\frac{1}{2}}u$, which can be combined with Eq.~\ref{eq:usolution} to yield
\begin{equation}\label{eq:hsolution}
h = \hat{\rho} + W^{-\frac{1}{2}}(R^TW^{-\frac{1}{2}})^{\dagger}(\sqrt{I}e^{i\phi} - R^T\hat{\rho}).
\end{equation}

Now that we have expressed the optimal $h$ in terms of the variable $\phi$ in Eq.~\ref{eq:hsolution}, we now seek the value of $\phi$ that minimizes $||h - \hat{\rho}||_w$. This can be solved by noting that
\begin{equation}\label{eq:hsolutionnorm}
||h - \hat{\rho}||_w = ||W^\frac{1}{2}(h-\hat{\rho}))||_2 = ||(R^TW^{-\frac{1}{2}})^{\dagger} ||_2\, |\sqrt{I}e^{i\phi} - R^T\hat{\rho}|,
\end{equation}
which is minimized over $\phi$ when the two terms in the absolute value have the same phase, i.e.,  $e^{i\phi} = R^T\hat{\rho}/|R^T\hat{\rho}|$. Substituting this solution into Eq.~\ref{eq:hsolution} and simplifying yields
\begin{equation}\label{eq:hphisolution}
h = \hat{\rho} + W^{-\frac{1}{2}}(R^TW^{-\frac{1}{2}})^{\dagger}R^T\hat{\rho}\left(\frac{\sqrt{I}}{|R^T\hat{\rho}|} - 1\right).
\end{equation}
Since the pseudoinverse of a vector $a$ is $a^{\dagger} = a^*/\|a\|^2$, Eq.~\ref{eq:hphisolution} can be further simplified to
\begin{equation}\label{eq:hphisolution2}
h = \hat{\rho} + W^{-\frac{1}{2}}\frac{W^{-\frac{1}{2}}\overline{R}R^T\hat{\rho}}{\|R^TW^{-\frac{1}{2}}\|^2}\left(\frac{\sqrt{I}}{|R^T\hat{\rho}|} - 1\right).
\end{equation}

By expressing the inner products in Eq.~\ref{eq:hphisolution2} as summations and adding back the $j$ and $s$ dependencies, we arrive at
\begin{equation}
P_{\rm D}\hat{\rho}_{j,s} = h_{j,s} = \hat{\rho}_{j,s} + \frac{\overline{R_{j,s}}\sum_{s'} R_{j,s'}\hat{\rho}_{j,s'}}{w_{j,s}\sum_{s'} w_{j,s'}^{-1}|R_{j,s'}|^2}\left(\frac{\sqrt{I_j}}{|\sum_{s'} R_{j,s'}{\hat\rho_{j,s'}}|}-1\right),
\end{equation}
which is the expression given in Eq.~\ref{eq:PD} in the main text.

Note that if $\sum_{s} R_{j,s}(\qz)\hat{\rho}_{j,s}(\mathbf q) = 0$, then Eq.~\ref{eq:hsolutionnorm} becomes $||h - \hat{\rho}||_w  = ||(R^TW^{-\frac{1}{2}})^{\dagger} ||_2\, \sqrt{I}$, which is independent of $\phi$. Therefore, under this condition, $h = \hat{\rho} + W^{-\frac{1}{2}}(R^TW^{-\frac{1}{2}})^{\dagger}\sqrt{I}e^{i\phi}$ is a solution to Eq.~\ref{eq:constrained_min} for any choice of $\phi$. In this case, we choose the convention $\phi = 0$, which yields $P_{\rm D}\hat{\rho}_{j,s} = \hat{\rho}_{j,s}+\overline{R_{j,s}}\sqrt{I_j}/(w_{j,s}\sum_{s'} w_{j,s'}^{-1}|R_{j,s'}|^2)$, as given in the main text.

\section{S3.\hspace{.5em} Proof of the Maximum Specimen Rotation Angle Requirements}
Here we prove that, if Eq.~\ref{eq:phimax} in the main text holds, then Eq.~\ref{eq:phimaxorig} holds for all $(q_x,q_y)$ sampled by the diffraction pattern at the reference orientation. First, we show that Eq.~\ref{eq:phimaxorig} is maximized when $\theta_f = 0$ and either $\alpha_f = \alpha_f^{\min}$ or $\alpha_f = \alpha_f^{\max}$, or when $\theta_f = \theta_f^{\max}$ and $\alpha_f = \alpha_f^{\max}$. To simplify notation, we denote
\begin{align}
\tilde{D}      &= \frac{4 D_{||}}{\lambda},\\
Q_p^2(\tf,\af) &= \lambda^2|\qp(\tf,\af)|^2 \\ \nonumber
               &= \cos^2(\af)+\cos^2(\ai) - 2\cos(\tf)\cos(\af)\cos(\ai),\\
h(\tf,\af)     &= \pi+2\arctan\left(\left|\frac{q_y}{q_x}\right|\right) - 2\arcsin\left(\frac{1}{4D_{||}|\qp|}\right)\\ \nonumber
               &= \pi +2\arctan\left(\left|\frac{\cos(\alpha_i)-\cos(\theta_f)\cos(\alpha_f)}{\sin(\theta_f)\cos(\alpha_f)}\right|\right)-2\arcsin\left(\frac{1}{\tilde{D}Q_p(\tf,\af)}\right)\\ \nonumber
               &= \pi + 2h_1(\tf,\af) + 2h_2(\tf,\af),\\
h_1(\tf,\af)   &= \arctan\left(\left|\frac{\cos(\alpha_i)-\cos(\theta_f)\cos(\alpha_f)}{\sin(\theta_f)\cos(\alpha_f)}\right|\right)\\
h_2(\tf,\af)   &= -\arcsin\left(\frac{1}{\tilde{D}Q_p(\tf,\af)}\right).
\end{align}

Using the above notation, and noting that $h(-\tf,\af) = h(\tf,\af)$, Eq.~\ref{eq:phimaxorig} holding for each sampled $(q_x,q_y)$ satisfying $|q_y| > 1/(4D_{||})$ is equivalent to
\begin{equation}
    \phi^{\max} \geq \max_{\substack{|\cos(\alpha_i)-\cos(\tf)\cos(\af)| > 1/\tilde{D}\\ \af^{\min}\leq \af \leq \af^{\max}\\ 0\leq\tf\leq\tf^{\max}}} h(\tf,\af).
\end{equation}

Note that $h(\tf,\af)\geq \pi$ for all points in the admissible region, with equality $h(\tf,\af) = \pi$ for boundary points that
satisfy $|\cos(\alpha_i)-\cos(\tf)\cos(\af)| = 1/\tilde{D}$. Therefore, these boundary points are minima of $h$. The boundary result follows from applying the identity $\arctan(x) = \arcsin(x/\sqrt{1+x^2})$ to the definition of $h$.

Within the admissible region, we have the following partial derivatives:
\begin{align}\label{eq:h1tf}
\frac{\partial h_1}{\partial \theta_f} &= \frac{\cos(\alpha_f)(\cos(\alpha_f)-\cos(\theta_f)\cos(\alpha_i))}{Q_p^2(\tf,\af)} {\rm sgn}\left(\frac{\cos(\alpha_i)-\cos(\theta_f)\cos(\alpha_f)}{\sin(\theta_f)\cos(\alpha_f)}\right),\\[1em]
\label{eq:h2tf}
    \frac{\partial h_2}{\partial \theta_f} &= \frac{\sin(\theta_f)\cos(\alpha_f)\cos(\alpha_i)}{Q_p^2(\theta_f,\alpha_f)\sqrt{\tilde{D}^2Q_p^2(\theta_f,\alpha_f)-1}},\\[1em]
\label{eq:h1af}
\frac{\partial h_1}{\partial \alpha_f} &= \frac{\sin(\theta_f)\sin(\alpha_f)\cos(\alpha_i)}{Q_p^2(\tf,\af)}{\rm sgn}\left(\frac{\cos(\alpha_i)-\cos(\theta_f)\cos(\alpha_f)}{\sin(\theta_f)\cos(\alpha_f)}\right),\\[1em]
\label{eq:h2af}
\frac{\partial h_2}{\partial \alpha_f} &= \frac{\sin(\alpha_f)(\cos(\theta_f)\cos(\alpha_i)-\cos(\alpha_f))}{Q_p^2(\theta_f,\alpha_f)\sqrt{\tilde{D}^2Q_p^2(\theta_f,\alpha_f)-1}},
\end{align}
where ${\rm sgn}(x)$ is the sign function.

In the following calculations, we will often drop the $(\tf,\af)$ dependency for $Q_p$ to simplify notation. 

Consider the following four cases:

\vspace{1em}

\noindent\textbf{Case 1:} $\af \geq \ai$ and $\caf - \ctf\cai \geq 0$.

$\af \geq \ai$ implies that $\cai - \ctf\caf \geq \cai - \caf \geq 0$. Therefore, Eqs.~\ref{eq:h1tf}-\ref{eq:h2tf} are nonnegative, and so $\partial h/\partial \tf \geq 0$.

The conditions of this case also imply that Eq.~\ref{eq:h1af} is nonnegative and Eq.~\ref{eq:h2af} is nonpositive. Therefore, $\partial h/\partial \af \geq 0$ if and only if $(\partial h_1/\partial \af)^2 - (\partial h_2/\partial \af)^2 \geq 0$. We have that 
\begin{align}\label{eq:case1ineq}
    \left(\frac{\partial h_1}{\partial \af}\right)^2 - \left(\frac{\partial h_2}{\partial \af}\right)^2 &= \frac{\ssaf}{Q_p^4(\tilde{D}^2Q_p^2-1)}\Big(\sstf\csai(\tilde{D}^2Q_p^2 -1)\notag\\
    &\hspace{1em}-\left(\csaf + \cstf\csai -2\ctf\caf\cai\right)\Big)\notag\\
    &= \frac{\ssaf}{Q_p^4(\tilde{D}^2Q_p^2-1)}\Big(\sstf\csai(\tilde{D}^2Q_p^2 -1)\notag\\
    &\hspace{1em}-\left(\csaf + (1-\sstf)\csai -2\ctf\caf\cai\right)\Big) \notag\\
    &= \frac{\ssaf}{Q_p^4(\tilde{D}^2Q_p^2-1)}\Big(\sstf\csai(\tilde{D}^2Q_p^2 -1) \notag\\
    &\hspace{1em}- \left(Q_p^2 - \sstf\csai\right)\Big)\notag\\
    &= \frac{\ssaf}{Q_p^4(\tilde{D}^2Q_p^2-1)}\left(\tilde{D}^2Q_p^2\sstf\csai - Q_p^2 \right)\notag\\
    &= \frac{\tilde{D}^2\ssaf}{Q_p^2(\tilde{D}^2Q_p^2-1)}\left(\sstf\csai - \frac{1}{\tilde{D}^2} \right).
\end{align}

If $\stf\cai \geq 1/\tilde{D}$ then Eq.~\ref{eq:case1ineq} is nonnegative, and thus $\partial h /\partial \af \geq 0$. Otherwise, if $\stf\cai < 1/\tilde{D}$, we have that
\begin{align}
    \frac{1}{\tilde{D}^2} &> \sstf\csai\notag\\
                          &= \csai - \cstf\csai\notag\\
                          &= (\cai-\ctf\caf)^2 - \cstf\csai - \cstf\csaf\notag\\
                          &\hspace{1em}+2\ctf\caf\cai\notag\\
                          &\geq (\cai-\ctf\caf)^2 - \cstf\csai - \cstf\csaf\notag\\
                          &\hspace{1em}+2\cstf\csai\notag\\
                          &= (\cai-\ctf\caf)^2 + \cstf(\csai-\csaf)\notag\\
                          &\geq (\cai-\ctf\caf)^2,
\end{align}
where the second inequality follows from the condition $\caf -\ctf\cai \geq 0$, and the third inequality follows from $\af\geq \ai$. Therefore, if $\stf\cai < 1/\tilde{D}$, then $(\tf,\af)$ is not in the admissible region. Conversely, if $(\tf,\af)$ is in the admissible region, then $\stf\cai \geq 1/\tilde{D}$, and thus $(\partial h/\partial \af)(\tf,\af) \geq 0$. 

Since $\partial h/\partial \tf \geq 0$ and $\partial h/\partial \af \geq 0$ in the admissible region for this case, and since boundary points satisfying $|\cos(\alpha_i)-\cos(\tf)\cos(\af)| = 1/\tilde{D}$ are minimizers, the maximum of $h$ for this case must occur at $(\tf^{\max},\af^{\max})$, assuming the admissible region in this case is nonempty.

\vspace{1em}

\noindent\textbf{Case 2:} $\af \geq \ai$ and $\caf - \ctf\cai \leq 0$.

These conditions imply that Eq.~\ref{eq:h1tf} is nonpositive and Eq.~\ref{eq:h2tf} is nonnegative. In this case, $\partial h/\partial \tf \geq 0$ if and only if $(\partial h_1/\partial \tf)^2 - (\partial h_2/\partial \tf)^2 \leq 0$. We have that 
\begin{align}
    \left(\frac{\partial h_1}{\partial \tf}\right)^2 - \left(\frac{\partial h_2}{\partial \tf}\right)^2 &= \frac{\csaf}{Q_p^4(\tilde{D}^2Q_p^2-1)}\Big((\caf-\ctf\cai)^2(\tilde{D}^2Q_p^2 -1)\notag\\
    &\hspace{1em}-\sstf\csai\Big)\notag\\
    &= \frac{\csaf}{Q_p^4(\tilde{D}^2Q_p^2-1)}\Big((\caf-\ctf\cai)^2\tilde{D}^2Q_p^2 -\csaf\notag\\
    &\hspace{1em}-\cstf\csai +2\ctf\caf\cai-\sstf\csai\Big)\notag\\
    &= \frac{\csaf}{Q_p^4(\tilde{D}^2Q_p^2-1)}\Big((\caf-\ctf\cai)^2\tilde{D}^2Q_p^2 -\csaf\notag\\
    &\hspace{1em}-\csai +2\ctf\caf\cai\Big)\notag\\
    &= \frac{\csaf}{Q_p^4(\tilde{D}^2Q_p^2-1)}\Big((\caf-\ctf\cai)^2\tilde{D}^2Q_p^2 -Q_p^2\Big)\notag\\
    &= \frac{\tilde{D}^2\csaf}{Q_p^2(\tilde{D}^2Q_p^2-1)}\left((\caf-\ctf\cai)^2 - 1/\tilde{D}^2\right).\label{eq:case2ineq}
\end{align}
Therefore, $(\partial h/\partial \tf)(\tf,\af) \geq 0$ if $\ctf\cai-\caf \leq 1/\tilde{D}$, and $(\partial h/\partial \tf)(\tf,\af) \leq 0$ if $\ctf\cai-\caf \geq 1/\tilde{D}$.

Note that the latter inequality above implies that 
\begin{equation}
    \frac{1}{\tilde{D}} \leq \ctf\cai-\caf \leq \cai-\caf = |\cai-\cos(0)\caf|,
\end{equation}
meaning that $(0,\af)$ is in the admissible region. 

The above arguments imply that, for a fixed  $\af$, $h(\tf,\af)$ is monotonically decreasing for $0\leq \tf\leq \arccos((\caf+1/\tilde{D})/\cai)$ and monotonically increasing for $\tf \geq \arccos((\caf+1/\tilde{D})/\cai)$. Therefore, the maximum of $h(\tf,\af)$ within the admissible region (assuming it is nonempty) for this case occurs at either $\tf = 0$ or $\tf = \min(\arccos(\caf/\cai),\tf^{\max})$.

The conditions of this case also imply that Eqs.~\ref{eq:h1af}-\ref{eq:h2af} are nonnegative, and hence $\partial h/\partial \af \geq 0$.

Therefore, the maximum in the admissible region for this case occurs at $\af = \af^{\max}$ and either $\tf = 0$ or $\tf = \min(\arccos(\cos(\af^{\max})/\cai),\tf^{\max})$, assuming the admissible region in this case is nonempty.

\vspace{1em}

\noindent\textbf{Case 3:} $\af \leq \ai$ and $\cai-\ctf\caf \geq 0$.

The first condition implies that $\caf - \ctf\cai \geq \caf - \cai \geq 0$. Therefore, in this case, we have that Eqs.~\ref{eq:h1tf}-\ref{eq:h2tf} are nonnegative and hence $\partial h/\partial \tf \geq 0$.

These conditions also imply that Eq.~\ref{eq:h1af} is nonnegative and Eq.~\ref{eq:h2af} is nonpositive. Therefore, as in Case 1, $\partial h/\partial \af \geq 0$ if and only if $(\partial h_1/\partial \af)^2 - (\partial h_2/\partial \af)^2 \geq 0$. As shown in Eq.~\ref{eq:case1ineq}, if $\stf\cai \geq 1/\tilde{D}$ then $(\partial h/\partial \af)(\tf,\af) \geq 0$ . If $\stf\cai < 1/\tilde{D}$ then
\begin{align}
    \frac{1}{\tilde{D}^2} &> \sstf\csai\notag\\
                          &= \csai - \cstf\csai\notag\\
                          &= (\cai-\ctf\caf)^2 - \cstf\csai - \cstf\csaf\notag\\
                          &\hspace{1em}+2\ctf\caf\cai\notag\\
                          &\geq (\cai-\ctf\caf)^2 - \cstf\csai - \cstf\csaf\notag\\
                          &\hspace{1em}+2\cstf\csaf\notag\\
                          &= (\cai-\ctf\caf)^2 + \cstf(\csaf-\csai)\notag\\
                          &\geq (\cai-\ctf\caf)^2,
\end{align}
where the second inequality follows from the condition $\cai -\ctf\caf \geq 0$, and the third inequality follows from $\af\leq \ai$. Therefore, if $\stf\cai < 1/\tilde{D}$, then $(\tf,\af)$ is not in the admissible region.

Therefore, the maximum of $h$ within the admissible region for this case occurs at $\tf^{\max}$ and $\min(\af^{\max},\ai)$, assuming the admissible region in this case is nonempty.

\vspace{1em}

\noindent\textbf{Case 4:} $\af \leq \ai$ and $\cai - \ctf\caf \leq 0$.

These conditions imply that Eq.~\ref{eq:h1tf} is nonpositive and Eq.~\ref{eq:h2tf} is nonnegative. As in Case 2, $\partial h/\partial \tf \geq 0$ if and only if $(\partial h_1/\partial \tf)^2 - (\partial h_2/\partial \tf)^2 \leq 0$. Similarly, via Eq.~\ref{eq:case2ineq}, if $\caf-\ctf\cai \leq 1/\tilde{D}$ then $(\partial h/\partial \tf)(\tf,\af) \geq 0$ , and if $\caf-\ctf\cai \geq 1/\tilde{D}$ then $(\partial h/\partial \tf)(\tf,\af) \leq 0$. 

If $(\tf,\af)$ is in the admissible region, then 
\begin{equation}
\begin{aligned}
    \caf-\ctf\cai &\geq \ctf\caf-\cai\\
    &= |\cai-\ctf\caf|\\
    &\geq \frac{1}{\tilde{D}},
\end{aligned}
\end{equation}
and therefore $(\partial h/\partial \tf)(\tf,\af) \leq 0$.

The conditions of this case also imply that Eqs.~\ref{eq:h1af}-\ref{eq:h2af} are nonpositive, and thus $\partial h/\partial \af \leq 0$.

Therefore, the maximum in the admissible region for this case occurs at $\theta_f = 0$ and $\af = \af^{\min}$, assuming the admissible region in this case is nonempty.

\vspace{1em}

\noindent\textbf{Case Summary:} The maximum of $h$ in the admissible regions for the above cases occur at 
\begin{itemize}
    \item Case 1: $(\tf^{\max},\af^{\max})$,
    \item Case 2: $(0,\af^{\max})$ or $(\arccos(\cos(\af^{\max})/\cai),\af^{\max})$ (or $(\tf^{\max},\af^{\max})$ if it satisfies the conditions of Case 2),
    \item Case 3: $(\tf^{\max}, \min(\af^{\max}, \ai))$,
    \item Case 4: $(0, \af^{\min})$.
\end{itemize}

Since $(\arccos(\cos(\af^{\max})/\cai),\af^{\max})$ is on the boundary between Case 1 and Case 2, and since $(\tf^{\max},\af^{\max})$ is a maximizer of $h$ for Case 1, we have that if $\af^{\max} \geq \ai$ then $h(\tf^{\max},\af^{\max}) \geq h(\arccos(\cos(\af^{\max})/\cai),\af^{\max})$. Similarly, since $(\tf^{\max}, \ai)$ occurs at the boundary between Case 1 and Case 3, we have that if $\af^{\max} \geq \ai$ then $h(\tf^{\max},\af^{\max}) \geq h(\tf^{\max}, \ai)$. Therefore, the possible maxima in the admissible region are located at $(\tf^{\max},\af^{\max})$, $(0,\af^{\max})$, or $(0, \af^{\min})$, assuming the admissible region is nonempty.

For the potential maximum at $(\tf^{\max},\af^{\max})$, since $h_2(\tf^{\max},\af^{\max})$ is negative and typically very small when $\tf^{\max}$ is large enough that the diffraction patterns span several Shannon pixels in the $\qp$ dimensions, we can drop it to yield
\begin{equation}\label{eq:phimaxS1}
\begin{aligned}
\phi^{\max} &\geq \pi + 2h_1(\tf^{\max},\af^{\max})\\
&=\pi + 2\arctan\left(\frac{|\cos(\alpha_i)-\cos(\theta_f^{\max})\cos(\alpha_f^{\max})|}{\sin(\theta_f^{\max})\cos(\alpha_f^{\max})}\right).
\end{aligned}
\end{equation}

For the potential maxima at $(0, \af^{\min})$ and $(0,\af^{\max})$ (assuming they are in the admissible region), since $h_1(0,\alpha_f) = \pi$, we can write 
\begin{equation}\label{eq:phimaxS2}
\begin{aligned}
\phi^{\max} &\geq 2\pi+2\max\left(h_2(0,\af^{\min}),h_2(0,\af^{\max})\right)\\
&= 2\pi - 2\arcsin\left(\frac{1}{4D_{||}q_{||,0}^{\max}}\right), \text{ where}\\
     & q_{||,0}^{\max}\hspace*{-.095em}\!=\!\frac{\!\max\!\left(\cos(\alpha_f^{\min})-\cos(\alpha_i), \cos(\alpha_i)-\cos(\alpha_f^{\max})\right)}{\lambda}.
\end{aligned}
\end{equation}
To address the case when these points are not in the admissible region, we can instead express Eq.~\ref{eq:phimaxS2} as
\begin{equation}\label{eq:phimaxS3}
\phi^{\max} \geq 2\pi - 2\arcsin\left(\min\left(\frac{1}{4D_{||}q_{||,0}^{\max}},1\right)\right).
\end{equation}

Combining Eqs.~\ref{eq:phimaxS1} and \ref{eq:phimaxS3} yields Eq.~\ref{eq:phimax} in the main text.

\section{S4.\hspace{.5em} Proof of the Missing Candlestick Region}
Here we show that none of the Fourier coordinates sampled by the DWBA in a CSSI experiment lie within the missing candlestick region defined in Eq.~\ref{eq:dhg} of the main text. To simplify the calculations, we assume that no backscattering is measured, i.e., that all angles are less than $\pi/2$, and we multiply the inequalities in Eq.~\ref{eq:dhg} by $\lambda^2$. 

Recall that at the reference orientation $\phi = 0$, the $s$-th term of the DWBA for an intensity measured at a pixel with incident angle $\alpha_i$, exit angle $\alpha_f$, and signed horizontal scattering angle $\theta_f$, samples the Fourier transform at $\mathbf q = (\qp, q_z^s)$, where
\begin{equation}\label{eq:SI_qp_qzs}
\begin{aligned}
\qp &= \frac{1}{\lambda}\big(\sin(\tf)\cos(\af), \cos(\tf)\cos(\af)-\cos(\ai)\big),\\
q_z^s &= \begin{cases}\frac{1}{\lambda}\big(\saf+\sai\big), &\text{if } s=1,\\
                    \frac{1}{\lambda}\big(\saf-\sai\big), &\text{if } s =2,\\
                    \frac{1}{\lambda}\big(-\saf+\sai\big), &\text{if } s =3,\\
                    \frac{1}{\lambda}\big(-\saf-\sai\big), &\text{if } s =4.
\end{cases}
\end{aligned}
\end{equation}

Since Eq.~\ref{eq:dhg} depends only on the magnitude $|\qp|$ of the horizontal component and $|R_{\phi}\qp| = |\qp|$, we can assume, without loss of generality, that the in-plane rotation angle is $\phi = 0$. 

We have the following lower bound on the magnitude of the horizontal component of the Fourier coordinate:
\begin{equation}\label{eq:lambdaqp}
\begin{aligned}
\lambda^2|\qp|^2 &= \sin^2(\theta_f)\cos^2(\alpha_f) + \cos^2(\theta_f)\cos^2(\alpha_f) - 2\cos(\theta_f)\cos(\alpha_f)\cos(\alpha_i)+\cos^2(\alpha_i)\\
&= \cos^2(\alpha_f) + \cos^2(\alpha_i) - 2\cos(\theta_f)\cos(\alpha_f)\cos(\alpha_i)\\
&\geq \cos^2(\alpha_f) + \cos^2(\alpha_i) - 2\cos(\alpha_f)\cos(\alpha_i)\\
&= |\cos(\alpha_f)-\cos(\alpha_i)|^2.
\end{aligned}
\end{equation}

Alternatively, we can form the following lower bound:
\begin{equation} \label{eq:case1theta}
\begin{aligned}
\lambda^2|\qp|^2 &= \cos^2(\alpha_f) + \cos^2(\alpha_i) - 2\cos(\theta_f)\cos(\alpha_f)\cos(\alpha_i)\\
&\geq \cos^2(\theta_f)\cos^2(\alpha_f) + \cos^2(\alpha_i) - 2\cos(\theta_f)\cos(\alpha_f)\cos(\alpha_i)\\
&= |\cos(\theta_f)\cos(\alpha_f) - \cos(\alpha_i)|^2.\\
\end{aligned}
\end{equation}

The magnitudes of the vertical components of the Fourier coordinates in the DWBA are 
\begin{equation}
\begin{aligned}
|q_z^1| &= |q_z^4| &=\ &\frac{1}{\lambda}\big(\sin(\alpha_f)+\sin(\alpha_i)\big),\\
|q_z^2| &= |q_z^3| &=\ &\frac{1}{\lambda}\left|\sin(\alpha_f)-\sin(\alpha_i)\right|.
\end{aligned}
\end{equation}

We now show that for all $\alpha_i,\alpha_f \in [0,\pi/2)$ and $\tf \in(-\pi/2,\pi/2)$ the corresponding $\mathbf q$ coordinates given by the DWBA in Eq.~\ref{eq:SI_qp_qzs} do not satisfy Eq.~\ref{eq:dhg}. Consider the following four cases for $|q_z^s|$ and $s$. 

\noindent \textbf{Case 1:} Suppose $|q_z^s| \geq \frac{2}{\lambda}\sin(\alpha_i)$ and $s \in \{1,4\}$. This inequality implies that
\begin{equation}\label{eq:case1ineq_candle}
\begin{aligned}
&\sin(\alpha_f)+ \sin(\alpha_i) \geq 2\sin(\alpha_i)\\
\Leftrightarrow &\sin(\alpha_f) \geq \sin(\alpha_i)\\
\Leftrightarrow &\alpha_f \geq \alpha_i\\
\Rightarrow &|\cos(\theta_f)\cos(\alpha_f)-\cos(\alpha_i)| = \cos(\alpha_i)-\cos(\theta_f)\cos(\alpha_f).
\end{aligned}
\end{equation}

Using Eqs.~\ref{eq:case1theta} and \ref{eq:case1ineq_candle}, we can derive the following upper bound:
\begin{equation}
\begin{aligned}
    &\left(\lambda|\qp| - \cos(\alpha_i)\right)^2+ \left(\lambda|q_z^s| - \sin(\alpha_i)\right)^2\\
    &= \left(\lambda|\qp| - \cos(\alpha_i)\right)^2+ \left(\sin(\alpha_i) + \sin(\alpha_f) -  \sin(\alpha_i)\right)^2\\
    &= \lambda^2|\qp|^2 + \cos^2(\alpha_i)-2\lambda|\qp|\cos(\alpha_i) + \sin^2(\alpha_f)\\
    &= \cos^2(\alpha_f) + \cos^2(\alpha_i) - 2\cos(\theta_f)\cos(\alpha_f)\cos(\alpha_i) + \cos^2(\alpha_i)-2\lambda|\qp|\cos(\alpha_i) + \sin^2(\alpha_f)\\
    &= 1 + 2\cos^2(\alpha_i) - 2\cos(\theta_f)\cos(\alpha_f)\cos(\alpha_i) -2\lambda|\qp|\cos(\alpha_i)\\
        &\leq 1 + 2\cos^2(\alpha_i) - 2\cos(\theta_f)\cos(\alpha_f)\cos(\alpha_i) - 2|\cos(\theta_f)\cos(\alpha_f)-\cos(\alpha_i)|\cos(\alpha_i)\\
        &= 1 + 2\cos^2(\alpha_i) - 2\cos(\theta_f)\cos(\alpha_f)\cos(\alpha_i) + 2(\cos(\theta_f)\cos(\alpha_f)-\cos(\alpha_i))\cos(\alpha_i)\\
    &= 1.
\end{aligned}
\end{equation}

\noindent \textbf{Case 2:} Suppose $|q_z^s| \geq \frac{2}{\lambda}\sin(\alpha_i)$ and $s \in \{2,3\}$. This inequality implies that

\begin{equation}\label{eq:case2alpha}
\begin{aligned}
&|\sin(\alpha_f)- \sin(\alpha_i)| \geq 2\sin(\alpha_i)\\
\Leftrightarrow &\max\left(\sin(\alpha_f)-\sin(\alpha_i), \sin(\alpha_i)-\sin(\alpha_f)\right)  \geq \sin(\alpha_i)\\
\Leftrightarrow &\max\left(\sin(\alpha_f)-2\sin(\alpha_i), -\sin(\alpha_f)\right) \geq 0\\
\Leftrightarrow &\sin(\alpha_f)-2\sin(\alpha_i) \geq 0\\
\Leftrightarrow &\sin(\alpha_f) \geq 2\sin(\alpha_i)\\
\Rightarrow &\alpha_f \geq \alpha_i.\\
\end{aligned}
\end{equation}

Using Eqs.~\ref{eq:case2alpha} and \ref{eq:case1theta}, we can derive the following upper bound:
\begin{align}
    &\left(\lambda|\qp| - \cos(\alpha_i)\right)^2+ \left(\lambda|q_z^s| - \sin(\alpha_i)\right)^2\notag\\
    &= \left(\lambda|\qp| - \cos(\alpha_i)\right)^2+ \left(\sin(\alpha_f) - \sin(\alpha_i) -  \sin(\alpha_i)\right)^2\notag\\
    &= \lambda^2|\qp|^2 + \cos^2(\alpha_i)-2\lambda|\qp|\cos(\alpha_i) + \sin^2(\alpha_f)+4\sin^2(\alpha_i)-4\sin(\alpha_f)\sin(\alpha_i)\notag\\
    &\leq \lambda^2|\qp|^2 + \cos^2(\alpha_i)-2\lambda|\qp|\cos(\alpha_i) + \sin^2(\alpha_f)+4\sin^2(\alpha_i)-4\sin(\alpha_i)\sin(\alpha_i)\notag\\
    &= \cos^2(\alpha_f) + \cos^2(\alpha_i) - 2\cos(\theta_f)\cos(\alpha_f)\cos(\alpha_i) + \cos^2(\alpha_i)-2\lambda|\qp|\cos(\alpha_i) + \sin^2(\alpha_f)\notag\\
    &= 1 + 2\cos^2(\alpha_i) - 2\cos(\theta_f)\cos(\alpha_f)\cos(\alpha_i) -2\lambda|\qp|\cos(\alpha_i)\notag\\
    &\leq 1 + 2\cos^2(\alpha_i) - 2\cos(\theta_f)\cos(\alpha_f)\cos(\alpha_i) + 2(\cos(\theta_f)\cos(\alpha_f)-\cos(\alpha_i))\cos(\alpha_i)\notag\\
    &= 1.
\end{align}

\noindent \textbf{Case 3:} Suppose $|q_z^s| < \frac{2}{\lambda}\sin(\alpha_i)$ and $s \in \{1,4\}$. This inequality implies that
\begin{equation}
\begin{aligned}
&\sin(\alpha_f)+ \sin(\alpha_i) < 2\sin(\alpha_i)\\
\Leftrightarrow &\sin(\alpha_f) < \sin(\alpha_i)\\
\Leftrightarrow &\alpha_f < \alpha_i.\\
\end{aligned}
\end{equation}
We can then use Eq.~\ref{eq:lambdaqp} to derive the following lower bound:
\begin{equation}
\begin{aligned}
    &\left(\lambda|\qp| + \cos(\alpha_i)\right)^2+ \left(\lambda|q_z^s| - \sin(\alpha_i)\right)^2\\
        &\geq \left(|\cos(\alpha_f)-\cos(\alpha_i)| + \cos(\alpha_i)\right)^2+ \left(\sin(\alpha_f) + \sin(\alpha_i) -  \sin(\alpha_i)\right)^2\\
    &= \left(\cos(\alpha_f)-\cos(\alpha_i) + \cos(\alpha_i)\right)^2+ \left(\sin(\alpha_f) + \sin(\alpha_i) -  \sin(\alpha_i)\right)^2\\
    &= \cos^2(\alpha_f) + \sin^2(\alpha_f)\\
    &= 1.
\end{aligned}
\end{equation}

\noindent \textbf{Case 4:} Suppose $|q_z^s| < \frac{2}{\lambda}\sin(\alpha_i)$ and $s \in \{2,3\}$.

If $\alpha_f < \alpha_i $, then we can derive the following lower bound:
\begin{equation}
\begin{aligned}
    &\left(\lambda|\qp| + \cos(\alpha_i)\right)^2+ \left(\lambda|q_z^s| - \sin(\alpha_i)\right)^2\\
    &\geq \left(\cos(\alpha_f)-\cos(\alpha_i) + \cos(\alpha_i)\right)^2+ \left(\sin(\alpha_i) - \sin(\alpha_f) -  \sin(\alpha_i)\right)^2\\
    &= \cos^2(\alpha_f) + \sin^2(\alpha_f)\\
    &= 1.
\end{aligned}
\end{equation}
If $\alpha_f \geq \alpha_i$, then we can show that the same lower bound holds:
\begin{equation}
\begin{aligned}
    &\left(\lambda|\qp| + \cos(\alpha_i)\right)^2+ \left(\lambda|q_z^s| - \sin(\alpha_i)\right)^2\\
    &\geq \left(\cos(\alpha_i)-\cos(\alpha_f) + \cos(\alpha_i)\right)^2+ \left(\sin(\alpha_f) - \sin(\alpha_i) -  \sin(\alpha_i)\right)^2\\
    &= \cos^2(\alpha_f) + 4\cos^2(\alpha_i) - 4\cos(\alpha_f)\cos(\alpha_i) + \sin^2(\alpha_f) + 4\sin^2(\alpha_i) - 4\sin(\alpha_f)\sin(\alpha_i)\\
    &= 5 - 4\cos(\alpha_f-\alpha_i)\\
    &\geq 1.
\end{aligned}
\end{equation}

\noindent \textbf{Case Summary:} Combining the above four cases and multiplying by $1/\lambda^2$, we have that the coordinates $(\qp,\qz)$ of each of the four DWBA terms satisfy
\begin{equation}
\begin{aligned}
    &\left(|\qp| + \frac{1}{\lambda}\cos(\alpha_i)\right)^2\hspace*{-.5em} + \left(|\qz| - \frac{1}{\lambda}\sin(\alpha_i)\right)^2\hspace*{-.5em} \geq \left(\frac{1}{\lambda}\right)^2\hspace*{-.5em},\quad \text{ if }|\qz| < \frac{2}{\lambda}\sin(\alpha_i),\\
    &\left(|\qp| - \frac{1}{\lambda}\cos(\alpha_i)\right)^2\hspace*{-.5em} + \left(|\qz| - \frac{1}{\lambda}\sin(\alpha_i)\right)^2\hspace*{-.5em} \leq \left(\frac{1}{\lambda}\right)^2\hspace*{-.5em},\quad \text{ if } |\qz| \geq \frac{2}{\lambda}\sin(\alpha_i).
    \end{aligned}
\end{equation}
Therefore, the DWBA is unable to sample the specimen's Fourier transform in the region
\begin{equation}
\begin{aligned}
    &\left(|\qp| + \frac{1}{\lambda}\cos(\alpha_i)\right)^2\hspace*{-.5em} + \left(|\qz| - \frac{1}{\lambda}\sin(\alpha_i)\right)^2\hspace*{-.5em} < \left(\frac{1}{\lambda}\right)^2\hspace*{-.5em},\quad \text{ if }|\qz| < \frac{2}{\lambda}\sin(\alpha_i),\\
    &\left(|\qp| - \frac{1}{\lambda}\cos(\alpha_i)\right)^2\hspace*{-.5em} + \left(|\qz| - \frac{1}{\lambda}\sin(\alpha_i)\right)^2\hspace*{-.5em} > \left(\frac{1}{\lambda}\right)^2\hspace*{-.5em},\quad \text{ if } |\qz| \geq \frac{2}{\lambda}\sin(\alpha_i).
    \end{aligned}
\end{equation}

Note that these bounds are sharp, with equality attained at $\tf=0$: for $\qz \geq \sai/\lambda$ by term $s=1$ with $0\leq\af<\pi/2$; for $0\leq\qz \leq \sai/\lambda$ by term $s=3$ with $0\leq\af\leq\ai$; for $-\sai/\lambda\leq\qz \leq 0$ by term $s=2$ with $0\leq\af\leq\ai$; and for $\qz \leq -\sai/\lambda$ by term $s=4$ with $0\leq\af<\pi/2$.

\section{S5.\hspace{.5em} Reconstruction from CSSI Data Sampled on a Standard Uniform Grid versus a Nonuniform Staggered Grid}

\vspace{-1em}

Here we present results highlighting the advantages of reducing CSSI intensity data to a nonuniform staggered grid rather than a standard uniform grid. As discussed in Section~\ref{sec:Cartesian} of the main text, the staggered grid ensures that the Fourier transform of the specimen density is sampled at $\qz = 0$ by at least one of the four terms in the DWBA for some intensity values on this grid, provided the incident-angle conditions in Section~\ref{sec:incident} are met. Since the $\qz=0$ region of the Fourier domain governs the bulk shape of the reconstruction, constraining this region is vital for overall stability. Moreover, by employing the acceleration techniques in Section~\ref{sec:CartAccel}, reconstruction from a CSSI intensity sampled on the staggered grid requires negligible additional computational cost compared to a uniform grid.

To illustrate the instabilities introduced by using a uniform grid and how the staggered grid overcomes them, we perform reconstructions of the porous medium from complete CSSI intensity volumes simulated on uniform and nonuniform staggered grids for incident angles $\alpha_i = 0.257,\,0.260,\,0.263$. Although the angular differences are small, they have a significant effect on reconstruction stability with a uniform grid. This effect arises from the distances between the shift $\qzi$ in the DWBA and the nearest $\qz$ values sampled by the uniform grids. For the above incident angles, these distances are $0.5\, h_z$, $0.29\, h_z$, and $0.0\, h_z$, respectively, where $h_z$ is the voxel length in the $z$ dimension. The lowest $\qz$ values that can be sampled by the DWBA for the intensities on the uniform grids correspond to these distances. 

\begin{figure}[h!] 
 \centering
 \includegraphics[width = .77\textwidth]{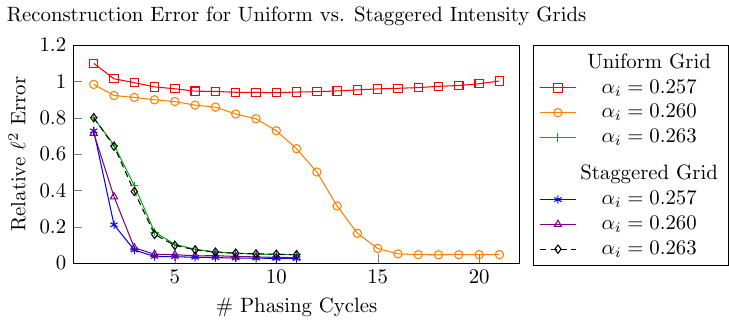}
\vspace{-1.5em}
\caption{Relative errors in porous medium reconstructions using a standard uniform intensity grid versus a nonuniform staggered intensity grid for three different incident angles, shown as a function of iterative phasing cycle. For each reconstruction, the rightmost point on the line indicates the error after refinement, which is performed after the last phasing cycle.}\label{fig:stag_error}
\end{figure}

We used $N_{\rm cycle} = 10$ phasing cycles, except for the uniform grid examples with $\alpha_i = 0.257$ and $\alpha_i=0.260$, where we used $N_{\rm cycle} = 20$ to accommodate their slower convergence. For each reconstruction, we plot the relative $\ell^2$ error as a function of phasing cycle in Fig.~\ref{fig:stag_error}.

The reconstructions using intensity data sampled on a uniform grid become increasingly unstable as the distance between $\qzi$ and the nearest $\qz$ on the grid increases. While the $\alpha_i=0.260$ uniform grid reconstruction ultimately reaches an accurate result, it converges much more slowly than those using staggered grids. For $\alpha_i=0.257$, which has the largest distance between $\qzi$ and the grid, the error for the uniform grid case remains high throughout the reconstruction. However, for $\alpha_i=0.263$, where $\qzi$ is an exact multiple of $h_z$, the uniform grid case shows excellent convergence, comparable to the results on a staggered grid. 

Unlike the uniform grid results, the reconstructions using intensity data sampled on staggered grids achieve excellent convergence and accuracy across all three incident angles, demonstrating that the staggered grids overcome the instabilities and parameter sensitivities associated with the uniform grid. Also, note that there is a small improvement in convergence between the $\alpha_i = 0.257$ and $\alpha_i = 0.260$ reconstructions over the $\alpha_i = 0.263$ reconstruction, since the staggered grids associated with the first two angles oversample the intensity more.

\vspace*{-.5em}

\newpage

\section{S6.\hspace{.5em} Example of a Local Minimum Induced by Partial 180-Degree Rotation Invariance for Unitary Reflection Coefficients}

\vspace{-1em}

\begin{figure}[H] 
 \centering
\includegraphics[width = .85\textwidth]{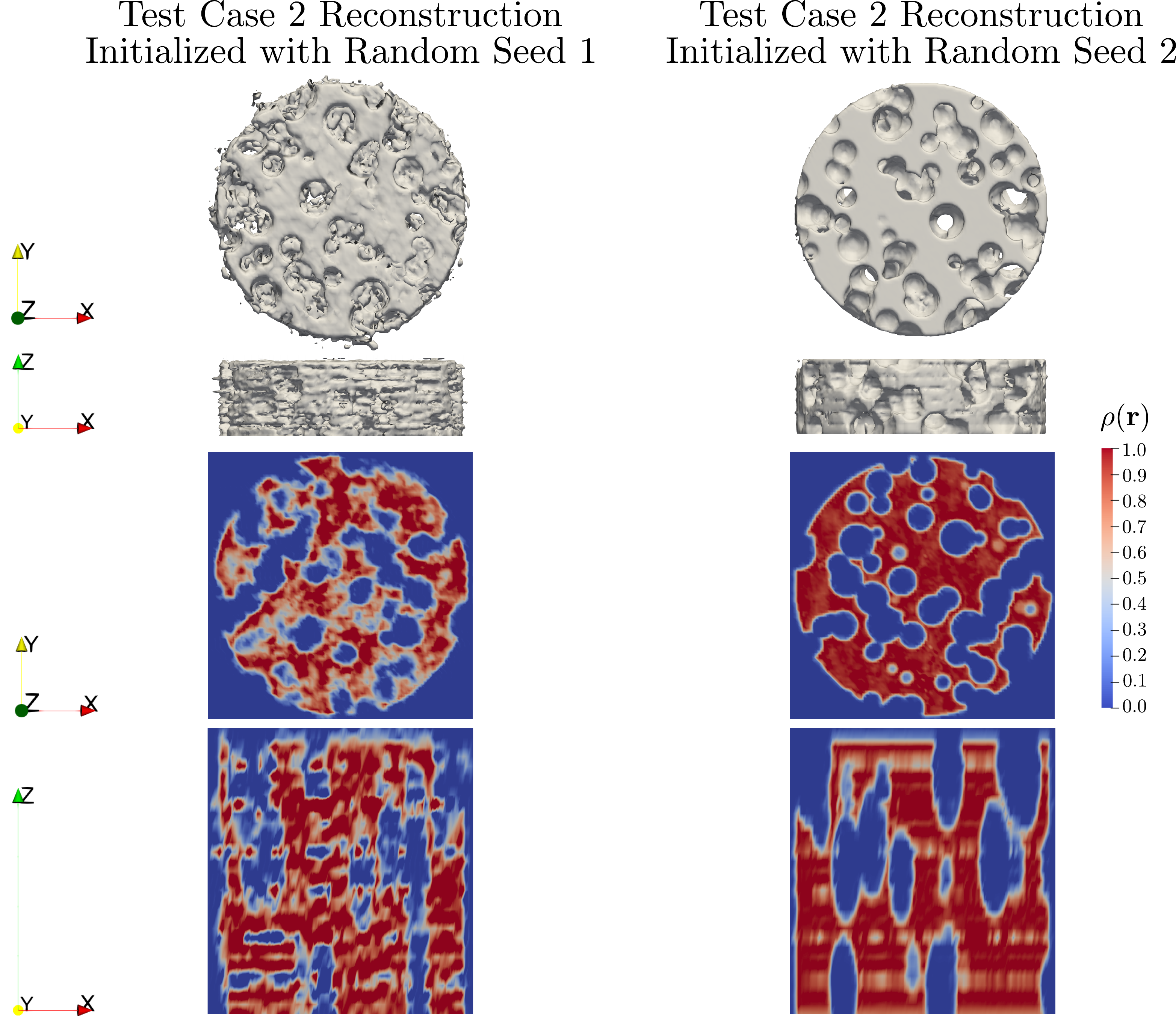}
\caption{Reconstructions from test case 2 using different random seeds. The reconstruction on the left gets trapped in a local minimum induced by the partial 180-degree rotation invariance of the CSSI intensity for unitary reflection coefficients, whereas the reconstruction on the right recovers the correct overall structure apart from artifacts induced by the interference function.}\label{fig:local_minimum}
\end{figure}

In Fig.~\ref{fig:local_minimum}, we show an example of a local minimum induced by the partial 180-degree rotation invariance of the CSSI intensity for unitary reflection coefficients, discussed in Section~\ref{sec:invariance} of the main text. The reconstructions are from test case 2 in Section~\ref{sec:results_volume}, using two different random seeds to generate the starting densities. The first reconstruction becomes trapped in a local minimum, resembling a low-resolution 180-degree rotation of the true object, but failing to capture high-resolution details. In contrast, the second reconstruction avoids this local minimum, capturing the correct orientation and finer details, except for the horizontal-line artifacts induced by the zeros of the interference function. 

\section{S7.\hspace{.5em} Validation of NUFFT-Based DWBA Forward Model Against BornAgain}

\vspace{-1em}
 
We verified the consistency of our physical units and DWBA formulation by comparing our NUFFT-based forward model against the BornAgain software \cite{pospelov2020bornagain}. As an example, we computed CSSI diffraction patterns for a cylinder of radius 320 nm and height 200 nm, using the same experimental parameters as the $\alpha_i = 0.26^{\circ}$ porous-medium example described in Table~\ref{table:rot_series_params} of Section~\ref{sec:rot_series} in the main text.

\begin{figure}[H] 
 \centering
\includegraphics[width = .7\textwidth]{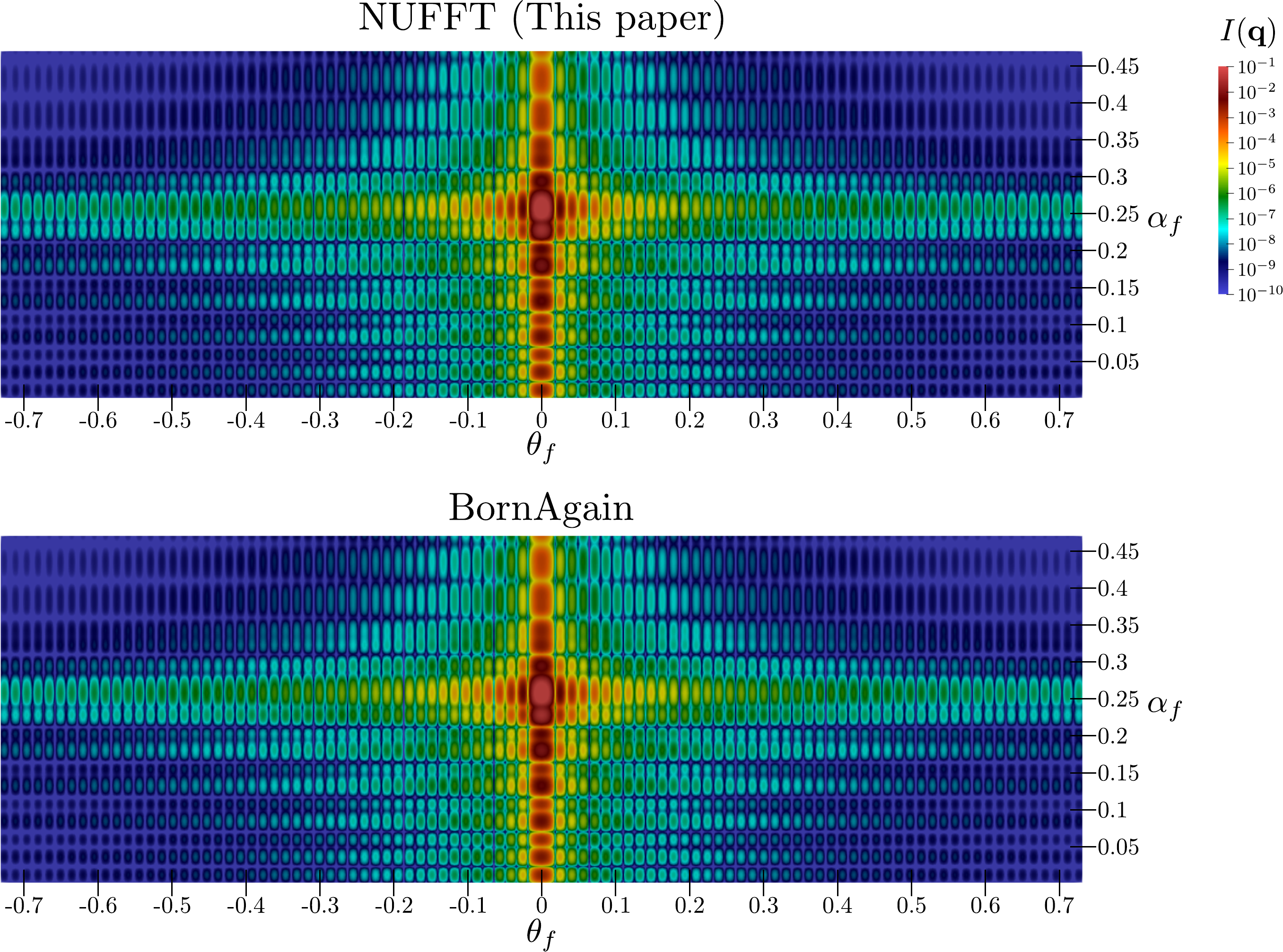}

\vspace{-1em}

\caption{Simulated CSSI diffraction patterns of a cylinder (radius 320 nm, height 200 nm) using our NUFFT-based forward model (top) and BornAgain (bottom). The relative $\ell^2$ error between the two patterns is approximately $0.005$.}\label{fig:Bornagain}
\end{figure}

While BornAgain evaluates the DWBA using exact analytical expressions for geometric objects, our NUFFT-based approach represents the object as a discretized density and therefore cannot exactly reproduce a geometric shape. To minimize the resulting discretization error, we used a large 3D grid of dimensions $1021 \times 1021 \times 301$ for the NUFFT-based simulation. The BornAgain output was rescaled by a constant factor to account for differing normalization conventions. The resulting diffraction patterns are shown in Fig.~\ref{fig:Bornagain}, with a relative $\ell^2$ error of approximately $0.005$, confirming agreement of our physical units and formulation. We note that, unlike BornAgain, our approach is capable of modeling general 3D densities and is not confined to geometric shapes.

\bibliography{references}